\documentclass[12pt, draftclsnofoot, onecolumn]{IEEEtran}
\usepackage{graphicx}
\usepackage{amsmath,bbm,amssymb,amsfonts,amstext,amsopn,sansmath}
\usepackage{cite}
\usepackage{balance}
\usepackage{url}
\usepackage{epsfig}
\usepackage{setspace}
\usepackage{stmaryrd}
\usepackage{psfrag}	
\usepackage{multirow}
\usepackage{float}
\usepackage[process=auto]{pstool}
\usepackage{etoolbox}
\usepackage{algorithm}
\usepackage{algorithmic}
\allowdisplaybreaks
\usepackage[nolist]{acronym}
\usepackage{makecell}

\newcommand{\x}{\mathsf{x}}
\newcommand{\y}{\mathsf{y}}
\newcommand{\z}{\mathsf{z}}
\newcommand{\jj}{\mathsf{j}}
\newcommand{\e}{\mathsf{e}}
\newcommand{\Herm}{\mathsf{H}}

\def\by{\mathbf{y}}

\def\bH{\mathbf{H}}

\def\bA{\mathbf{A}}

\def\bq{\mathbf{q}}
\def\bh{\mathbf{h}}

\def\bx{\mathbf{x}}

\def\bG{\mathbf{G}}
\def\bD{\mathbf{D}}
\def\ba{\mathbf{a}}
\def\bd{\mathbf{d}}
\def\bz{\mathbf{z}}

\def\bPsi{\boldsymbol{\Psi}}
\def\bOmega{\boldsymbol{\Omega}}
\def\bSigma{\boldsymbol{\Sigma}}

\def\Cset{\mathbb{C}}

\newcommand{\diag}{\mathrm{diag}}

\newtheorem{lem}{Lemma}
\newtheorem{remk}{Remark}

\newtheorem{prop}{Proposition}
\newtheorem{corol}{Corollary}

\newtoggle{OneColumn}
\toggletrue{OneColumn}

\newtoggle{arXiv}
\toggletrue{arXiv}

\newcommand{\BLUE}{\color{black}}
\newcommand{\RED}{\color{black}}

\EndPreamble
\begin{document}
\title{Physics-based Modeling and Scalable Optimization of Large Intelligent Reflecting Surfaces}

\author{Marzieh Najafi, \textit{Student Member, IEEE}, Vahid Jamali, \textit{Member, IEEE}, \\  Robert Schober, \textit{Fellow, IEEE}, and H. Vincent Poor, \textit{Life Fellow, IEEE}
	\thanks{This paper will be presented in part at the Asilomar Conference on Signals, Systems, and Computers 2020 \cite{najafi2020asilomar}.}
	\thanks{M. Najafi, V. Jamali, and R. Schober are with the Institute for Digital Communications at Friedrich-Alexander University Erlangen-N\"urnberg (FAU) (e-mail:
	marzieh.najafi@fau.de;	vahid.jamali@fau.de; 	robert.schober@fau.de).}
	\thanks{H. Vincent Poor is with the Department of Electrical Engineering, Princeton University, Princeton, NJ 08544 USA (e-mail: poor@princeton.edu).}
}

\maketitle

\vspace{-1.5cm}

\begin{abstract}
Intelligent reflecting surfaces (IRSs) have the potential to transform wireless communication channels into smart reconfigurable propagation environments. To realize this new paradigm, the passive IRSs have to be large, especially for communication in far-field scenarios,  so that they can compensate for the large end-to-end path-loss, which is caused by the multiplication of the individual path-losses of the transmitter-to-IRS and IRS-to-receiver channels. However, optimizing a large number of sub-wavelength IRS elements imposes a significant challenge for online transmission. To address this issue, in this paper, we develop a physics-based model and a scalable optimization framework for large IRSs. The basic idea is to partition the IRS unit cells into several subsets, referred to as tiles, model the impact of each tile on the wireless channel, and then optimize each tile in two stages, namely an offline design stage  and an online optimization stage. For physics-based modeling, we borrow concepts from the radar literature, model each tile as an \textit{anomalous} reflector, and derive its impact on the wireless channel for a given phase shift by solving the corresponding integral equations for the electric and magnetic vector fields. In the offline design stage, the IRS unit cells of each tile are jointly designed for the support of different transmission modes, where each transmission mode effectively corresponds to a given configuration of the phase shifts that the unit cells of the tile apply to an impinging electromagnetic wave.  In the online optimization stage, the best transmission mode of each tile is selected such that a desired quality-of-service (QoS) criterion is maximized.  We consider an exemplary downlink system and study the minimization of the base station (BS) transmit power subject to QoS constraints for the users. Since the resulting mixed-integer programming problem for joint optimization of the BS beamforming vectors and the tile transmission modes is non-convex, we derive two efficient suboptimal solutions, which are based on alternating optimization and  a greedy approach, respectively. We show that the proposed modeling and optimization framework can be used to efficiently optimize large IRSs comprising thousands of unit cells. 
\end{abstract}

\acresetall
\section{Introduction}

Smart wireless environments are a newly emerging concept in wireless communications where intelligent reflecting surfaces (IRSs) are deployed to influence the propagation characteristics of the wireless channel \cite{di2019smart,liaskos2019novel,di2020smart,huang2020holographic}. IRSs consist of a large number of programmable sub-wavelength elements, so-called unit cells or meta atoms, that can change the properties of an impinging electromagnetic (EM) wave while reflecting it. For instance,  a properly designed unit cell phase distribution across the surface enables the IRS to alter the direction of the wavefront of the reflected wave, thereby realizing the generalized Snell's law \cite{kaina2014shaping,zhu2013active}. Moreover,  since the unit cells are passive and cost-efficient, it is expected that \textit{large} IRSs comprising hundreds if not thousands of unit cells can be manufactured and deployed to assist a link \cite{wu2019intelligent,jamali2018scalable,tang2019wireless}. Proof-of-concept implementations confirming the benefits of IRSs have been reported in \cite{yang2016programmable,tang2019wireless,dai2020reconfigurable}.

\subsection{State-of-the-Art Approach to IRS Optimization}

Let us consider an IRS with $Q$ unit cells. A widely-adopted model for IRSs in the literature is to assume that each unit cell individually acts as a diffusive scatterer which is able to change the phase of the impinging EM wave, i.e., $E_{r,q}=E_{t,q}\e^{\jj\beta_q}$, where $E_{t,q}$ and $E_{r,q}$ denote the incident and reflected scalar electric fields of the $q$-th unit cell, respectively, and $\beta_q\in\mathcal{B}\subset[0,2\pi)$ is the phase change applied by the $q$-th unit cell taking values from set $\mathcal{B}$  \cite{wu2019intelligent,jamali2018scalable,yu2020robust}. The IRS is configured by optimizing the $\beta_q$ which leads to non-convex optimization problems due to the unit-modulus constraint imposed by $|\e^{\jj\beta_q}|=1$, see \cite{wu2019intelligent,jamali2018scalable,jamali2019intelligent,yu2020robust,huang2019reconfigurable,pan2020multicell,guo2019weighted,wang2019intelligent} for different approaches to cope with this non-convex constraint. Unfortunately, these optimization methods are not scalable for large IRSs as the number of optimization variables becomes unmanageably large. For example, assuming a unit-cell spacing of half a wavelength,  a $1$~m-by-$1$~m IRS would comprise $Q=1100$ unit cells for a carrier frequency of $5$~GHz.  Therefore, the direct optimization of $\beta_q,\forall q$, may not be a feasible approach for the online design of large~IRSs. 

Moreover, as the physics-based models in \cite{ozdogan2019intelligent,bjornson2020power,di2020analytical} suggest, the path-loss of the end-to-end IRS-assisted links is significant for \textit{far-field} scenarios and indeed  a \textit{very large} IRS is needed to overcome it in practice. To see this, let $\rho_d$, $\rho_t$, and $\rho_r$ denote  the transmitter-to-receiver, transmitter-to-IRS, and IRS-to-receiver distances, respectively. {\BLUE Thereby, the smallest area of an IRS for which the free-space path-loss of the end-to-end  IRS-assisted link can be equal to the path-loss of the unobstructed direct link is $ \frac{\lambda\rho_t\rho_r}{\rho_d}$, where $\lambda$ denotes the wavelength, see Corollaries~\ref{Corol:MinArea} and \ref{Corol:MinQ} for details.} For example, assuming $\rho_d=200$~m, $\rho_t=\rho_r=100$~m, and a unit-cell spacing of half a wavelength, $Q\propto\frac{200}{\lambda}\approx 3300$ and $6600$  unit cells are needed for carrier frequencies of $5$ and $10$~GHz, respectively. {\BLUE In fact, the need for such large IRSs is confirmed by recent experimental systems \cite{tang2019wireless} which feature large IRSs with $1700$ and $10200$ unit cells for a carrier frequency of $10.5$~GHz.} We note that the main reason why many existing works, e.g., \cite{wu2019intelligent,yu2020robust,bai2020latency}, report performance gains for IRSs with much smaller numbers of unit cells (e.g., on the order of  tens of unit cells) is the use of link models which do not include all the losses for far-field scenarios and/or an exceedingly weak direct link, see also \cite{ozdogan2019intelligent}.  
Other existing works consider a small number of IRS elements but assume that each element is able to provide an effective \textit{constant gain} that is \textit{implicitly} incorporated in the channel gain \cite{guo2019weighted,wang2019intelligent,cao2019delay}. Such elements, which we refer to as tiles, can be realized and are larger than the wavelength or are comprised of several sub-wavelength unit cells \cite{guo2019weighted}. However, the gain provided by these tiles is not constant and depends on several factors such as the angle of incident, the angle of reflection/observation, and the polarization of the impinging wave. A physics-based model that accounts for these factors is essential to properly model the signals and interference at the receivers of a wireless system. In particular, a large gain implies highly directive tiles which means that a receiver that is not located close to the main lobe of the radiation pattern of these tiles will receive negligible power. Hence, any further attempt to constructively combine the negligible powers arriving from different tiles by configuring the phase shifts of the tiles will lead to an inefficient solution. 
 Therefore, the  development of a physics-based end-to-end channel model for \textit{large IRSs}  that accounts for all relevant effects and allows for the \textit{scalable optimization} of the IRS configuration is of utmost importance. 

\subsection{Proposed Modeling and Optimization Framework for IRS}

In this paper, we develop a scalable optimization framework for large IRSs, which is rooted in a physics-based IRS channel model and provides a tunable tradeoff between performance and complexity.  The basic idea is to partition the IRS unit cells into $N\ll Q$ tiles. We then model the impact of a tile on the wireless channel for a given phase-shift configuration of the unit cells of the tile. Using this model, we optimize the tile configuration in two stages, namely an offline design  and an online optimization~stage.

\textbf{Physics-based Model for IRS-Assisted Channel:}   Borrowing an analogy from the radar literature, we model each tile as an \textit{anomalous} reflector. Hence, for a given configuration of the unit cell phase shifts, which we refer to as a transmission mode, and assuming a far-field scenario,  a tile is characterized  by a response function $g(\boldsymbol{\Psi}_t,\boldsymbol{\Psi}_r)$. The tile response function is essentially a generalization of the radar cross section (RCS) of an object \cite{balanis2012advanced} and accounts for both the amplitude and phase of the reflected EM wave. More specifically, the tile response function determines how a plane wave impinging from direction $\boldsymbol{\Psi}_t$  with a given polarization is reflected in direction $\boldsymbol{\Psi}_r$ for a given phase-shift configuration of the unit cells of the tile.  We derive  $g(\boldsymbol{\Psi}_t,\boldsymbol{\Psi}_r)$ by solving the corresponding integral equations for the electric and magnetic vector fields \cite{balanis2012advanced}. In particular, we first derive  $g(\boldsymbol{\Psi}_t,\boldsymbol{\Psi}_r)$  for ideal continuous tiles with programmable surface impedance. Exploiting this result, we then derive $g(\boldsymbol{\Psi}_t,\boldsymbol{\Psi}_r)$ for discrete tiles consisting of sub-wavelength unit cells. We show that discrete tiles with a unit-cell spacing of less than $\lambda/2$ can accurately approximate continuous tiles. {\BLUE Furthermore, as an example, we consider  a downlink communication system comprising a base station (BS), an IRS, and multiple users  and model the end-to-end channels of the users  as functions of the response functions $g(\boldsymbol{\Psi}_t,\boldsymbol{\Psi}_r)$ of all tiles of the IRS where each tile can be configured to support a number of different transmission modes.} 

We note that the physics-based model derived in this paper generalizes the models in \cite{ozdogan2019intelligent,bjornson2020power,di2020analytical} which provide interesting insights, but were obtained under more restrictive assumptions. For instance, in  \cite{ozdogan2019intelligent}, the scatter field was characterized for a specific polarization and the angles of the impinging and reflected waves were in the same plane, see Remark~\ref{Remk:Special}. However, in practice, several waves may impinge on the same IRS from different directions and with different polarizations and will be redirected in different directions.  In \cite{bjornson2020power}, the authors studied the power scaling laws for asymptotically large IRSs; however, similar to \cite{ozdogan2019intelligent}, general incident and reflection directions were not considered. Furthermore, in \cite{di2020analytical}, the authors modeled an IRS in a \textit{two-dimensional} system using the \textit{scalar} theory of diffraction and the Huygens-Fresnel principle. {\BLUE Moreover, an empirical path-loss model was derived in \cite{tang2019wireless} and its accuracy was verified via experimental data. However, this model employs an empirical response for the IRS unit cells and was derived based on the scalar electric/magnetic field assumption \cite{salem2014manipulating}, i.e., it does not explicitly account for wave polarization.} In contrast, in this paper, we consider a three-dimensional system and by analyzing the electric and magnetic vector fields, we characterize the reflected wave for all observation angles when a plane wave with arbitrary incident angle and arbitrary polarization impinges on the IRS. {\BLUE We note that while in \cite{ozdogan2019intelligent,bjornson2020power,di2020analytical,tang2019wireless} path-loss models for IRS-assisted links were proposed,  scalable end-to-end channel models that can be exploited for efficient optimization of large IRSs were not provided.}

\textbf{Offline Design:}  The unit cells of each tile are jointly designed offline to provide $M$ different transmission modes, i.e., $M$ different phase-shift configurations. In general, the capabilities of the transmission modes may range from changing the direction of the wavefront to splitting the incoming wave into multiple directions or scattering the wave. However, in this paper, we focus on  transmission modes that enable the generalized Snell's law, i.e.,  change the direction of the reflected wavefront \cite{kaina2014shaping,zhu2013active,najafi2019intelligent}. To this end, we design an offline codebook of transmission modes, which is the product of three component codebooks, namely two \textit{reflection} codebooks that jointly enable the tile to reflect an incident wave along desired elevation and azimuth angles and a \textit{wavefront phase} codebook which enables constructive/destructive superposition of the waves that arrive from different tiles at the receivers.  

\textbf{Online Optimization:} In the online optimization stage, the objective is to select for each fading realization the best transmission mode for each tile  such that a desired performance criterion is maximized. The resulting optimization problems are in the form of \textit{mixed integer programming} \cite{lee2011mixed,ng2014robust,ghanem2019resource}. We emphasize that, for a given channel realization, it may not be necessary to use all transmission modes in the offline codebook for online optimization. In fact, to reduce the complexity of online optimization, we propose to select a subset $\mathcal{M}$ of the transmission modes, e.g., those modes that yield a non-negligible signal power at the intended receivers. This reduces the search space for online optimization from $|\mathcal{B}|^Q$ for the naive approach which directly optimizes the phase shifts of the unit cells, $\beta_q\in\mathcal{B},\forall q$, to $|\mathcal{M}|^N$, where the number of modes $|\mathcal{M}|$ and the number of tiles $N$ are design parameters that can be selected by the system designer to trade performance for complexity. 


For concreteness, we focus on a downlink communication system  where a multiple-antenna BS serves multiple single-antenna users. We assume that the direct link exists but may be severely shadowed/blocked and hence an IRS is deployed to improve the communication. For this system, we jointly optimize the IRS and the precoder at the BS to minimize the transmit power of the BS while guaranteeing a minimum quality-of-service (QoS) for the users \cite{wu2019intelligent,cao2019delay}. Since the formulated problem is non-convex, we develop two efficient suboptimal solutions. The first solution exploits well-known techniques such as alternating optimization (AO) and semidefinite programming, whereas the second solution is greedy and configures the tiles and the precoder in an iterative manner while ensuring that the transmit power is reduced in each iteration. Finally, we use computer simulations to quantify the impact of the number of unit cells, number of tiles, blockage of the direct link, and the position of the IRS on the system performance. In particular, we show that the proposed optimization framework can be used to configure IRSs with thousands of unit cells by grouping them into a small number of tiles (on the order of a few tens) without significant performance degradation.

\begin{table*}\vspace{-0.5cm}
	\BLUE
	\caption{Symbols used frequently throughout the paper and their Definitions.\vspace{-0.3cm}} \label{Table:Symbols}
	\begin{center}
		\scalebox{0.6}{
			\begin{tabular}{|| c | c ||  c | c ||}\hline
				Symbol & Definition & Symbol & Definition \\ \hline 
				$L_\x^{\rm tot},L_\y^{\rm tot}$ & Length of the IRS along $x$- and $y$-axes, respectively
				&
				$N$ & Number of tiles
				\\ \hline 
				$L_\x,L_\y$ & Length of each tile along $x$- and $y$-axes, respectively
				&
				$M$ & Number of transmission modes
				\\ \hline 
				$Q$ & Total number of IRS unit cells 
				& 
				$\mathcal{B}_\x,\mathcal{B}_\y, \mathcal{B}_0$ & \makecell{Reflection codebooks along $x$- and $y$-axes \\ and wavefront phase codebook, respectively}
				\\ \hline 
				$Q_\x,Q_\y$ & \makecell{Number of tile unit cells along \\ $x$- and $y$-axes, respectively}
				&
				$\rho_d,\rho_t,\rho_r$ &\makecell{Distances of the BS-to-user, BS-to-IRS, and \\ IRS-to-user links, respectively}
				\\ \hline 
				$d_\x,d_\y$ & Unit cell spacing along $x$- and $y$-axes, respectively
				&
				$\theta,\phi,\varphi$ & Elevation, azimuth, and polarization angles, respectively
				\\ \hline 
				$L_{\rm uc}$ & Length and width of square unit cells
				&
				$g(\boldsymbol{\Psi}_t,\boldsymbol{\Psi}_r)$ &\makecell{Tile response function along  reflection direction $\boldsymbol{\Psi}_r$ \\ for an incident wave from direction  $\boldsymbol{\Psi}_t$}
				\\ \hline 
				$\eta$ & Characteristic impedance
				&
				$g^{\mathsf{c}},g^{\mathsf{d}},g_{(n_\x,n_\y)}$ &\makecell{Response function for continuous tile, discrete tile, and \\ $(n_\x,n_\y)$-th unit cell of the discrete tile, respectively}
				\\ \hline 
				$\Gamma,\tau,\beta$& \makecell{Reflection coefficient, its amplitude, and\\  its phase, respectively}
				&
				$g_{n,m},g_{m}$ &\makecell{Tile response functions of the $n$-th tile and the reference \\ tile $n=0$ for the $m$-th transmission mode, respectively}
				\\ \hline 
				$\lambda,\kappa$ & Wavelength and wavenumber, respectively
				& 
				$g_{\rm uc},\bar{g}_{\rm uc}$ &\makecell{Unit-cell factor (in meter) and unit-less unit-cell factor}
				\\ \hline 
			\end{tabular}
		} 
	\end{center}
	
\end{table*}

 The remainder of this paper is organized as follows. In Section~\ref{Sec:Model}, we present the proposed physics-based model for continuous and discrete tiles as well as the end-to-end channel model for IRS-assisted wireless systems. In Section~\ref{Sec:Optimization}, we first design an offline codebook for the transmission modes of the tiles. Subsequently, we formulate and solve an online optimization problem. Simulation results are presented in Section~\ref{Sec:Sim}, and  conclusions are drawn in Section~\ref{Sec:Conclusions}. {\BLUE A list of symbols that are used  throughout the paper is provided in Table~\ref{Table:Symbols}.}

\section{End-to-End Channel Model for IRS-Assisted Wireless Systems}\label{Sec:Model}

{\BLUE In this section, we first present the considered IRS structure and the proposed division into multiple tiles. Subsequently, we focus on one IRS tile and derive the tile response functions for both continuous and discrete tiles for a given transmission mode, incident wave (characterized by its incident and polarization angles), and reflection direction. Exploiting these results, we then develop the proposed end-to-end channel model for IRS-assisted wireless systems, which accounts for the impact of all IRS tiles, the transmission modes of all tiles, and the incident, reflection, and polarization angles.}

\subsection{IRS Structure}\label{Sec:IRSstruct}

\begin{figure}[t]\vspace{-0.5cm}
	\centering
	\includegraphics[width=0.8\textwidth]{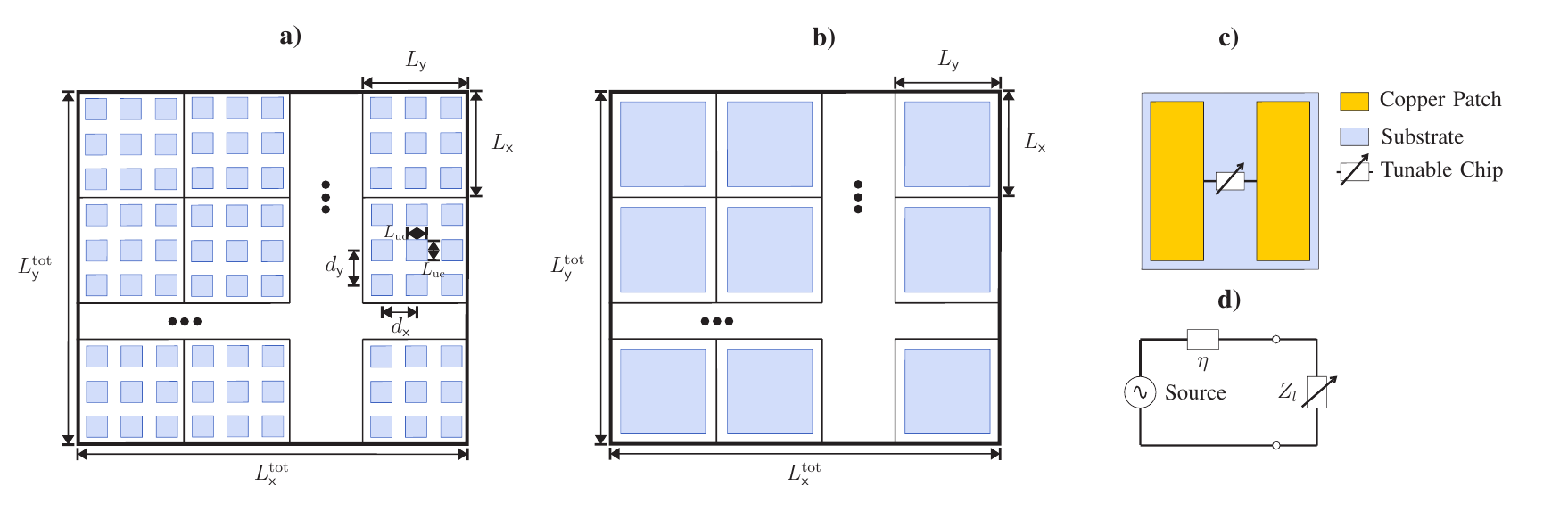}\vspace{-0.3cm}
	\caption{Schematic illustration of planar IRS of size $L_{\x}^{\mathrm{tot}}\times L_{\y}^{\mathrm{tot}}$ partitioned into tiles of size $L_\x \times L_\y$. a) Each tile is composed of square unit cells of size $L_{\rm uc}\times L_{\rm uc}$ which are spaced by $d_\x$ and $d_\y$ along the $x$ and $y$ directions, respectively. b) Each tile may be approximated as a continuous surface. c) Schematic illustration of tunable unit cells. d) Equivalent circuit model for the unit cells, see \cite{liu2019intelligent,zhu2013active,abeywickrama2020intelligent} for various implementations of the tunable chip.\vspace{-0.5cm}} 
	\label{Fig:IRS}
\end{figure}

We consider a large rectangular IRS of size $L_{\x}^{\mathrm{tot}}\times L_{\y}^{\mathrm{tot}}$ placed in the $x-y$ plane where $L_\x^{\mathrm{tot}},L_\y^{\mathrm{tot}}\gg \lambda$, see Fig.~\ref{Fig:IRS}.  The IRS is composed of many sub-wavelength unit cells (also known as meta atoms) of size $L_{\rm uc}\times L_{\rm uc}$ that are able to change the properties of an impinging EM wave when reflecting it. Typically, each unit cell contains programmable components  (such as tunable varactor diodes or switchable positive-intrinsic-negative (PIN) diodes) that can change the reflection coefficient of the surface, which we denote by $\Gamma$, see Fig.~\ref{Fig:IRS} c). 

We assume that the IRS is partitioned into tiles of size $L_\x \times L_\y$. For notational simplicity, let us assume that $L_{\x}^{\mathrm{tot}}/L_{\x}$ and $L_{\y}^{\mathrm{tot}}/L_{\y}$ are integers and in total, there are $N=L_{\x}^{\mathrm{tot}}L_{\y}^{\mathrm{tot}}/(L_{\x}L_{\y})$ tiles. Each tile consists of several programmable sub-wavelength unit cells. Here, assuming a unit-cell spacing of $d_\x$ and $d_\y$ along the $x$ and $y$ axes, respectively, the total number of unit cells of the IRS is given by $Q=NQ_\x Q_\y$, where $Q_\x=L_\x/d_\x$ and $Q_\y=L_\y/d_\y$. When $d_\x=d_\y\approx L_{\rm uc}\ll \lambda$ and $L_\x,L_\y\gg \lambda$, the collection of all unit cells on one tile acts as a continuous programmable surface \cite{estakhri2016wave,asadchy2016perfect}, cf. Fig.~\ref{Fig:IRS} b). In this paper, an ideal tile that acts as a continuous programmable surface is referred to as a \textit{continuous tile}. In contrast, a practical tile that comprises a discrete number of unit cells  is referred to as a \textit{discrete tile}. We use the notion of continuous tiles in Section~\ref{Sec:Cont} to rigorously analyze the reflected EM field. 

\begin{figure}[t]\vspace{-0.5cm}
	\centering
	\includegraphics[width=0.4\textwidth]{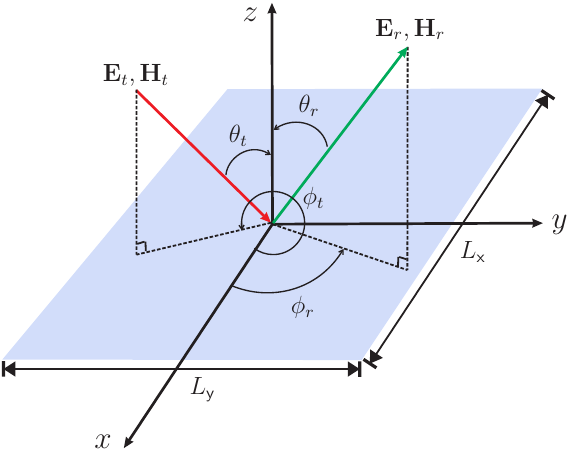}\vspace{-0.3cm}
	\caption{Uniform plane wave impinges on a rectangular conducting tile with incident angle $\boldsymbol{\Psi}_t=(\theta_t,\phi_t,\varphi_t)$ and is reflected with a desired reflection angle $\boldsymbol{\Psi}_r=(\theta_r,\phi_r)$. \vspace{-0.5cm}}
	\label{Fig:CoordinatesContinuous}
\end{figure}

{\BLUE\subsection{Tile Response Function}}
We focus on the far-field scenario where the curvature of the wavefront originating from a distant source on the tile can be neglected. Therefore, the incident EM wave can be modeled as a plane wave impinging on the tile and is characterized by $\boldsymbol{\Psi}_t=(\theta_t,\phi_t,\varphi_t)$, see Fig.~\ref{Fig:CoordinatesContinuous}. Here, $\theta_t$ and $\phi_t$ denote the elevation and azimuth angles of the direction from which the incident wave impinges on the IRS, respectively, and $\varphi_t$ determines the polarization of the incident wave. The unit cells of the tile act as secondary sources and reflect the incident EM wave. The signal observed at a receiver in the far field of the tile can be characterized by the overall complex tile response function $g(\boldsymbol{\Psi}_t,\boldsymbol{\Psi}_r)$, where  $\boldsymbol{\Psi}_r=(\theta_r,\phi_r)$ denotes the reflection angle  at which the receiver is located\footnote{The polarization of the reflected wave with respect to the orientation of the receive antenna determines the amount of received power induced at a receiver. However, since the orientations of the receivers are random, the power loss due to the mismatch between the receive antenna orientation and the reflected wave polarization can be absorbed into the IRS-to-receiver channel gains. Hence, the polarization of the reflected wave is not explicitly included in $\boldsymbol{\Psi}_r$, although it can be derived from our analysis, see Appendix~\ref{App:PropContinuous}.}.  The {\RED square} of the tile response function, $|g(\boldsymbol{\Psi}_t,\boldsymbol{\Psi}_r)|^2$ {\RED (in meter$^2$)}, is  referred to as the RCS of an object \cite[p.~584]{balanis2012advanced}. Here, to be able to model the superposition of multiple waves at a receiver in the far field, we generalize the concept of RCS to also include the phase information, i.e., 
\begin{IEEEeqnarray}{rll}\label{Eq:Gdef}
	g(\boldsymbol{\Psi}_t,\boldsymbol{\Psi}_r)	= \underset{\rho_r\to\infty}{\mathrm{lim}} \,\, \sqrt{4\pi \rho_r^2} \e^{-\jj \kappa\rho_r} \frac{E_r(\boldsymbol{\Psi}_r)}{E_t(\boldsymbol{\Psi}_t)},
\end{IEEEeqnarray}
where $\kappa=\frac{2\pi}{\lambda}$ is the wave number, $E_t(\boldsymbol{\Psi}_t)$ is a phasor denoting the complex amplitude of the incident electric field impinging from angle $\boldsymbol{\Psi}_t$ on the tile center (i.e., $(x,y)=(0,0)$),  and $E_r(\boldsymbol{\Psi}_r)$ is a phasor denoting the complex amplitude of  the reflected electric field in direction $\boldsymbol{\Psi}_r$ and at distance $\rho_r$ from the tile center.  

{\BLUE
	Before further analyzing the tile response function $g$, we first derive the free-space path-loss of an IRS-assisted link, denoted by $\mathrm{PL}_{\mathrm{IRS}}$, for a given $g$. In principle, the free-space path-loss $\mathrm{PL}$ characterizes the received power $P_{\rm rx}$ for a given  transmit power $P_{\rm tx}$ as  $\frac{P_{\rm rx}}{P_{\rm tx}} = D_{\rm tx}D_{\rm rx} \mathrm{PL}$, where $D_{\rm tx}$ and $D_{\rm rx}$ denote the directivities of the transmit and receive antennas, respectively \cite{balanis2005antenna}. Let $\rho_d$, $\rho_t$, and $\rho_r$ denote  the transmitter-to-receiver, transmitter-to-IRS, and IRS-to-receiver distances, respectively. For an unobstructed direct link, the free-space path-loss is given by $\mathrm{PL}_d=\left(\frac{\lambda}{4\pi\rho_d}\right)^2$ \cite{balanis2005antenna}. The following lemma provides $\mathrm{PL}_{\mathrm{IRS}}$ in terms of $g$.
\begin{lem}\label{Lem:Pathloss}
	 The free-space path-loss of an IRS-assisted link with tile response function $g$ is given~by
	\begin{IEEEeqnarray}{lll}\label{Eq:Pathloss}
		\mathrm{PL}_{\mathrm{IRS}} = \frac{4\pi |g|^2}{\lambda^2}\mathrm{PL}_t \mathrm{PL}_r,
	\end{IEEEeqnarray}
	where  $\mathrm{PL}_t=\left(\frac{\lambda}{4\pi\rho_t}\right)^2$ and $\mathrm{PL}_r=\left(\frac{\lambda}{4\pi\rho_r}\right)^2$ are the free-space path-losses of the transmitter-to-IRS and IRS-to-receiver links, respectively.
	\end{lem}
	\begin{IEEEproof}
		The proof is provided in Appendix~\ref{App:LemPathloss}.
	\end{IEEEproof}

		Lemma~\ref{Lem:Pathloss} suggests that the end-to-end path-loss of an IRS-assisted link can be decomposed into three parts, namely the path-loss of the transmitter-to-IRS link, $\mathrm{PL}_t$, the path-loss of the IRS-to-receiver link, $\mathrm{PL}_r$,  and the term $\frac{4\pi |g|^2}{\lambda^2}$ accounting for the impact of the IRS. We note that this observation is in agreement with the path-loss model  in \cite{tang2019wireless} except that in \cite{tang2019wireless}, an empirical model was used for $g$, whereas  in this paper,  we derive $g$ from physical principles in Sections~\ref{Sec:Cont} and \ref{Sec:Discrete}.
}

In the following, we derive $g(\boldsymbol{\Psi}_t,\boldsymbol{\Psi}_r)$ for both continuous and discrete tiles for a given transmission mode. Whenever necessary, we use the superscripts $\mathsf{c}$ and $\mathsf{d}$ for the tile response function to distinguish between continuous and discrete tiles, respectively.


\subsection{Continuous Tiles}\label{Sec:Cont}

In order to study the impact of a continuous tile on an impinging EM wave, we first explicitly define the incident electric and magnetic fields. Here, we assume the following incident fields with arbitrary polarization and incident angle \cite[Ch.~11]{balanis2012advanced}\footnote{In \cite{balanis2012advanced},  the incident wave is always assumed to be in the $y-z$ plane and the polarization is either transverse electric $x$ (TE$^x$) or transverse magnetic $x$ (TM$^x$) to facilitate the analysis. While these assumptions are without loss of generality when analyzing one impinging wave, in this paper, we deal with scenarios where multiple waves may arrive from different angles and with different polarizations, and hence, these simplifying assumptions cannot simultaneously hold for all impinging waves. Therefore, we generalize the formulation of the electric and magnetic fields in \cite{balanis2012advanced} to arbitrary incident angles.}   
\begin{IEEEeqnarray}{lll}\label{Eq:IncidentFields}
	\mathbf{E}_t(\boldsymbol{\Psi}_t) = E_0  \e^{\jj \kappa \mathbf{a}_t \cdot (\mathbf{e}_\x x+ \mathbf{e}_\y y + \mathbf{e}_\z z)} \mathbf{a}_E
	\quad\text{and}\quad
	\mathbf{H}_t(\boldsymbol{\Psi}_t) =  \frac{E_0}{\eta}  \e^{\jj \kappa \mathbf{a}_t \cdot (\mathbf{e}_\x x+ \mathbf{e}_\y y + \mathbf{e}_\z z)} \mathbf{a}_H,
\end{IEEEeqnarray}
where $E_0$ is the magnitude of the incident electric field, $\mathbf{a}\cdot\mathbf{b}$ denotes the inner product of vectors $\mathbf{a}$ and $\mathbf{b}$, $\eta=\sqrt{\frac{\mu}{\epsilon}}$ is the characteristic impedance, $\mu$ is the magnetic permeability, and $\epsilon$ is the electric permittivity. Moreover, $\mathbf{e}_\x$, $\mathbf{e}_\y$, and $\mathbf{e}_\z$ denote the unit vectors in the $x$, $y$, and $z$ directions, respectively, and $\mathbf{a}_E$ and $\mathbf{a}_H$ denote the directions of the electric field and the magnetic field, respectively,  and {\RED $\mathbf{a}_t$ is the direction from which the wave impinges on the IRS.} Note that $\mathbf{a}_E$, $\mathbf{a}_H$, and $\mathbf{a}_t$  are mutually orthogonal. 
In spherical coordinates, the  direction of the incident wave is defined as
\begin{IEEEeqnarray}{lll}\label{Eq:Incident}
\mathbf{a}_t=(\sin(\theta_t)\cos(\phi_t),\sin(\theta_t)\sin(\phi_t),\cos(\theta_t))\triangleq(A_\x(\boldsymbol{\Psi}_t),A_\y(\boldsymbol{\Psi}_t),A_\z(\boldsymbol{\Psi}_t)).
\end{IEEEeqnarray}
Note that $\mathbf{a}_E$ and $\mathbf{a}_H$ lie in the plane orthogonal to $\mathbf{a}_t$. Let $(H_\x,H_\y)$ denote the components of the magnetic field in the $x-y$ plane. Defining $\varphi_t=\tan^{-1}(\frac{H_\y}{H_\x})$, which determines  the polarization of the wave, we obtain $\mathbf{a}_E$ and $\mathbf{a}_H$  as
\begin{IEEEeqnarray}{lll} 
	 \mathbf{a}_E= \mathbf{a}_t\times \mathbf{a}_H \quad \text{and} \quad \mathbf{a}_H = b \big(c(\boldsymbol{\Psi}_t)\cos(\varphi_t),c(\boldsymbol{\Psi}_t)\sin(\varphi_t),\sqrt{1-c^2(\boldsymbol{\Psi}_t)}\big), 	 
\end{IEEEeqnarray}
where  $c(\boldsymbol{\Psi}_t)=\frac{A_\z(\boldsymbol{\Psi}_t)}{\sqrt{A_{\x,\y}^2(\boldsymbol{\Psi}_t)+A_\z^2(\boldsymbol{\Psi}_t)}}$,  $A_{\x,\y}(\boldsymbol{\Psi}_t) = \cos(\varphi_t) A_\x(\boldsymbol{\Psi}_t) + \sin(\varphi_t) A_\y(\boldsymbol{\Psi}_t)$,  $b=\mathrm{sign}\big(\frac{H_\x}{c(\boldsymbol{\Psi}_t)\cos(\varphi_t)}\big)$, $\mathrm{sign}(\cdot)$ denotes the sign of a real number, and $\mathbf{a}\times\mathbf{b}$ denotes the cross product between vectors $\mathbf{a}$ and $\mathbf{b}$. Note that the reference complex amplitude of the incident electric field in \eqref{Eq:Gdef} can be obtained from the electric vector field in  \eqref{Eq:IncidentFields} as $E_t(\boldsymbol{\Psi}_t)=E_0  \e^{\jj \kappa \mathbf{a}_t \cdot (\mathbf{e}_\x x+ \mathbf{e}_\y y + \mathbf{e}_\z z)} \big|_{(x,y,z)=(0,0,0)}=E_0$.


We assume that the surface impedance is suitably designed to realize reflection coefficient $\Gamma=\tau\e^{\jj\beta(x,y)}$, where $\beta(x,y)$ is the phase shift applied at point $(x,y)$ on the tile  and {\RED $\tau$ is the amplitude of the reflection coefficient which is assumed to be constant across the tile.  We note that the exact value of $\tau$  depends on the specific realization of the unit cells. For example, $\tau$ can be chosen to enforce the passiveness of the surface (see Remark~\ref{Remk:Passive}) and to account for potential power losses in the unit cells \cite{estakhri2016wave,asadchy2016perfect}.}
The tangential components of the scattered electric and magnetic fields are given by $\mathbf{E}_r=\Gamma\mathbf{E}_t$ and $\mathbf{H}_r=-\Gamma\mathbf{H}_t$, respectively. In order to determine the scattered fields, we employ the Electromagnetic Equivalence Theorem \cite[Ch.~7]{balanis2012advanced} and assume that \textit{only} scattered fields $(\mathbf{E}_r,\mathbf{H}_r)$  exist in the  environment and that the IRS is replaced by a perfectly magnetically conducting (PMC) surface. To compensate for the field discontinuity across the boundaries, an electric current $\mathbf{J}_r^{\mathrm{pmc}} = \mathbf{n}\times\mathbf{H}_r\big|_{z=0}$ should be introduced on the surface, where $\mathbf{n}$ is the normal vector of the surface \cite[Ch.~7, eq. (7.42)]{balanis2012advanced}. Next, using the Image Theory for large flat surfaces \cite[Ch.~7.4]{balanis2012advanced}, an equivalent \textit{obstacle-free}  system is obtained by removing the PMC and replacing $\mathbf{J}_r^{\mathrm{pmc}}$ with an equivalent electric current $\mathbf{J}_r$  obtained as  \cite[Ch.~7.4, 7.8]{balanis2012advanced}
\begin{IEEEeqnarray}{lll}\label{Eq:Current_r}
	\mathbf{J}_r &=2\mathbf{J}_r^{\mathrm{pmc}} = 2\mathbf{n}\times\mathbf{H}_r\big|_{z=0}
	= -2\Gamma\mathbf{n}\times\mathbf{H}_t\big|_{z=0}\nonumber\\
	&\overset{(a)}{=}  -2\tau\frac{E_0}{\eta} \e^{\jj \kappa[A_\x(\boldsymbol{\Psi}_t)x+A_\y(\boldsymbol{\Psi}_t)y]+\jj\beta(x,y)} \mathbf{e}_\z \times \mathbf{a}_H  \overset{(b)}{=}    \e^{\jj \kappa[A_\x(\boldsymbol{\Psi}_t)x+A_\y(\boldsymbol{\Psi}_t)y]+\jj\beta(x,y)} (J_\x\mathbf{e}_\x + J_\y\mathbf{e}_\y),
\end{IEEEeqnarray}
where equality $(a)$ follows from the assumption that the incident EM wave is a plane wave, and for equality $(b)$, we used the definitions
$J_\x = 2 \frac{E_0}{\eta}\tau c(\boldsymbol{\Psi}_t)\sin(\varphi_t)$ and $J_\y = -2 \frac{E_0}{\eta}\tau c(\boldsymbol{\Psi}_t)\cos(\varphi_t)$. The magnitude of the equivalent electric current is given by $\|\mathbf{J}_r\| = \sqrt{J_\x^2+J_\y^2} = 2 \tau\frac{E_0}{\eta} c(\boldsymbol{\Psi}_t)$, where depending on the incident angle and the polarization, we have $c(\boldsymbol{\Psi}_t)\in[\cos(\theta_t),1]$.

The reflected electric and magnetic fields induced by electric current $\mathbf{J}_r$ in an obstacle-free environment are found as follows \cite[Ch.~6]{balanis2012advanced} 
\begin{IEEEeqnarray}{lll}\label{Eq:HEV}
	\mathbf{E}_r = \frac{1}{\jj\omega\epsilon}\nabla \times \mathbf{H}_r\quad\text{and}\quad \mathbf{H}_r  = \frac{1}{\mu} \nabla \times \mathbf{V}, 
\end{IEEEeqnarray}
where $\nabla \times$ is the curl operator, $\omega=\kappa/\sqrt{\mu\epsilon}$, and $\mathbf{V}$ is an auxiliary vector potential, which assuming a far-field scenario is given by
\begin{IEEEeqnarray}{lll}\label{Eq:EintegralFF}
	\mathbf{V}(\boldsymbol{\Psi}_r) =  \frac{\mu \e^{-\jj \kappa\rho_r}}{4\pi \rho_r} \int_{x=-L_\x/2}^{L_\x/2}\int_{y=-L_\y/2}^{L_\y/2} \mathbf{J}_r(x,y) \e^{\jj \kappa \sqrt{x^2+y^2} \cos(\alpha)} \mathrm{d}x\mathrm{d}y.
\end{IEEEeqnarray}
Here, $\alpha$ is the angle between the vector specified by angle $\boldsymbol{\Psi}_r$ and the line that connects $(x,y)$ with the origin. 

In order to solve the integral equation in \eqref{Eq:EintegralFF}, we have to assume a given phase-shift profile, $\beta(x,y)$, for the tile surface, i.e., a transmission mode. One criterion to design a tile transmission mode is to facilitate reflection in a certain direction, i.e., the generalized Snell's law~\cite{kaina2014shaping,zhu2013active}. In particular, we design the tile to reflect an EM wave impinging from direction $\boldsymbol{\Psi}_t^*$ towards direction $\boldsymbol{\Psi}_r^*$ and analyze the tile response function $g(\boldsymbol{\Psi}_t,\boldsymbol{\Psi}_r)$ caused by the corresponding phase-shift profile for an EM wave impinging from an arbitrary direction $\boldsymbol{\Psi}_t$ (including $\boldsymbol{\Psi}_t^*$) and observed at an arbitrary observation angle $\boldsymbol{\Psi}_r$ (including $\boldsymbol{\Psi}_r^*$). For ease of presentation, let us define the amplitude and phase of $g(\boldsymbol{\Psi}_t,\boldsymbol{\Psi}_r)$ as $g_{||}(\boldsymbol{\Psi}_t,\boldsymbol{\Psi}_r)$ and $g_{\angle}(\boldsymbol{\Psi}_t,\boldsymbol{\Psi}_r)$, respectively, up to a sign, i.e., $g_{||}(\boldsymbol{\Psi}_t,\boldsymbol{\Psi}_r)=\pm |g(\boldsymbol{\Psi}_t,\boldsymbol{\Psi}_r)|$ and $g_{\angle}(\boldsymbol{\Psi}_t,\boldsymbol{\Psi}_r)=\angle g(\boldsymbol{\Psi}_t,\boldsymbol{\Psi}_r)\pm \pi$ such that $g(\boldsymbol{\Psi}_t,\boldsymbol{\Psi}_r)=g_{||}(\boldsymbol{\Psi}_t,\boldsymbol{\Psi}_r)\e^{\jj g_{\angle}(\boldsymbol{\Psi}_t,\boldsymbol{\Psi}_r)}$. Here, $|\cdot|$ and $\angle$ denote the absolute value and phase of a complex number, respectively. Moreover, let $A_i(\boldsymbol{\Psi}_t,\boldsymbol{\Psi}_r)=A_i(\boldsymbol{\Psi}_t)+A_i(\boldsymbol{\Psi}_r),i\in\{\x,\y\}$. 

\begin{prop}\label{Prop:Continuous}
	For given $\boldsymbol{\Psi}_t^*$ and $\boldsymbol{\Psi}_r^*$, let us impose the following linear phase-shift function $\beta(x,y)=\beta(x)+\beta(y)$ with
	\begin{IEEEeqnarray}{rll}\label{Eq:BetaContinuous}
		\beta(x) = -\kappa A_\x(\boldsymbol{\Psi}_t^*,\boldsymbol{\Psi}_r^*)x+\frac{\beta_0}{2}
		\quad\text{and}\quad
		\beta(y) = -\kappa A_\y(\boldsymbol{\Psi}_t^*,\boldsymbol{\Psi}_r^*)y+\frac{\beta_0}{2}. 
	\end{IEEEeqnarray}
	Then, the amplitude of the corresponding tile response function for an EM wave impinging from an arbitrary direction $\boldsymbol{\Psi}_t$  and observed at arbitrary reflection direction $\boldsymbol{\Psi}_r$ is obtained as  
	\begin{IEEEeqnarray}{lll}\label{Eq:AmplContinuous}
		g_{||}^{\mathsf{c}}(\boldsymbol{\Psi}_t,\boldsymbol{\Psi}_r) = \frac{\sqrt{4\pi} \tau L_\x L_\y}{\lambda}\widetilde{g}(\boldsymbol{\Psi}_t,\boldsymbol{\Psi}_r) \nonumber \\
		\times \mathrm{sinc}\left(\frac{\kappa L_\x [A_\x(\boldsymbol{\Psi}_t,\boldsymbol{\Psi}_r) - A_\x(\boldsymbol{\Psi}_t^*,\boldsymbol{\Psi}_r^*)]}{2}\right) \mathrm{sinc}\left(\frac{\kappa L_\y [A_\y(\boldsymbol{\Psi}_t,\boldsymbol{\Psi}_r) - A_\y(\boldsymbol{\Psi}_t^*,\boldsymbol{\Psi}_r^*)]}{2}\right),
	\end{IEEEeqnarray}
	where $\mathrm{sinc}(x)=\sin(x)/x$ and
	\begin{IEEEeqnarray}{lll} \label{Eq:Gtilde}
		\widetilde{g}(\boldsymbol{\Psi}_t,\boldsymbol{\Psi}_r) = c(\boldsymbol{\Psi}_t)
		\left\|
		\begin{bmatrix}
		\cos(\varphi_t)\cos(\theta_r)\sin(\phi_r)-\sin(\varphi_t)\cos(\theta_r)\cos(\phi_r)\\
		 \sin(\varphi_t)\sin(\phi_r)+\cos(\varphi_t)\cos(\phi_r)
		\end{bmatrix}
	\right\|_2.
	\end{IEEEeqnarray}
	The phase of the tile response function is obtained as  $g_{\angle}^{\mathsf{c}}(\boldsymbol{\Psi}_t,\boldsymbol{\Psi}_r) = \frac{\pi}{2}+\beta_0$. 
\end{prop}
\begin{IEEEproof}
	The proof is provided in Appendix~\ref{App:PropContinuous}.
\end{IEEEproof}

\begin{corol}\label{Corol:MaxContinuous}
	Assuming large $L_\x,L_\y\gg \lambda$ and $\boldsymbol{\Psi}_t^*=\boldsymbol{\Psi}_t$, the maximum value of $|g(\boldsymbol{\Psi}_t,\boldsymbol{\Psi}_r)|$ is observed at $\boldsymbol{\Psi}_r=\boldsymbol{\Psi}_r^*$ and is given by
	\begin{IEEEeqnarray}{rll}\label{Eq:max_cont}
		\underset{\boldsymbol{\Psi}_r\to\boldsymbol{\Psi}_r^*}{\lim}|g^{\mathsf{c}}(\boldsymbol{\Psi}_t,\boldsymbol{\Psi}_r)| 
		= \frac{\sqrt{4\pi} \tau L_\x L_\y}{\lambda}\widetilde{g}(\boldsymbol{\Psi}_t^*,\boldsymbol{\Psi}_r^*) 
		\overset{(a)}{\leq} \frac{\sqrt{4\pi} L_\x L_\y}{\lambda},
	\end{IEEEeqnarray}
	where $(a)$ holds with equality when $\theta_t=\theta_r=0$ and $\tau=1$.
\end{corol}
\begin{IEEEproof}
	For $L_\x,L_\y\gg \lambda$, the sinc functions in \eqref{Eq:AmplContinuous} decay fast when $|A_i(\boldsymbol{\Psi}_t,\boldsymbol{\Psi}_r) - A_i(\boldsymbol{\Psi}_t^*,\boldsymbol{\Psi}_r^*)|>0,\,\,i\in\{\x,\y\}$, whereas $\widetilde{g}(\boldsymbol{\Psi}_t,\boldsymbol{\Psi}_r)\neq0,\,\,\forall \boldsymbol{\Psi}_t,\boldsymbol{\Psi}_r$, is not a function of $L_\x,L_\y$ and does not vary as quickly as the sinc functions. Hence,    $|g^{\mathsf{c}}(\boldsymbol{\Psi}_t,\boldsymbol{\Psi}_r)| $  attains its maximum value when the sinc functions have their maximum values, i.e., $A_i(\boldsymbol{\Psi}_t,\boldsymbol{\Psi}_r) = A_i(\boldsymbol{\Psi}_t^*,\boldsymbol{\Psi}_r^*)$ or equivalently at $\boldsymbol{\Psi}_r=\boldsymbol{\Psi}_r^*$ for  $\boldsymbol{\Psi}_t=\boldsymbol{\Psi}_t^*$. Moreover, for normal incident, we have $c(\boldsymbol{\Psi}_t)=1$ and for normal reflection,  the norm term in \eqref{Eq:Gtilde} is  one. This implies that  $\widetilde{g}(\boldsymbol{\Psi}_t^*,\boldsymbol{\Psi}_r^*)$  attains its maximum value, i.e., one, which leads to the inequality in \eqref{Eq:max_cont}.
\end{IEEEproof}
	
\begin{figure}[t]\vspace{-0.5cm}
	\centering
	\includegraphics[width=1\textwidth]{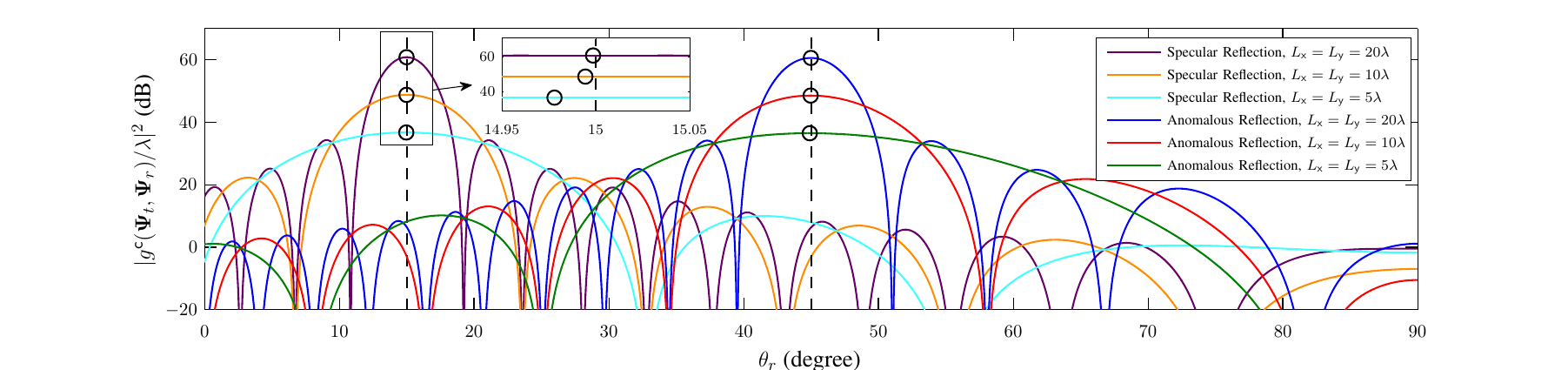}
	\vspace{-0.7cm}
	\caption{\BLUE Amplitude of response function, $|g^{\mathsf{c}}(\boldsymbol{\Psi}_t,\boldsymbol{\Psi}_r)/\lambda|^2$, in dB vs. $\theta_r$ for an $L_\x\times L_\y$ continuous tile, $(\theta_t,\phi_t,\varphi_t)=(15^{\circ},225^{\circ},22.5^{\circ})$, $\phi_r=45^{\circ}$, and $\tau=0.8$. We assume
		$(\theta_t^*,\phi_t^*)=(15^{\circ},225^{\circ})$ and $(\theta_r^*,\phi_r^*)=(15^{\circ},45^{\circ})$ for specular reflection and $(\theta_t^*,\phi_t^*)=(15^{\circ},225^{\circ})$ and $(\theta_r^*,\phi_r^*)=(45^{\circ},45^{\circ})$  for anomalous reflection.\vspace{-0.5cm}} 
	\label{Fig:FuncGcontinuous}
\end{figure}

{\BLUE
	\begin{corol}\label{Corol:MinArea}\BLUE
		The smallest area of a continuous IRS, $A_{\mathrm{IRS}}$, for which 
 the free-space path-loss of the IRS-assisted link is equal to the path-loss of the unobstructed direct link is given by
\begin{IEEEeqnarray}{lll}
 A_{\mathrm{req}}
	= 	\frac{\lambda\rho_t \rho_r }{\rho_d}.
\end{IEEEeqnarray}
\end{corol}
\begin{IEEEproof}
\BLUE	In order to maximize the power received at the intended receiver, we assume that all IRS tiles act as  a single tile which leads to the   effective IRS response function $g_{\mathrm{IRS}}$ whose value is upper bounded by $\frac{\sqrt{4\pi} N L_\x L_\y}{\lambda}$, cf. Corollary~\ref{Corol:MaxContinuous}. Substituting $g_{\mathrm{IRS}}$ in $\mathrm{PL}_{\mathrm{IRS}}$ given in   \eqref{Eq:Pathloss}, we obtain that  $A_{\mathrm{IRS}}=N L_\x L_\y$  has to be equal to $A_{\mathrm{req}}=\frac{\lambda\rho_t \rho_r }{\rho_d}$ for $\textrm{PL}_{\mathrm{IRS}}=\textrm{PL}_{d}$ to hold.
\end{IEEEproof}}

Fig.~\ref{Fig:FuncGcontinuous} shows the amplitude of the tile response function for both specular and anomalous reflections for several tile sizes. We now highlight some insights from Proposition~\ref{Prop:Continuous}, Corollaries~\ref{Corol:MaxContinuous}, \ref{Corol:MinArea}, and Fig.~\ref{Fig:FuncGcontinuous}: 

\textit{i)} Eq.~\eqref{Eq:AmplContinuous}  and Fig.~\ref{Fig:FuncGcontinuous} show that $|g^{\mathsf{c}}(\boldsymbol{\Psi}_t,\boldsymbol{\Psi}_r)|$ becomes narrower as $L_\x$ and $L_\y$ increase. However, even for large tiles of size $L_\x=L_\y=20\lambda$, the $10$-dB beamwidth\footnote{Here, the $10$-dB beamwidth is defined as the maximum range of $\theta_r$ around $\theta_r^*$ for which $|g(\boldsymbol{\Psi}_t,\boldsymbol{\Psi}_r)|^2$ is not more than $10$~dB smaller than its maximum value.} is around $6$ degree which can cause significant interference to unintended receivers in far-field scenarios. 

\textit{ii)} Fig.~\ref{Fig:FuncGcontinuous} suggests that for a given tile size, the peak and beamwidth of $|g^{\mathsf{c}}(\boldsymbol{\Psi}_t,\boldsymbol{\Psi}_r)|$ for anomalous reflection are in general different from those for specular reflection.

\textit{iii)} Let $\boldsymbol{\Psi}_r^\star={\mathrm{argmax}}_{\boldsymbol{\Psi}_r}\,|g^{\mathsf{c}}(\boldsymbol{\Psi}_t,\boldsymbol{\Psi}_r)|$. One expects to obtain $\boldsymbol{\Psi}_r^\star=(\theta_t,\pi+\phi_t)=\boldsymbol{\Psi}_r^*$ for specular reflection; however, this is generally not true, see also \cite[p.~596]{balanis2012advanced}. The reason for this behavior is that although the sinc functions in \eqref{Eq:AmplContinuous}  attain their maximum values at $\boldsymbol{\Psi}_r=\boldsymbol{\Psi}_r^*$, this is not necessarily the case for $\widetilde{g}(\boldsymbol{\Psi}_t,\boldsymbol{\Psi}_r)$ which can cause $\boldsymbol{\Psi}_r^\star$ to deviate from $\boldsymbol{\Psi}_r^*$. For instance, Fig.~\ref{Fig:FuncGcontinuous} shows that for $L_\x=L_\y=5\lambda$, the maximum value of $|g^{\mathsf{c}}(\boldsymbol{\Psi}_t,\boldsymbol{\Psi}_r)|$ occurs at $\theta_r=14.98^\circ$ instead of at $\theta_r^*=\theta_t=15^\circ$. {\BLUE Nevertheless, for large $L_\x,L_\y\gg\lambda$, the width of the sinc functions decreases and eventually they become the dominant factor for determining $\boldsymbol{\Psi}_r^\star$ which leads to $\boldsymbol{\Psi}_r^\star=\boldsymbol{\Psi}_r^*$, cf. Corollary~\ref{Corol:MaxContinuous}.} 

\textit{iv)} For the phase-shift function in \eqref{Eq:BetaContinuous}, the phase of $g^{\mathsf{c}}(\boldsymbol{\Psi}_t,\boldsymbol{\Psi}_r)$ is equal to $\beta_0$ up to a constant. In other words, if we change the phase induced on the \textit{entire tile surface} by a constant,  $|g^{\mathsf{c}}(\boldsymbol{\Psi}_t,\boldsymbol{\Psi}_r)|$ remains the same and $\angle g^{\mathsf{c}}(\boldsymbol{\Psi}_t,\boldsymbol{\Psi}_r)$ changes by that constant. 

{\BLUE \textit{v)} Corollary~\ref{Corol:MinArea} reveals that for given $\rho_t$, $\rho_r$, and $\rho_d$, a smaller IRS is needed for higher frequencies  for the strength of the IRS-assisted link to be identical to that of the unobstructed direct link.  Moreover, for a given $\rho_t+\rho_r$,  Corollary~\ref{Corol:MinArea} suggests that deploying the IRS close to the receiver (or transmitter) is advantageous as $A_{\mathrm{req}}$ becomes smaller, see \cite{ozdogan2019intelligent,di2020analytical,bjornson2020power} for similar conclusions. We further study the impact of the position of the IRS on performance in Fig.~\ref{Fig:PowerDistance} in Section~\ref{Sec:Sim}.}

\begin{remk}\label{Remk:Passive}
	{\RED As is shown in \cite{estakhri2016wave,asadchy2016perfect}, the value of $\tau$ can be set to ensure that the power of the incident wave is identical to the power of the reflected wave, which makes the surface globally passive. If both the incident and reflected waves are plane waves and the surface is large, the condition for surface passivity  is $\tau=\sqrt{\frac{\cos(\theta_t)}{\cos(\theta_r^\star)}}$ \cite{estakhri2016wave,asadchy2016perfect}. For large tiles, the reflected wave is a plane wave for  the linear phase-shift design in \eqref{Eq:BetaContinuous} and we obtain $\theta_r^\star= \sin^{-1}\!\big(\!\sqrt{(A_\x(\boldsymbol{\Psi}_t^*,\boldsymbol{\Psi}_r^*)-A_\x(\boldsymbol{\Psi}_t))^2+(A_\y(\boldsymbol{\Psi}_t^*,\boldsymbol{\Psi}_r^*)-A_\y(\boldsymbol{\Psi}_t))^2}\,\big)$. Specifically, for $\boldsymbol{\Psi}_t=\boldsymbol{\Psi}_t^*$, we obtain $\theta_r^\star=\theta_r^*$.}
\end{remk}

\begin{remk}\label{Remk:Special}
	{\BLUE The RCS derived  in \cite[Ch.~11.3.2]{balanis2012advanced} for specular reflection of TE$^x$ waves is a special case of Proposition~\ref{Prop:Continuous} with $\phi_t=\frac{3\pi}{2}$, $\varphi_t=0$, $\phi_r^*=\frac{\pi}{2}$,  $\tau=1$, and $A_i(\boldsymbol{\Psi}_t^*,\boldsymbol{\Psi}_r^*)=\beta_0=0,\,\,i\in\{\x,\y\}$, which implies $c(\boldsymbol{\Psi}_t)=1$, $\widetilde{g}(\boldsymbol{\Psi}_t,\boldsymbol{\Psi}_r) = \sqrt{\cos^2(\theta_r)\sin^2(\phi_r)+\cos^2(\phi_r)}$,  $A_\x(\boldsymbol{\Psi}_t,\boldsymbol{\Psi}_r)=\sin(\theta_r)\cos(\phi_r)$, and $A_\y(\boldsymbol{\Psi}_t,\boldsymbol{\Psi}_r)=\sin(\theta_r)\sin(\phi_r)-\sin(\theta_t)$.}
	Similarly, the result given in  \cite[Lemma~2]{ozdogan2019intelligent} is a special case of Proposition~\ref{Prop:Continuous} with $\phi_t=\frac{3\pi}{2}$, $\varphi_t=\frac{\pi}{2}$, $\phi_r^*=\phi_r=\frac{\pi}{2}$,  $\tau=1$, and $A_i(\boldsymbol{\Psi}_t^*,\boldsymbol{\Psi}_r^*)=\beta_0=0,\,\,i\in\{\x,\y\}$, which implies  $\widetilde{g}(\boldsymbol{\Psi}_t,\boldsymbol{\Psi}_r) = c(\boldsymbol{\Psi}_t)=\cos(\theta_t)$,  $A_\x(\boldsymbol{\Psi}_t,\boldsymbol{\Psi}_r)=0$ and $A_\y(\boldsymbol{\Psi}_t,\boldsymbol{\Psi}_r)=\sin(\theta_r)-\sin(\theta_t)$.
\end{remk}
	
\subsection{Discrete Tiles}\label{Sec:Discrete} 

Although Proposition~\ref{Prop:Continuous} provides insights regarding the impact of the system parameters, a continuous programmable tile surface is difficult to implement and, in practice, each tile is comprised of many discrete sub-wavelength unit cells, see Fig.~\ref{Fig:IRS}. Therefore, let us now assume that each tile consists of $Q_{\x}Q_{\y}$ unit cells of size $L_{\rm uc}\times L_{\rm uc}$ which are uniformly spaced along the $x$ and $y$ axes with a spacing of $d_\x$ and $d_\y$, respectively. For simplicity, we assume $Q_{\x}$ and $Q_{\y}$ are even numbers. Therefore, the position of the $(n_\x,n_\y)$-th unit cell is given by $(x,y)=(n_\x d_\x,n_\y d_\y)$ for $n_\x=-\frac{Q_\x}{2}+1,\dots,\frac{Q_\x}{2}$ and $n_\y=-\frac{Q_\y}{2}+1,\dots,\frac{Q_\y}{2}$. Moreover, we assume that the $(n_\x,n_\y)$-th unit cell applies a phase shift $\beta_{n_\x,n_\y}$ to the reflected electric field. Exploiting Proposition~\ref{Prop:Continuous},  we can characterize the response of an individual unit cell, denoted by $g_{(n_\x,n_\y)}(\boldsymbol{\Psi}_t,\boldsymbol{\Psi}_r),\,\,\forall n_\x,n_\y$, as {\BLUE
\begin{IEEEeqnarray}{rll} \label{Eq:DiscreteUC}
	g_{(n_\x,n_\y)}(\boldsymbol{\Psi}_t,\boldsymbol{\Psi}_r)	= 
	g_{\rm uc}(\boldsymbol{\Psi}_t,\boldsymbol{\Psi}_r) 
	 \e^{\jj\beta_{n_\x,n_\y}}
	 \e^{\jj \kappa d_\x [ A_\x(\boldsymbol{\Psi}_t)+ A_\x(\boldsymbol{\Psi}_r)]n_\x} 
	 \, \e^{\jj \kappa d_\y [A_\y(\boldsymbol{\Psi}_t)+ A_\y(\boldsymbol{\Psi}_r)]n_\y},\quad\,\,\,\,
\end{IEEEeqnarray}
where  $ g_{\rm uc}(\boldsymbol{\Psi}_t,\boldsymbol{\Psi}_r)=\frac{\jj\sqrt{4\pi} \tau L_{\rm uc}^2}{\lambda}\widetilde{g}(\boldsymbol{\Psi}_t,\boldsymbol{\Psi}_r)
 \mathrm{sinc}\left(\frac{\kappa L_{\rm uc} A_\x(\boldsymbol{\Psi}_t,\boldsymbol{\Psi}_r) }{2}\right) \mathrm{sinc}\left(\frac{\kappa L_{\rm uc} A_\y(\boldsymbol{\Psi}_t,\boldsymbol{\Psi}_r) }{2}\right)$ is referred to as the unit-cell factor and characterizes the unit-cell radiation pattern  as a function of the polarization of the incident wave, the incident angle $\boldsymbol{\Psi}_t$, the size of the unit cell, and the observation angle $\boldsymbol{\Psi}_r$ \cite{balanis2012advanced}.   Exploiting the identity $\lim_{x\to 0} \mathrm{sinc}(x)=1$, the unit-cell factor simplifies to $ g_{\rm uc}(\boldsymbol{\Psi}_t,\boldsymbol{\Psi}_r)=\frac{\jj\sqrt{4\pi} \tau L_{\rm uc}^2}{\lambda}\widetilde{g}(\boldsymbol{\Psi}_t,\boldsymbol{\Psi}_r)$ for $L_{\rm uc}\ll\lambda$.}
The tile response function of the entire tile, denoted by $g^{\mathsf{d}}(\boldsymbol{\Psi}_t,\boldsymbol{\Psi}_r)$,  is the superposition of the responses of all unit cells of the tile \cite[Ch. 3]{lau2012reconfigurable} and is obtained as
\begin{IEEEeqnarray}{rll}\label{Eq:Discrete}
	g^{\mathsf{d}}(\boldsymbol{\Psi}_t,\boldsymbol{\Psi}_r)	=  \sum_{n_\x=-\frac{Q_\x}{2}+1}^{\frac{Q_\x}{2}}  \sum_{n_\y=-\frac{Q_\y}{2}+1}^{\frac{Q_\y}{2}}  g_{(n_\x,n_\y)}(\boldsymbol{\Psi}_t,\boldsymbol{\Psi}_r).
\end{IEEEeqnarray}
The following proposition presents the phase-shift design needed to realize the generalized Snell's law  and the resulting tile response function. 


\begin{prop}\label{Prop:Discrete} 
	For given $\boldsymbol{\Psi}_t^*$ and $\boldsymbol{\Psi}_r^*$, let us impose the constant phase shifts $\beta_{n_\x,n_\y}=\beta_{n_\x}+\beta_{n_\y}$ with
	\begin{IEEEeqnarray}{rll}\label{Eq:Beta_discrete}
		\beta_{n_\x} = -\kappa d_\x A_\x(\boldsymbol{\Psi}_t^*,\boldsymbol{\Psi}_r^*)n_\x+\frac{\beta_0}{2}
		\quad\text{and}\quad
		\beta_{n_\y} = -\kappa d_\y A_\y(\boldsymbol{\Psi}_t^*,\boldsymbol{\Psi}_r^*)n_\y+\frac{\beta_0}{2}. 
	\end{IEEEeqnarray}
	Then, the amplitude of the corresponding tile response function for an EM wave impinging from an arbitrary direction $\boldsymbol{\Psi}_t$  and being observed at arbitrary reflection direction $\boldsymbol{\Psi}_r$ is given by  
	\begin{IEEEeqnarray}{lll}\label{Eq:Ampl_discrete}
		g^{\mathsf{d}}_{||}(\boldsymbol{\Psi}_t,\boldsymbol{\Psi}_r) = 
		|g_{\rm uc}(\boldsymbol{\Psi}_t,\boldsymbol{\Psi}_r)| \nonumber \\
		\times \frac{\sin\left(\frac{\pi Q_\x d_\x}{\lambda} [A_\x(\boldsymbol{\Psi}_t,\boldsymbol{\Psi}_r)-A_\x(\boldsymbol{\Psi}_t^*,\boldsymbol{\Psi}_r^*)]\right)}{\sin\left(\frac{\pi d_\x }{\lambda} [A_\x(\boldsymbol{\Psi}_t,\boldsymbol{\Psi}_r)-A_\x(\boldsymbol{\Psi}_t^*,\boldsymbol{\Psi}_r^*)]\right)}  
		\times \frac{\sin\left(\frac{\pi Q_\y d_\y}{\lambda} [A_\y(\boldsymbol{\Psi}_t,\boldsymbol{\Psi}_r)-A_\y(\boldsymbol{\Psi}_t^*,\boldsymbol{\Psi}_r^*)]\right)}{\sin\left(\frac{\pi d_\y }{\lambda} [A_\y(\boldsymbol{\Psi}_t,\boldsymbol{\Psi}_r)-A_\y(\boldsymbol{\Psi}_t^*,\boldsymbol{\Psi}_r^*)]\right)}\quad\,\,\,\,
	\end{IEEEeqnarray}
	and the phase of the tile response function is given by 
	\begin{IEEEeqnarray}{lll}\label{Eq:Phase_discrete}
		g^{\mathsf{d}}_{\angle}(\boldsymbol{\Psi}_t,\boldsymbol{\Psi}_r) = \beta_0+\frac{\pi}{2} \nonumber \\
		+\frac{\pi d_\x\left[A_\x(\boldsymbol{\Psi}_t,\boldsymbol{\Psi}_r)-A_\x(\boldsymbol{\Psi}_t^*,\boldsymbol{\Psi}_r^*)\right]}{\lambda} 
		+\frac{\pi d_\y\left[A_\y(\boldsymbol{\Psi}_t,\boldsymbol{\Psi}_r)-A_\y(\boldsymbol{\Psi}_t^*,\boldsymbol{\Psi}_r^*)\right]}{\lambda} 
		 .\quad
	\end{IEEEeqnarray}
\end{prop}
\begin{IEEEproof}
	The proof is provided in Appendix~\ref{App:PropDiscrete}.
\end{IEEEproof}

\begin{corol}\label{Corol:MaxDiscrete}
	Assuming large $Q_\x$ and $Q_\y$, and $\boldsymbol{\Psi}_t^*=\boldsymbol{\Psi}_t$, the maximum value of $|g^{\mathsf{d}}(\boldsymbol{\Psi}_t,\boldsymbol{\Psi}_r)|$ occurs at $\boldsymbol{\Psi}_r=\boldsymbol{\Psi}_r^*$ and is obtained as
	\begin{IEEEeqnarray}{rll}
		\underset{\boldsymbol{\Psi}_r\to\boldsymbol{\Psi}_r^*}{\lim}|g^{\mathsf{d}}(\boldsymbol{\Psi}_t,\boldsymbol{\Psi}_r)| 
		= |g_{\rm uc}(\boldsymbol{\Psi}_t^*,\boldsymbol{\Psi}_r^*)| Q_\x Q_\y \overset{(a)}{\leq} \frac{\sqrt{4\pi}  L_{\rm uc}^2}{\lambda} \,Q_\x Q_\y,
	\end{IEEEeqnarray}
	where $(a)$ holds with equality when  $\theta_t=\theta_r=0$ and $\tau=1$.
\end{corol}
\begin{IEEEproof}
	The proof is similar to the proof for Corollary~\ref{Corol:MaxContinuous}. 
\end{IEEEproof}

\begin{figure}[t]\vspace{-0.5cm}
	\centering
	\includegraphics[width=1\textwidth]{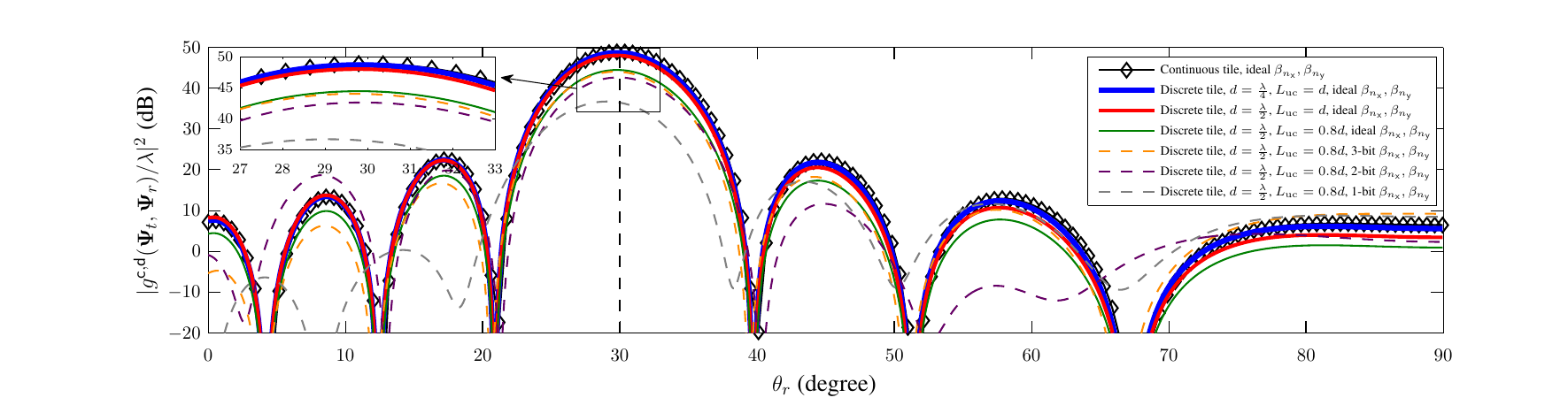}\vspace{-0.5cm}
	\caption{\BLUE Amplitude of the tile response function, $|g^{\mathsf{c,d}}(\boldsymbol{\Psi}_t,\boldsymbol{\Psi}_r)/\lambda|^2$, in dB vs. $\theta_r$  for $(\theta_t,\phi_t,\varphi_t)=(0,0,22.5^{\circ})$, $\phi_r=45^{\circ}$, $(\theta_t^*,\phi_t^*)=(0,0)$, $(\theta_r^*,\phi_r^*)=(30^{\circ},45^{\circ})$, $d_\x=d_\y=d$, and $\tau=0.8$. For a fair comparison, the sizes of the continuous and discrete tiles are identical with $L_\x=L_\y=10\lambda$. For discrete tiles, the number of unit cells along the $x$ and $y$ axes are $Q_\x=\frac{L_\x}{d_\x}$ and $Q_\y=\frac{L_\y}{d_\y}$, respectively, the unit cell size is $L_{\rm uc}\times L_{\rm uc}$, and we consider both ideal continuous and uniformly quantized phase shifts $\beta_{n_\x}$ and $\beta_{n_\y}$.\vspace{-0.5cm} }
	\label{Fig:FuncGdiscrete}
\end{figure}

{\BLUE
	\begin{corol}\label{Corol:MinQ}\BLUE
For discrete IRSs, the minimum number of IRS unit cells, $Q$, for which the free-space path-loss of the IRS-assisted link is equal to the path-loss of the unobstructed direct link is given by
\begin{IEEEeqnarray}{lll}
	 Q_{\mathrm{req}}
	= 	\frac{\lambda\rho_t \rho_r }{ L_{\rm uc}^2\rho_d }\overset{(a)}{=}\frac{4\rho_t \rho_r }{\lambda\rho_d},
\end{IEEEeqnarray}
where equality $(a)$ holds when $L_{\rm uc}=d_\x =d_\y=\frac{\lambda}{2}$.
\end{corol}
\begin{IEEEproof}\BLUE
	The proof is similar to that given for Corollary~\ref{Corol:MinArea} except that we now upper bound  $g_{\mathrm{IRS}}$  by $\frac{\sqrt{4\pi}  L_{\rm uc}^2NQ_\x Q_\y}{\lambda}$ based on Corollary~\ref{Corol:MaxDiscrete}.
\end{IEEEproof}
}

In Fig.~\ref{Fig:FuncGdiscrete}, we plot $\BLUE |g^{\mathsf{d}}(\boldsymbol{\Psi}_t,\boldsymbol{\Psi}_r)/\lambda|^2$ obtained from Proposition~\ref{Prop:Discrete}  vs.  $\theta_r$ for various scenarios and compare it with  $\BLUE |g^{\mathsf{c}}(\boldsymbol{\Psi}_t,\boldsymbol{\Psi}_r)/\lambda|$ obtained from Proposition~\ref{Prop:Continuous}. In the following, we highlight some insights that Proposition~\ref{Prop:Discrete}, Corollaries~\ref{Corol:MaxDiscrete} and \ref{Corol:MinQ}, and Fig.~\ref{Fig:FuncGdiscrete} provide: 

\textit{i)} For $d_\x,d_\y\to 0$ (extremely sub-wavelength unit cells) and $L_{\rm uc}=d_\x=d_\y$ (compact deployment of the unit cells), the tile response function of a discrete tile in Proposition~\ref{Prop:Discrete} becomes identical to that of a continuous tile in Proposition~\ref{Prop:Continuous}, i.e., 
$g^{\mathsf{d}}(\boldsymbol{\Psi}_t,\boldsymbol{\Psi}_r)=g^{\mathsf{c}}(\boldsymbol{\Psi}_t,\boldsymbol{\Psi}_r)$.

\textit{ii)} Fig.~\ref{Fig:FuncGdiscrete}  shows that for a discrete tile to accurately approximate a continuous tile,  it is sufficient that $L_{\rm uc}=d_\x=d_\y\leq \frac{\lambda}{2}$ holds.  In practice, there are gaps between the IRS unit cells, i.e., $L_{\rm uc}<d_\x,d_\y$. This leads to a decrease of the effective size of the tile, see Fig.~\ref{Fig:FuncGdiscrete} for $L_{\rm uc}=0.8d$. 

\textit{iii)} In both Propositions~\ref{Prop:Continuous} and \ref{Prop:Discrete}, we assumed that the phase shift introduced by the tile unit cells can assume any real value which is an idealized assumption as finite resolution phase shifts are typically applied in practice. Nevertheless, Fig.~\ref{Fig:FuncGdiscrete}  suggests that a 3-bit uniform quantization of the phase shifts $\beta_{n_\x}$ and $\beta_{n_\y}$ yields a tile response function which is very close to the one obtained for ideal real-valued phase shifts. Moreover, from Fig.~\ref{Fig:FuncGdiscrete}, we also observe that even for a 1-bit phase shift quantization, the tile response function has a shape similar to the ideal case although the peak is reduced and the side lobes  deviate from those for ideal real-valued phase shifts, which is consistent with  \cite{kaina2014shaping}.


{\BLUE
\textit{iv)} Assuming $(\rho_d,\rho_t,\rho_r)=(200~\text{m},100~\text{m},100~\text{m})$, $\tau =1$,  $d_\x =d_\y=L_{\rm uc}=\frac{\lambda}{2}$, and  carrier frequencies of $5$, $10$, and $28$~GHz, Corollary~\ref{Corol:MinQ} reveals that at least $Q_{\mathrm{req}}=3333,6666,18667$ unit cells are needed, respectively, for the IRS-assisted link to achieve the same free-space path-loss as the unobstructed direct link. Moreover, while Corollary~\ref{Corol:MinArea} shows that the required minimum IRS size $A_{\mathrm{req}}$ decreases with the carrier frequency, Corollary~\ref{Corol:MinQ} suggests that the required minimum number of IRS unit cells $Q_{\mathrm{req}}$ increases with the carrier frequency.}

\subsection{End-to-End System Model}

\begin{figure} 
	\centering
	\includegraphics[width=0.6\textwidth]{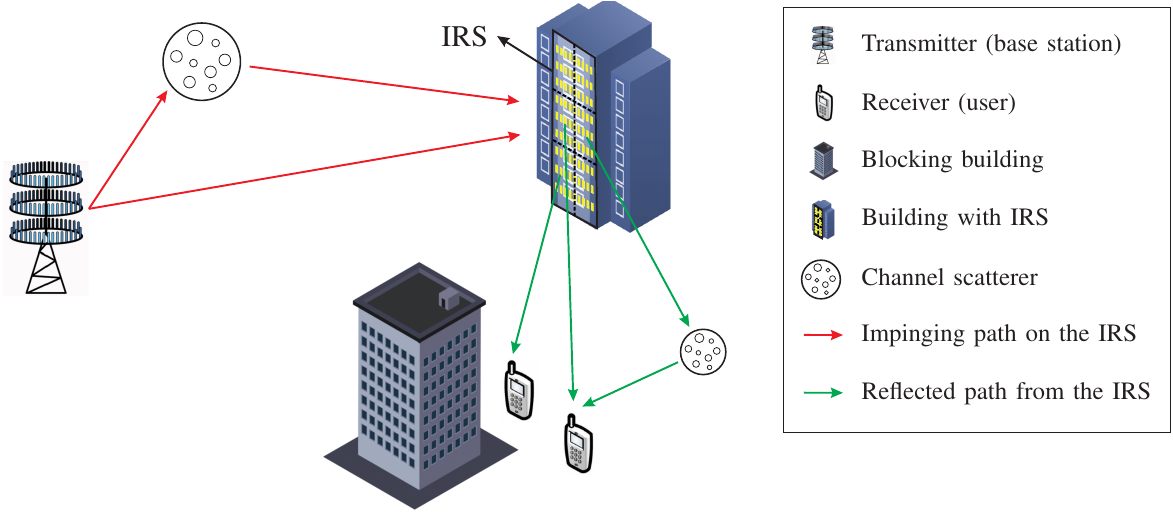}\vspace{-0.2cm}
	\caption{\BLUE Schematic illustration of an IRS-assisted wireless system. Only the IRS-guided paths are shown for clarity of presentation. \vspace{-0.5cm}}
	\label{Fig:SysMod}
\end{figure}

{\BLUE In this section, we first present the end-to-end system model that is widely used in the literature and characterizes the IRS in terms of the phase shifts applied by the individual unit cells \cite{guo2019weighted,yu2020robust,bai2020latency,wu2019intelligent,zhang2020joint,pan2020multicell,wang2019intelligent,cao2019delay}. We refer to this model as the phase-shift model. Subsequently, we highlight some drawbacks of the phase-shift model. To overcome these drawbacks,  we develop a new model that characterizes the IRS in terms of the tile response function $g(\boldsymbol{\Psi}_t,\boldsymbol{\Psi}_r)$ and the  corresponding transmission modes. We refer to this new model as the transmission-mode model.

 \textit{Phase-shift Model:} Let us consider a system comprising a multiple-antenna BS, an IRS, and $K$ multiple-antenna users. 
	The direct links between the transmitter and the users may exist but are severely shadowed (e.g., by a blocking building). The downlink communication is assisted by an IRS that comprises $Q$  unit cells. The end-to-end system model can be written as \cite{guo2019weighted,yu2020robust,bai2020latency,wu2019intelligent,zhang2020joint,pan2020multicell,wang2019intelligent,cao2019delay}
	\begin{IEEEeqnarray}{ll}\label{Eq:model_phase}
		\by_k & = (\bH_{d,k} + \bH_{r,k} \bOmega\bH_t)\bx  + \bz_k ,\quad k = 1,\ldots,K,
	\end{IEEEeqnarray}
	where $\bx\in\Cset^{N_t}$, $\by_k\in\Cset^{N_r}$, and $\bz_k\in\Cset^{N_r}$
	denote the BS's transmit signal, the received signal at the $k$-th user, and the additive white Gaussian noise (AWGN) 
	at the $k$-th user, respectively. Here, $\Cset$ denotes the set of complex numbers and $N_t$ and $N_r$ denote the numbers of antennas at the BS and each user, respectively. Moreover, $\bH_{d,k}\in\Cset^{N_r\times N_t}$, $\bH_t\in\Cset^{Q\times N_t}$, and $\bH_{r,k}\in\Cset^{N_r\times Q}$ denote the BS-to-user~$k$, BS-to-IRS, and IRS-to-user~$k$ channel matrices, respectively.  Furthermore, $\bOmega=\diag(\bar{g}_{\rm uc}\e^{\jj \beta_1},\dots,\bar{g}_{\rm uc}\e^{\jj \beta_Q})\in\Cset^{Q\times Q}$ is a diagonal matrix with main diagonal entries $\bar{g}_{\rm uc}\e^{\jj \beta_1},\dots,\bar{g}_{\rm uc}\e^{\jj \beta_Q}$, where $\beta_q$ is the phase-shift applied by the $q$-th unit cell and $\bar{g}_{\rm uc}$ denotes the unit-less unit-cell factor. The relation between $\bar{g}_{\rm uc}$ and the unit-cell factor $g_{\rm uc}$ (in unit of meters) defined after \eqref{Eq:DiscreteUC} is explained below.

 The phase-shift model in \eqref{Eq:model_phase} suffers from two drawbacks: \textit{i)} In the literature, the elements of $\bH_{r,k}$ (and $\bH_t$) are typically modeled as independent random variables \cite{guo2019weighted,yu2020robust,pan2020multicell,bai2020latency,wu2019intelligent,zhang2020joint}. However, for large values of $Q$ (which are needed to provide a sufficient link budget), the ranks of $\bH_{r,k}$ and $\bH_t$ are limited by the number of channel scatterers leading to dependencies. \textit{ii)} The model in \eqref{Eq:model_phase} assumes a constant unit-cell factor $\bar{g}_{\rm uc}$ \cite{yu2020robust,pan2020multicell,bai2020latency,wang2019intelligent}. In fact, most works assume the ideal case and set $\bar{g}_{\rm uc}=1$ \cite{guo2019weighted,wu2019intelligent,cao2019delay,zhang2020joint}. However, as will be shown below, $\bar{g}_{\rm uc}$ is related to $g_{\rm uc}$, and hence, depends on the physical properties of the unit cell (such as its area) as well as the  incident and reflected angles and the polarization of the waves. These two drawbacks of the phase-shift model are addressed in the transmission-mode model developed in the following.

\textit{Transmission-Mode Model:} In order to present the proposed model, we first need to establish the following assumptions and notation:
	\begin{itemize}
		\item \textbf{Low-Rank Channel:} We assume there are multiple scatterers in the environment which cause the signal of a given transmitter to arrive at the IRS potentially via multiple paths and the signal reflected from the IRS to potentially also arrive at a given receiver via multiple paths \cite{alkhateeb2014channel,ghanaatian2019feedback}. This leads to a  low-rank channel model which allows us to decompose the channel matrices as $\bH_t = \bA_{t} \bSigma_{t} \bD_{t}^\Herm$, $\bH_{r,k} = \bA_{r,k} \bSigma_{r,k} \bD_{r,k}^\Herm$, and $\bH_{d,k} = \bA_{d,k} \mathbf{\Sigma}_{d,k} \bD_{d,k}^\Herm$, respectively, where the components $\{\bA_{t} \in \Cset^{Q \times {L_t}} $, $\bA_{r,k} \in \Cset^{N_r \times {L_{r,k}}}$, $\bA_{d,k} \in \Cset^{N_r \times {L_{d,k}}}\} $, $\{\mathbf{D}_{t} \in \Cset^{N_t \times {L_t}}$, $\bD_{r,k} \in \Cset^{Q \times {L_{r,k}}}$, $\bD_{d,k}\in \Cset^{N_{t} \times {L_{d,k}}}\}$, and $\{\bSigma_{t}\in \Cset^{L_t \times {L_t}}$, $\bSigma_{r,k} \in \Cset^{L_{r,k} \times {L_{r,k}}}$, $\bSigma_{d,k}\in \Cset^{L_{d,k} \times {L_{d,k}}}\} $ represent the receive steering matrices (corresponding to the angles-of-arrival (AoAs)), transmit steering matrices (corresponding to the angles-of-departure (AoDs)), and channel gains of the scatters, respectively, where  $L_t, L_{r,k},$ and  $L_{d,k}$ denote the numbers of scatterers of the BS-to-IRS,  IRS-to-user~$k$, and BS-to-user~$k$ channels, respectively. Moreover, $(\cdot)^\Herm$ denotes the Hermitian transpose. For future reference,  let us define $\bA_{t}=[\bA_{t,1}^\Herm,\dots,\bA_{t,N}^\Herm]^\Herm$ and $\bD_{r,k}=[\bD_{r,k,1}^\Herm,\dots,\bD_{r,k,N}^\Herm]^\Herm$, where $\bA_{t,n}\in\Cset^{Q_\x Q_\y\times L_t}$ and $\bD_{r,k,n}\in\Cset^{Q_\x Q_\y\times L_{r,k}}$ denote the receive steering matrix for the $n$-th tile and the transmit steering matrix of the $n$-th tile to user~$k$, respectively. 
		\item \textbf{Transmission Modes:} We assume that the IRS can select for each tile one of the possible transmission modes corresponding to one set of unit-cell phase shifts in Proposition~2.  The design of a finite number of transmission modes, $M$, will be discussed in detail in Section~\ref{Sec:Codebook}. Let $\beta_{q,n,m},\,\,q=1,\dots,Q_\x Q_\y, n=1,\dots,N,m=1,\dots,M$, denote the phase shift applied by unit cell $q$ of tile $n$ for transmission mode $m$. Unlike the phase-shift model in \eqref{Eq:model_phase}, the amplitudes of the entries of the phase-shift matrix depend on the AoAs and AoDs. Thus, we denote the phase-shift matrix for tile $n$, transmission mode $m$, incident wave angle $\boldsymbol{\Psi}_t$, and reflection angle $\boldsymbol{\Psi}_r$ by $\bOmega_{n,m}(\bPsi_t,\bPsi_r)=\diag(\bar{g}_{\rm uc}(\bPsi_t,\bPsi_r)\e^{\jj \beta_{1,n,m}},\dots,\bar{g}_{\rm uc}(\bPsi_t,\bPsi_r)\e^{\jj \beta_{Q_\x Q_\y,n,m}})\in\Cset^{Q_\x Q_\y\times Q_\x Q_\y}$. Moreover, using Lemma~\ref{Lem:Pathloss}, the unit-less unit-cell factor $\bar{g}_{\rm uc}(\bPsi_t,\bPsi_r)$ is related to the unit-cell factor $g_{\rm uc}(\bPsi_t,\bPsi_r)$ (in meters) defined in Section~\ref{Sec:Discrete} as $\bar{g}_{\rm uc}(\bPsi_t,\bPsi_r)=\frac{\sqrt{4\pi}}{\lambda}g_{\rm uc}(\bPsi_t,\bPsi_r)$.
		\item \textbf{Selection Variable:} Let $s_{n,m}\in\{0,1\}$ denote a binary variable which is equal to one if the $m$-th transmission mode  is selected for the $n$-th tile; otherwise, it is equal to zero. Moreover, let $\mathcal{M}\subset\{1,\dots,M\}$ denote the subset of transmission modes that is used for resource allocation in a given frame. We will discuss the design of $\mathcal{M}$ in Section~\ref{Sec:PreSelect}. Since, at any given time, the IRS can select only one transmission mode for each tile, $\sum_{m\in\mathcal{M}} s_{n,m}=1,\,\,\forall n$, has to hold.
	\end{itemize} 
	Using the above notation, the end-to-end channel model can be compactly written as follows 
	\begin{IEEEeqnarray}{ll}\label{Eq:E2Emodel}
		\by_k =  \Big(\bA_{d,k} \bSigma_{d,k} \bD_{d,k}^\Herm + \bA_{r,k} \bSigma_{r,k}  \bG_k  \bSigma_{t} \bD_{t}^\Herm 
		\Big)\bx  + \bz_k ,\quad k = 1,\ldots,K,
	\end{IEEEeqnarray}
	where $ \bG_k \in \Cset^{L_{r,k} \times L_{t}}$ is given by 
	\begin{IEEEeqnarray}{ll}\label{Eq:response}
		\bG_k 
		= \sum_{n=1}^N\sum_{m\in\mathcal{M}} s_{n,m} \bG_{n,m,k}. 
	\end{IEEEeqnarray}
	Here, the $n_r$-th row and $n_t$-th column of matrix $\bG_{n,m,k}\in \Cset^{L_{r,k} \times L_{t}}$ is given by 
	\begin{IEEEeqnarray}{ll} 
		\left[ \bG_{n,m,k}\right]_{n_r,n_t} = \bd _{r,k,n}^\Herm(\bPsi_r^{(n_r)})\bOmega_{n,m}(\bPsi_t^{(n_t)},\bPsi_r^{(n_r)})\ba_{t,n}(\bPsi_t^{(n_t)}) \triangleq \frac{\sqrt{4\pi}}{\lambda} g_{n,m}(\bPsi_t^{(n_t)},\bPsi_r^{(n_r)}),
	\end{IEEEeqnarray}
	where $\ba_{t,n}(\bPsi_t^{(n_t)})$ and $\bd _{r,k,n}(\bPsi_r^{(n_r)})$   are the $n_t$-th column of $\bA_{t,n}$ and the $n_r$-th column of $\bD_{r,k,n}$, respectively.} In other words, $g_{n,m}(\boldsymbol{\Psi}_t^{(n_t)},\boldsymbol{\Psi}_r^{(n_r)})$ denotes the response function of the $n$-th tile for the $m$-th transmission mode evaluated at the $n_t$-th AoA, specified by angle $\boldsymbol{\Psi}_t^{(n_t)}$, and the $n_r$-th AoD, specified by angle $\boldsymbol{\Psi}_r^{(n_r)}$. Assuming that the center of the $n$-th tile is placed at point $(x,y)=(u_\x^{(n)}L_\x,u_\y^{(n)}L_\y)$, where $u_\x^{(n)}$ and $u_\x^{(n)}$ are integer numbers,  $g_{n,m}(\boldsymbol{\Psi}_t,\boldsymbol{\Psi}_r)$ for AoA $\boldsymbol{\Psi}_t$ and AoD $\boldsymbol{\Psi}_r$ can be expressed as
\begin{IEEEeqnarray}{rll}\label{Eq:TilePhaseDiff}
	g_{n,m}(\boldsymbol{\Psi}_t,\boldsymbol{\Psi}_r) = \e^{\jj \kappa u_\x^{(n)}L_\x A_\x(\boldsymbol{\Psi}_t,\boldsymbol{\Psi}_r) + \jj \kappa u_\y^{(n)}L_\y A_\y(\boldsymbol{\Psi}_t,\boldsymbol{\Psi}_r)}  g_{m}(\boldsymbol{\Psi}_t,\boldsymbol{\Psi}_r), \quad
\end{IEEEeqnarray}
where $g_{m}(\boldsymbol{\Psi}_t,\boldsymbol{\Psi}_r)$ is the response function of the reference tile centered at the origin with $u_\x^{(n)}=u_\y^{(n)}=0$. {\BLUE The value of $g_{m}(\boldsymbol{\Psi}_t,\boldsymbol{\Psi}_r)$ can be computed from the analytical expressions given in  Proposition~1 and  Proposition~2 for continuous and discrete tiles, respectively.}

Defining $\widetilde{\bH}_t =  \bSigma_t \bD^\Herm$ and  $\widetilde{\bH}_{r,k} = \bA_{r,k} \bSigma_{r,k}$ the channel model in \eqref{Eq:E2Emodel} can be rewritten as follows
\begin{IEEEeqnarray}{rll}\label{Eq:E2Emodelsimple}
	\mathbf{y}_k
	&=  \Big[\bH_{d,k}+ \sum_{n=1}^N  \sum_{m\in\mathcal{M}}  s_{n,m} \bH_{n,m,k} \Big]\bx+\bz_k, \quad k=1,\dots,K,
\end{IEEEeqnarray}
where $\bH_{n,m,k}=\widetilde{\bH}_{r,k}\bG_{n,m,k}\widetilde{\bH}_t $ denotes the end-to-end channel for the $n$-th tile and the $m$-th  transmission mode. Furthermore, defining $\mathbf{H}_k^{e2e}=\bH_{d,k}+ \sum_{n=1}^N  \sum_{m\in\mathcal{M}}  s_{n,m} \bH_{n,m,k}$ as the effective IRS-assisted end-to-end channel between the BS and user $k$,  the channel model in \eqref{Eq:E2Emodelsimple} explicitly shows that unlike conventional systems, for which the channel matrix is fixed, for IRS-assisted communications, we can choose among $|\mathcal{M}|^N$ different end-to-end channel matrices which implies the realization of a smart reconfigurable wireless environment \cite{di2019smart,liaskos2019novel}.

\section{Two-Stage Optimization Framework for IRS-Assisted Communications}\label{Sec:Optimization}

Based on the end-to-end model developed in Section~\ref{Sec:Model}, in this section, we propose a two-stage framework for IRS optimization, namely an offline design stage and an online optimization~stage.

\subsection{Offline Tile Transmission Mode Codebook Design} \label{Sec:Codebook}

In the following, we design a tile transmission mode codebook comprised of a finite number of predefined phase-shift configurations according to  {\RED\eqref{Eq:Beta_discrete} in Proposition~\ref{Prop:Discrete}, i.e.,
\begin{IEEEeqnarray}{rll} \label{Eq:PhaseShiftCodebook}
\e^{\jj\beta_{n_\x,n_\y} } = \e^{\jj 2\pi  \bar{\beta}_\x n_\x}
\times \e^{\jj 2\pi\bar{\beta}_\y n_\y}
\times \e^{\jj2\pi\bar{\beta}_0},
\end{IEEEeqnarray}
where $\bar{\beta}_\x=-\frac{d_\x A_\x(\boldsymbol{\Psi}_t^*,\boldsymbol{\Psi}_r^*)}{\lambda}$, $\bar{\beta}_\y=-\frac{d_\y A_\y(\boldsymbol{\Psi}_t^*,\boldsymbol{\Psi}_r^*)}{\lambda}$, and $\bar{\beta}_0=\frac{\beta_0}{2\pi}$.}
The above phase-shift profile  $\beta_{n_\x,n_\y}$  is an affine function in variable $(n_\x,n_\y)$. Thereby, the slope of the phase-shift function determines the direction of the reflected wavefront and the affine  constant $\bar{\beta}_0$ determines the phase of the wavefront. In the following, instead of discretizing the expected angle of the incident EM wave, $\boldsymbol{\Psi}_t^*$, and the desired angle of the reflected EM wave, $\boldsymbol{\Psi}_r^*$, to construct the codebook, we directly discretize $\bar{\beta}_i,\,\,i\in\{\x,\y\}$. This is a more efficient design since different $(\boldsymbol{\Psi}_t^*,\boldsymbol{\Psi}_r^*)$ may yield the same value of $A_i(\boldsymbol{\Psi}_t^*,\boldsymbol{\Psi}_r^*)$ and consequently the same $\bar{\beta}_i,\,\,i\in\{\x,\y\}$. From \eqref{Eq:Incident}, we have $A_i(\boldsymbol{\Psi}_t^*,\boldsymbol{\Psi}_r^*)\in[-2,2]$.  {\RED Therefore, in principle, we have $ \bar{\beta}_i\in[-\frac{2d_i}{\lambda},\frac{2d_i}{\lambda}],\,\,\forall i\in\{\x,\y\}$. However, due to the periodicity of the complex exponential  functions in \eqref{Eq:PhaseShiftCodebook}, $\bar{\beta}_i$ and $\bar{\beta}_i+1$ yield the same tile response function, which leads to an effective support of $ \bar{\beta}_i$ of $[-\bar{\beta}_i^{\rm eff},\bar{\beta}_i^{\rm eff}]$, where $\bar{\beta}_i^{\rm eff}=\min\{\frac{2d_i}{\lambda},\frac{1}{2}\},\,\,\forall i\in\{\x,\y\}$. Similarly, we have $\bar{\beta}_0\in[-\frac{1}{2},\frac{1}{2}]$.} 
For convenience, we represent the transmission mode codebook as the product of three component codebooks, namely $\bar{\beta}_\x\in\mathcal{B}_\x$, $\bar{\beta}_\y\in\mathcal{B}_\y$,  and $\bar{\beta}_0\in\mathcal{B}_0$, where  $\mathcal{B}_\x$  and $\mathcal{B}_\y$  are referred to as the reflection codebooks and $\mathcal{B}_0$ is referred to as the wavefront phase codebook. Assuming that the size of the entire codebook is $M$, we have $M=|\mathcal{B}_\x|\times|\mathcal{B}_\y|\times|\mathcal{B}_0|$.

\begin{figure}\vspace{-0.5cm}
	\begin{minipage}[c]{0.5\linewidth}
		\centering
		\includegraphics[width=1\textwidth]{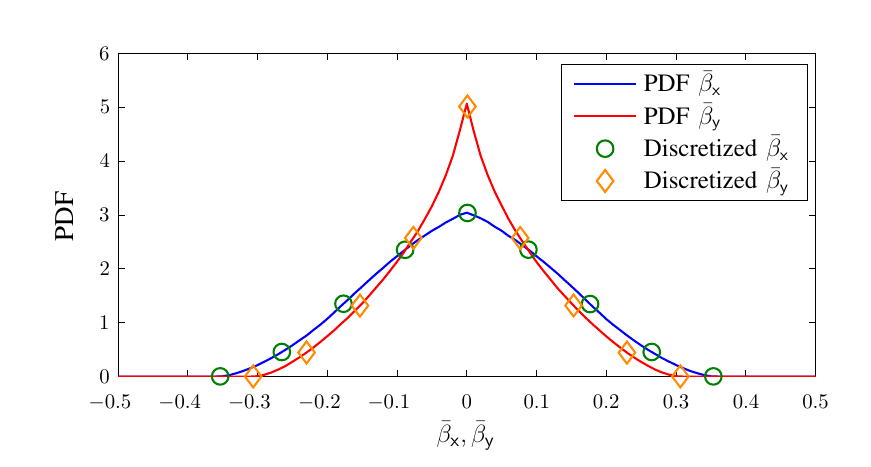}
	\end{minipage}
	\begin{minipage}[c]{0.5\linewidth}
		\centering
		\includegraphics[width=1\textwidth]{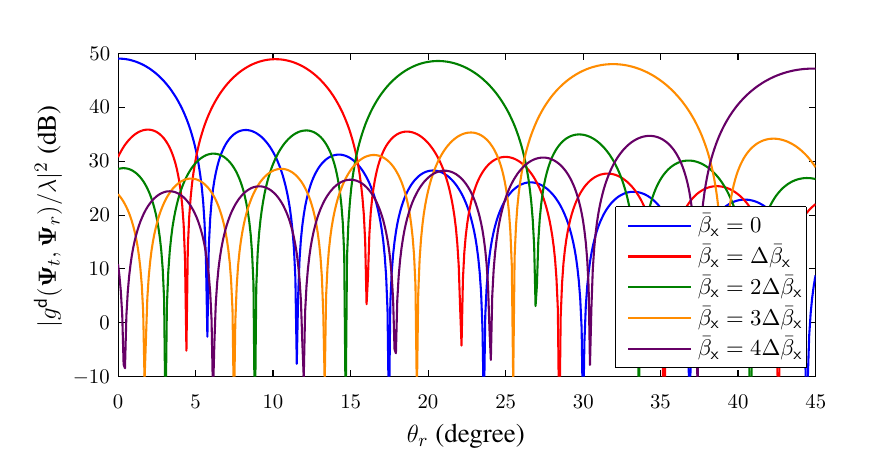}
	\end{minipage}
	\vspace{-0.3cm}
	\caption{\BLUE \textit{Left-hand side:} Distribution of $\bar{\beta}_\x$ and $\bar{\beta}_\y$ and discretized values with $\Delta \bar{\beta}_\x=\frac{\sqrt{2}}{16}$, $\bar{\beta}_\x^{\max}=-\bar{\beta}_\x^{\min}=\frac{\sqrt{2}}{4}$, $\Delta \bar{\beta}_\y=\frac{\sqrt{6}}{32}$, and $\bar{\beta}_\y^{\max}=-\bar{\beta}_\y^{\min}=\frac{\sqrt{6}}{8}$, where the elevation and azimuth angles are uniformly distributed in intervals $\theta_t^*,\theta_r^*\in[0,\pi/4]$, $\phi_t^*\in[0,\pi/3]$, and $\phi_r^*\in[\pi,\pi+\pi/3]$,  \textit{Right-hand side:} $|g^{\mathsf{d}}(\boldsymbol{\Psi}_t,\boldsymbol{\Psi}_r)|$ in dB vs. $\theta_r$  for $(\theta_t,\phi_t,\varphi_t)=(0,0,0)$,  $\phi_r=\pi$, $\bar{\beta}_\y=0$, $L_\x=L_\y=10\lambda$, $d_\x=d_\y=L_{\rm uc}=\frac{\lambda}{2}$,  $\tau =0.8$, $\Delta \bar{\beta}_\x=\frac{\sqrt{2}}{16}$, and $ \bar{\beta}_\y=0$. \vspace{-0.5cm}} 
	\label{Fig:Codebook}
\end{figure}

\textbf{Reflection Codebooks:} We assume all tiles use the same reflection codebook $\mathcal{B}_i,\,\,i\in\{\x,\y\}$. One simple option for construction of $\mathcal{B}_i$ is the uniform discretization of $\bar{\beta}_i,\,\,i\in\{\x,\y\}$, i.e., {\RED$\mathcal{B}_i=\{\bar{\beta}_i^{\min},\bar{\beta}_i^{\min}+\Delta\bar{\beta}_i,\dots, \bar{\beta}_i^{\max}\}$, where $\Delta\bar{\beta}_i=\frac{\bar{\beta}_i^{\max}-\bar{\beta}_i^{\min}}{|\mathcal{B}_i|-1}$ and $[\bar{\beta}_i^{\min},\bar{\beta}_i^{\max}]\subset [-\bar{\beta}_i^{\rm eff},\bar{\beta}_i^{\rm eff}]$ is the  range of parameter $\bar{\beta}_i$. In particular, choosing $[\bar{\beta}_i^{\min},\bar{\beta}_i^{\max}]\neq [-\bar{\beta}_i^{\rm eff},\bar{\beta}_i^{\rm eff}]$ may be beneficial} if the EM waves impinging and reflected at the IRS exhibit a limited range of AoAs and AoDs, respectively. For example, the left-hand side of {\RED Fig.~\ref{Fig:Codebook} shows the distribution of $\bar{\beta}_\x$ and $\bar{\beta}_\y$ for unit-cell spacing of $d_\x=d_\y=\frac{\lambda}{2}$ when  the elevation and azimuth angles  are uniformly distributed in the intervals $\theta_t^*,\theta_r^*\in[0,\pi/4]$, $\phi_t^*\in[0,\pi/3]$, and $\phi_r^*\in[\pi,\pi+\pi/3]$. In this case, from \eqref{Eq:Incident} and the definitions of $\bar{\beta}_\x$ and $\bar{\beta}_\y$, it follows that  $\bar{\beta}_\x^{\max}=-\bar{\beta}_\x^{\min}=\frac{\sin(\pi/4)}{2}=\frac{\sqrt{2}}{4}\approx0.3536$ and $\bar{\beta}_\y^{\max}=-\bar{\beta}_\y^{\min}=\frac{\sin(\pi/4)\sin(\pi/3)}{2}=\frac{\sqrt{6}}{8}\approx0.3062$. Therefore, as an example codebook, we may adopt $|\mathcal{B}_\x|=|\mathcal{B}_\y|=9$ leading to codebooks $\mathcal{B}_\x=\{0,\pm \frac{\sqrt{2}}{16},\pm \frac{\sqrt{2}}{8},\pm \frac{3\sqrt{2}}{16},\pm \frac{\sqrt{2}}{4}\}$ and $\mathcal{B}_\y=\{0,\pm \frac{\sqrt{6}}{32},\pm \frac{\sqrt{6}}{16},\frac{3\sqrt{6}}{32},\pm \frac{\sqrt{6}}{8}\}$.} The right-hand side of Fig.~\ref{Fig:Codebook} shows  $\BLUE|g^{\mathsf{d}}(\boldsymbol{\Psi}_t,\boldsymbol{\Psi}_r)/\lambda|^2$ for the phase-shift configurations generated by $\bar{\beta}_\x=0,\frac{\sqrt{2}}{16},\frac{\sqrt{2}}{8},\frac{3\sqrt{2}}{16},\frac{\sqrt{2}}{4}$ and $\bar{\beta}_\y=0$ when the incident wave is normal to the surface  and the observation point lies in the $x-z$ plane, i.e., $(\theta_t,\phi_t,\varphi_t)=(0,0,0)$,  $\phi_r=\pi$. As can be observed from this figure, the resulting tile response functions cover the considered range of elevation angles for the reflected EM wave, i.e., $\theta_r\in[0,\pi/4]$. 

\textbf{Wavefront Phase Codebook:} The role of $\bar{\beta}_0$ is to control the superposition of the waves reflected from different tiles at the receiver. In particular, if different tiles employ the same $\bar{\beta}_\x$ and $\bar{\beta}_\y$, they can cancel the phase differences caused by their different positions by choosing appropriate values for $\bar{\beta}_0$, see \eqref{Eq:TilePhaseDiff}. 
In other words, for two tiles $n$ and $n'\neq n$ to reflect an EM wave impinging from incident direction $\boldsymbol{\Psi}_t^*$ along reflection direction $\boldsymbol{\Psi}_r^*$, they not only have to choose appropriate and \textit{identical}  elements from the reflection codebooks but also phase shifts, denoted by $\bar{\beta}_0^{(n)}$ and $\bar{\beta}_0^{(n')}$, respectively, that meet
\begin{IEEEeqnarray}{rll}
	\RED \bar{\beta}_0^{(n)}\!-\!\bar{\beta}_0^{(n')} \!=\! \frac{1}{2\pi}\mathrm{mod}\left(\! -\kappa\!\left[(u_\x^{(n)}-u_\x^{(n')})L_\x A_\x(\boldsymbol{\Psi}_t^*,\boldsymbol{\Psi}_r^*) +(u_\y^{(n)}-u_\y^{(n')})L_\y A_\y(\boldsymbol{\Psi}_t^*,\boldsymbol{\Psi}_r^*)\right]\!,2\pi\right)\!, \, \forall n'\neq n,\quad\,\,\,
\end{IEEEeqnarray}
where  $\mathrm{mod}(x,y)$ denotes the $y$-modulus of real number $x$ which is needed due to the periodicity of the complex exponential function in \eqref{Eq:PhaseShiftCodebook}. Similar to the reflection codebooks, one simple option is to employ uniform discretization, i.e., {\RED$\mathcal{B}_0=\{\bar{\beta}_0^{\min},\bar{\beta}_0^{\min}+\Delta\bar{\beta}_0,\dots, \bar{\beta}_0^{\max}\}$, where $\Delta\bar{\beta}_0=\frac{\bar{\beta}_0^{\max}-\bar{\beta}_0^{\min}}{|\mathcal{B}_0|-1}$ and $[\bar{\beta}_0^{\min},\bar{\beta}_0^{\max}]\subset [-\frac{1}{2},\frac{1}{2}]$ is the  support of parameter $\bar{\beta}_0$, which can be chosen as a subset of  $[-\frac{1}{2},\frac{1}{2}]$  if prior knowledge about the range of $(\boldsymbol{\Psi}_t^*,\boldsymbol{\Psi}_r^*)$ is available.} 


\subsection{Mode Pre-selection for Online Optimization}\label{Sec:PreSelect}

For the end-to-end channel model in \eqref{Eq:E2Emodelsimple}, a given channel realization corresponds to a given number of scatterers in the environment. Therefore, due to the limited number of AoAs and AoDs pointing from the IRS to these scatterers, especially at high frequencies, only a limited number of phase-shift configurations in the reflection codebooks are suitable for reflecting the EM wave impinging from one of the scatterers/transmitters towards one of the other scatterers/receivers, see Fig.~\ref{Fig:SysMod}. In other words, for a given transmitter-receiver pair, the value of $\|\mathbf{H}_{n,m,k}\|_F$, where $\|\cdot\|_F$ denotes the Frobenius norm, is non-negligible only for a few of the transmission modes in the codebook. Therefore, before online optimization, we can first pre-select a subset of the possible transmission modes, whose indices are collected in set $\mathcal{M}$.  The exact criterion for this pre-selection depends on the application, of course. For instance, one may determine $\mathcal{M}$ based on the strength of the corresponding channel, i.e.,
\begin{IEEEeqnarray}{rll}\label{Eq:SelectedMode}
	\mathcal{M} = \big\{m|\exists (k,n): \|\mathbf{H}_{n,m,k}\|_F\geq \delta\big\},
\end{IEEEeqnarray}
where $\delta$ is a threshold. {\BLUE The value of $\delta$ is a design parameter which can be chosen to trade performance with complexity. In particular,  the smaller $\delta$ is chosen, the more modes are selected, which implies potentially higher performance at the cost of higher complexity of the subsequent online optimization.} Alternatively, one may consider more sophisticated criteria that account for the resulting interference at other users (e.g., for multi-user communications) or the information leakage to eavesdroppers (e.g., for secure communications).

\begin{figure}[t]\vspace{-0.5cm}
	\centering
	\includegraphics[width=0.8\textwidth]{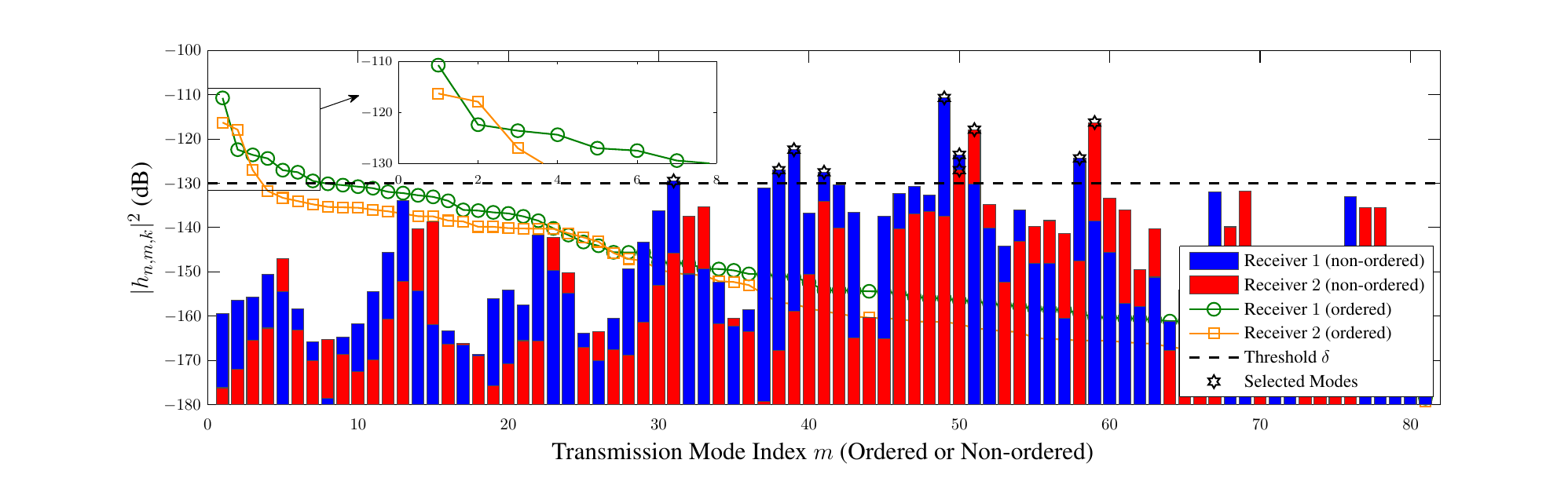}\vspace{-0.3cm}
	\caption{Illustration of codebook size reduction for online optimization. Channel gain $|h_{n,m,k}|^2$ for one tile and one channel realization versus the transmission mode index. The same IRS parameters and reflection codebook as for Fig.~\ref{Fig:Codebook} are assumed. \vspace{-0.5cm}}
	\label{Fig:CodewordReduction}
\end{figure}

 Fig.~\ref{Fig:CodewordReduction} shows an example for transmission mode selection for a system consisting of one single-antenna transmitter, an IRS, and two single-antenna receivers. The IRS parameters and the offline codebook are identical to those used in Fig.~\ref{Fig:Codebook}, i.e., the reflection codebooks have $\RED 9\times 9$ elements. We assume that there exist two paths (e.g., the direct path and one path via a scatterer) for each of the transmitter-to-IRS and IRS-to-receiver links. The channel gains are modeled as free space  path-loss where the transmitter-to-IRS and IRS-to-receiver distances are $\RED 1000\lambda$ (e.g., $\RED 60$~m for carrier frequency $5$~GHz). Fig.~\ref{Fig:CodewordReduction} shows $\|\mathbf{H}_{n,m,k}\|_F^2=|h_{n,m,k}|^2$ of receiver $k\in\{1,2\}$ for one IRS tile versus the index of the adopted transmission mode. For comparison,  we also plot  $|h_{n,m,k}|^2$ sorted in a descending order. Fig.~\ref{Fig:CodewordReduction} confirms that for a given channel realization, not all transmission modes corresponding to the $9\times9=81$ reflection codebook elements are useful. In fact, for threshold $\delta=-130$ dB (almost $20$ dB below the maximum $|h_{n,m,k}|^2$), only $10$ modes out of the $81$ possible modes are included in $\mathcal{M}$ according to \eqref{Eq:SelectedMode}, which significantly reduces the complexity of the subsequent online optimization stage.

\subsection{Online Optimization}\label{Sec:ExampleOpt}

The online configuration of the IRS requires the optimization of the binary mode selection variables, i.e., $s_{n,m},\,\,\forall n,m$. Moreover, there are typically other system parameters, e.g.,  the precoder at the transmitter, that have to be jointly optimized with $s_{n,m},\,\,\forall n,m$. Therefore, in general, the IRS online optimization problem formulated based on the proposed channel model in \eqref{Eq:E2Emodelsimple} belongs to the class of \textit{mixed integer programming} problems \cite{lee2011mixed}. For this class of optimization problems, efficient solution methods have been developed in the literature  \cite{lee2011mixed,ng2014robust,ghanem2019resource}. Nevertheless, the specific solution approach required for optimizing an IRS-assisted system depends on the considered system model and the adopted QoS criterion, of course. Therefore,  for concreteness, we focus on an IRS-assisted downlink system, {\BLUE where  $K$ single-antenna users are served by a BS with $N_t\geq K$ antennas  via linear precoding}. For this special case, the received signal of the $k$-th user in \eqref{Eq:E2Emodelsimple} simplifies as follows 
\begin{IEEEeqnarray}{rll}\label{Eq:E2Edownlink}
	y_k = \sum_{n=0}^N \sum_{m\in\mathcal{M}} s_{n,m} \mathbf{h}_{n,m,k}^\Herm\mathbf{Q}\mathbf{u} +z_k, \quad k=1,\dots,K,
\end{IEEEeqnarray}
where $\mathbf{u}\in\mathbb{C}^{N_r}$ contains the data for the $K$ users and satisfies $\mathsf{E}\{\mathbf{u}\mathbf{u}^\mathsf{H}\}=\mathbf{I}_{K}$ with $\mathsf{E}\{\cdot\}$ and $\mathbf{I}_{n}$ denoting  expectation and the $n\times n$ identity matrix, respectively. Moreover, $\mathbf{Q}=[\mathbf{q}_1,\dots,\mathbf{q}_{K}]\in\mathbb{C}^{N_t\times K}$ is a linear precoder matrix. Furthermore, $z_k\sim\mathcal{CN}(0,\sigma^2)$ denotes AWGN with mean zero and variance $\sigma^2$ impairing the $k$-th user and $\mathbf{h}_{n,m,k}=\mathbf{H}_{n,m,k}^\Herm$ represents the effective end-to-end channel between the BS and user $k$ via the $n$-th tile of the IRS for the $m$-th transmission mode. With a slight abuse of notation, we introduce the additional index $n=0$ and set $\mathbf{h}_{0,m,k}=\mathbf{H}_{d,k}^\Herm,\,\,\forall m$, to include the channel gains of the direct~link.

{\RED Let us define $\mathbf{S}=[\mathbf{s}_1,\dots,\mathbf{s}_N]$ where $\mathbf{s}_n\in\{0,1\}^{|\mathcal{M}|}$ is a binary vector with $s_{n,m}$ as  its $m$-th entry.} We wish to find the optimal $\mathbf{S}$ and $\mathbf{Q}$ that minimize the BS transmit power while guaranteeing a minimum signal-to-interference-noise ratio (SINR) for each user, i.e., 
\begin{IEEEeqnarray}{cll}\label{Eq:Prob}
	\underset{\mathbf{S}\in\{0,1\}^{|\mathcal{M}|\times N},\mathbf{Q}\in\mathbb{C}^{N_t\times K}}{\mathrm{minimize}}& \mathrm{tr}(\mathbf{Q}\mathbf{Q}^\mathsf{H}) \nonumber\\
	\mathrm{subject\,\,to}
		&\,\mathrm{C1:}\,\, \gamma_k(\mathbf{S},\mathbf{Q}) \geq \gamma_k^{\mathrm{thr}},\quad 
		&\,\mathrm{C2:}\,\, \sum_{m\in\mathcal{M}} s_{n,m}=1,\quad\forall n, 
\end{IEEEeqnarray}
where $\mathrm{tr}(\cdot)$ denotes the trace operator, $\gamma_k^{\mathrm{thr}}$ is the minimum required SINR of the $k$-th user, and $\gamma_k(\mathbf{S},\mathbf{Q})$ denotes the SINR of the $k$-th user, which is given by
\begin{IEEEeqnarray}{rll}\label{Eq:SNR}
	\gamma_k(\mathbf{S},\mathbf{Q}) 
	=\frac{\displaystyle\Big|\sum_{n,m}  s_{n,m} \bh_{n,m,k}^\Herm\bq_k\Big|^2}{\displaystyle\sum_{k'\neq k} \Big|\sum_{n,m} s_{n,m} \bh_{n,m,k}^\Herm\bq_{k'}\Big|^2+\sigma^2}.\quad
\end{IEEEeqnarray}

Problem \eqref{Eq:Prob} is non-convex due to 1) the product term  $s_{n,m}\mathbf{q}_k$, 2) the quadratic form of the denominator and numerator of \eqref{Eq:SNR}, and 3) the binary constraint $s_{n,m}\in\{0,1\}$. Nevertheless, various approaches have been proposed in the literature to cope with such non-convex optimization problems \cite{lee2011mixed,ng2014robust,ghanem2019resource}. In the following, we propose two efficient suboptimal solutions to \eqref{Eq:Prob}. {\BLUE We note that, since we assume $N_t\geq K$, for AoAs, AoDs, and channel gains generated randomly from continuous probability density functions, the problem in \eqref{Eq:Prob} is almost surely feasible since the end-to-end channel matrix $\mathbf{H}^{e2e}=[\bar{\mathbf{h}}_1,\dots,\bar{\mathbf{h}}_K]$,  where $\bar{\mathbf{h}}_k=\sum_{n,m}  s_{n,m} \mathbf{h}_{n,m,k}$, is full rank (i.e., ${\rm rank}(\mathbf{H}^{e2e})=K$) with probability one. Nevertheless, for ill-conditioned end-to-end channel matrices $\mathbf{H}^{e2e}$, the required transmit power may be high. In fact, IRSs provide a means to improve the well-conditioness of the end-to-end channel matrix $\mathbf{H}^{e2e}$.}

\subsubsection{\textbf{Alternating Optimization-Based Solution}}
{\RED Hereby, we develop a suboptimal solution based on  AO of $ \mathbf{s}_1,\dots,\mathbf{s}_N$, and $\mathbf{Q}$.}  However, straightforward AO of  $\mathbf{s}_n,\, \forall n$, and $\mathbf{Q}$ is not efficient as optimizing \eqref{Eq:Prob} with respect to $\RED\mathbf{s}_n$ for a given $\mathbf{Q}$ returns the solution to which the problem was initialized. To address this issue, we decompose  the precoder as $\mathbf{Q}=\sqrt{p}\widetilde{\mathbf{Q}}$, where $p=\mathrm{tr}(\mathbf{Q}\mathbf{Q}^\mathsf{H})$, i.e.,  $\mathrm{tr}(\widetilde{\mathbf{Q}}\widetilde{\mathbf{Q}}^{\mathsf{H}})=1$. Then, in Subproblem 1, we solve  \eqref{Eq:Prob}  jointly in $p$ and $\RED\mathbf{s}_n$ for given  $\RED\mathbf{s}_{n'},\,\forall n'\neq n$ and $\widetilde{\mathbf{Q}}$, and in Subproblem~2, we solve \eqref{Eq:Prob} for $\mathbf{Q}$ (or equivalently jointly for $p$ and  $\widetilde{\mathbf{Q}}$) for given $\RED\mathbf{s}_n,\,\,\forall n$. With this approach, transmit power $p$ is reduced in both subproblems while guaranteeing the users' SINR requirements.

\textbf{Subproblem~1 ({\RED Joint Power Minimization and Tile Configuration}):} {\RED First, we solve  \eqref{Eq:Prob}  jointly in $p$ and $\mathbf{s}_{n}$ for given $\mathbf{s}_{n'},\,\,\forall n'\neq n$, and $\widetilde{\mathbf{Q}}$. Exploiting the binary nature of $s_{n,m}$ and since $\mathbf{s}_{n'},\,\,\forall n'\neq n$, and $\widetilde{\mathbf{Q}}$ are fixed,  the SINR  in  \eqref{Eq:SNR} can be reformulated in terms of  $s_{n,m}$ and $p$ as follows
\begin{IEEEeqnarray}{rll}\label{Eq:SNR_snm}
	\gamma_k(\mathbf{S},\mathbf{Q}) 
	=\frac{ \sum_{m}   s_{n,m} p f_{n,m,k,k}}{\sum_{k'\neq k} \sum_{m}   s_{n,m} p f_{n,m,k,k'}+\sigma^2}
	\quad
	\text{with}
	\quad
	f_{n,m,k,k'} = \Big|\bh_{n,m,k}^\Herm\tilde{\bq}_{k'}+\sum_{n'\neq n,m}  s_{n',m} \bh_{n',m,k}^\Herm\tilde{\bq}_{k'}\Big|^2.\quad
\end{IEEEeqnarray}
Using the above notations, constraint $\mathrm{C1}$ is reformulated~as
\begin{IEEEeqnarray}{rll} 
	\widetilde{\mathrm{C1}}: \quad\sum_{m} s_{n,m} p
	\Big[f_{n,m,k}
	- \gamma_k^{\mathrm{thr}}\sum_{k'\neq k} f_{n,m,k'}
	\Big]	\geq \gamma_k^{\mathrm{thr}}\sigma^2.
\end{IEEEeqnarray}
Therefore, the first subproblem is given by
\begin{IEEEeqnarray}{cll}\label{Eq:ProbSubS}
	\text{P1:}\,\,\underset{p\geq 0,\,\, \mathbf{s}_{n}\in\{0,1\}^{|\mathcal{M}|}}{\mathrm{minimize}}\,\,p\quad
	\mathrm{subject\,\,to}\quad\widetilde{\mathrm{C1}},\mathrm{C2}.\quad\,\,
\end{IEEEeqnarray}
The solution of P1 is provided in the following lemma.}

{\RED\begin{lem}\label{Lem:ProbP1}
		The solution $(p^*,\mathbf{s}_{n}^*)$ to P1 is given by
		\begin{IEEEeqnarray}{cll}\label{Eq:P1sol}
			p^*=\max_{k=1,\dots,K} p_{m^*,k}
			\quad\text{and}\quad
			s_{n,m}^*=\begin{cases}
				1,&\mathrm{if}\,\,m=m^*\\
				0,&\mathrm{otherwise},
				\end{cases}
		\end{IEEEeqnarray}
		where 
		\begin{IEEEeqnarray}{cll}\label{Eq:Pmk} 
			m^*=\underset{m\in\mathcal{M}}{\mathrm{argmin}}
			\max_{k=1,\dots,K} p_{m,k},
			\quad
			p_{m,k} =  \begin{cases}
				\frac{\gamma_k^{\mathrm{thr}}\sigma^2}{f_{n,m,k}
					- \gamma_k^{\mathrm{thr}}\sum_{k'\neq k} f_{n,m,k'}},&\mathrm{if}\,\,f_{n,m,k}
				> \gamma_k^{\mathrm{thr}}\sum_{k'\neq k} f_{n,m,k'}\\
				+\infty,&\mathrm{otherwise}.
			\end{cases}
		\end{IEEEeqnarray}
	\end{lem}
\begin{IEEEproof}
	First, note that if $f_{n,m,k}
	\leq\gamma_k^{\mathrm{thr}}\sum_{k'\neq k} f_{n,m,k'}$  holds for any $k$, the considered transmission mode does not provide a feasible solution. Hence, let us focus on transmission modes for which  $f_{n,m,k}
>\gamma_k^{\mathrm{thr}}\sum_{k'\neq k} f_{n,m,k'}$ holds for all $k$. The left-hand side of constraint $\widetilde{\mathrm{C1}}$ is a monotonically increasing function in $p$. Thus, $p=p_{m,k}$ with $p_{m,k}$ given in \eqref{Eq:Pmk}  ensures that the inequality in $\widetilde{\mathrm{C1}}$ holds with equality for user $k$ using transmission mode $m$. The optimal $m^*$ is that $m$ which minimizes the power while ensuring that $\widetilde{\mathrm{C1}}$ holds for all users. This completes the proof.
\end{IEEEproof}
}

\textbf{Subproblem~2 ({\RED Joint Power Minimization and Beamforming Design}):} {\RED In this subproblem, we  solve \eqref{Eq:Prob} jointly for $p$ and  $\widetilde{\mathbf{Q}}$ (or equivalently $\mathbf{Q}$) for given $\mathbf{s}_n,\,\,\forall n$.}
Defining the positive semi-definite auxiliary matrix ${\mathbf{Q}}_k={\mathbf{q}}_k{\mathbf{q}}_k^\mathsf{H}$, constraint ${\mathrm{C1}}$ is reformulated as follows
\begin{IEEEeqnarray}{rll} 
\widetilde{\widetilde{\mathrm{C1}}}:\quad\mathrm{tr}\big({\bar{\mathbf{h}}_k}{\bar{\mathbf{h}}_k^{\mathsf{H}}}\mathbf{Q}_k\big)- \gamma_k^{\mathrm{thr}}\sum_{k'\neq k} \mathrm{tr}\big(\bar{\mathbf{h}}_k {\bar{\mathbf{h}}_k^{\mathsf{H}}}\mathbf{Q}_{k'}\big)
	\geq \gamma_k^{\mathrm{thr}}\sigma^2.
\end{IEEEeqnarray}
 Note that, by definition, $\mathbf{Q}_k$ has rank one which imposes a non-convex constraint, i.e., $\mathrm{rank}(\mathbf{Q}_k)= 1,\,\,\forall k$, has to hold where $\mathrm{rank}(\cdot)$ denotes the rank of a matrix. Nevertheless, the following lemma formally states that we can recover the optimal solution after dropping the rank~constraint.

\begin{lem}\label{Lem:Rank}
	There exists a $\mathbf{Q}_k,\forall k$, as the solution to problem
	\begin{IEEEeqnarray}{cll}\label{Eq:ProbRelaxedrank}
		\text{P2:} \,\,\underset{\mathbf{Q}_k\in\mathbb{C}^{N_t\times N_t}}{\mathrm{minimize}}\,\,&\sum_{k=1}^{K}\mathrm{tr}(\mathbf{Q}_k) \quad
		\mathrm{subject\,\,to}\quad\widetilde{\widetilde{\mathrm{C1}}},\,\,\mathrm{C3}:\mathbf{Q}_k\succeq\boldsymbol{0},
	\end{IEEEeqnarray}
	that has a rank equal to one, i.e., $\mathrm{rank}(\mathbf{Q}_k)= 1,\,\,\forall k$.
\end{lem}
\begin{IEEEproof}
	The proof follows similar steps as that in \cite[Appendix~B]{ng2014robust} and is provided in Appendix~\ref{App:Lem_Rank} for completeness.
\end{IEEEproof}


\begin{figure}[!t]
			\vspace{-0.8cm}
	\begin{minipage}{0.43\linewidth}
		\begin{algorithm}[H]
			\footnotesize
			\caption{\small AO-based Online Optimization}
			\begin{spacing}{1}\RED
				\begin{algorithmic}[1]\label{Alg:AOdesign}
					\STATE \textbf{input:} Number of iterations $N^{\rm itr}$, initial point $(\mathbf{S}^{(0)},\mathbf{Q}^{(0)})$, channel coefficients $\mathbf{h}_{n,m,k},\,\,\forall n,m,k$, and required SINRs $\gamma_k^{\mathrm{thr}},\,\,\forall k$. 
					\FOR{$i=1,\dots,N^{\rm itr}$}		
					\FOR{$n=1,\dots,N$}		
					\STATE Find $(\mathbf{s}_n^*,p^*)$ as the solution of P1 in \eqref{Eq:P1sol} for  $\mathbf{Q}^{(i-1)}$, $\mathbf{s}_{n'}^{(i)},\forall n'<n$, $\mathbf{s}_{n'}^{(i-1)},\forall n'>n$, and update $\mathbf{s}_n^{(i)}=\mathbf{s}_n^*$.		
					\ENDFOR		
					\STATE Find $\mathbf{Q}_k^*,\forall k$, as the solution of P2 in \eqref{Eq:ProbRelaxedrank} for $\mathbf{S}^{(i)}$ and update $\mathbf{Q}_k^{(i)}=\mathbf{Q}_k^*$.
					\STATE Set $\mathbf{Q}^{(i)}=[\mathbf{q}_1^{(i)},\dots,\mathbf{q}_{K}^{(i)}]$ with $\mathbf{q}_k^{(i)}=\sum_j\lambda_j\boldsymbol{\nu},\,\,\forall k$, where $\lambda_j,\,\,\forall j$, are the non-zero eigenvalues of $\mathbf{Q}_k^{(i)}$ and $\boldsymbol{\nu}$ is the eigenvector corresponding to the largest eigenvalue.
					\ENDFOR		
					\STATE \textbf{output:} $\mathbf{Q}=\mathbf{Q}^{(N^{\rm itr})}$ and $\mathbf{S}=\mathbf{S}^{(N^{\rm itr})}$.
				\end{algorithmic}
			\end{spacing}
		\end{algorithm}
	\end{minipage}%
	\begin{minipage}{0.02\linewidth}
		\quad
	\end{minipage}%
	\begin{minipage}{0.55\linewidth}
		\begin{algorithm}[H]
			\footnotesize
			\caption{\small Greedy Iterative Online Optimization}
			\begin{spacing}{1}
			\begin{algorithmic}[1]\label{Alg:GIdesign}
				\STATE \textbf{input:} Channel coefficients $\mathbf{h}_{n,m,k},\,\,\forall n,m,k$, and required SINRs $\gamma_k^{\mathrm{thr}},\,\,\forall k$. 
				\STATE Set $[\mathbf{S}]_{0,1}=1$ and $[\mathbf{S}]_{n,m}=0,\,\,\forall\,(n,m)\neq(0,1)$.
				\FOR{$n=1,\dots,N$}				
				\STATE Find $\mathbf{Q}_k^*,\forall k$, as the solution of P2 in \eqref{Eq:ProbRelaxedrank} for $\mathbf{S}$ and update $\mathbf{Q}_k^{(n)}=\mathbf{Q}_k^*$.
				\STATE Set $\mathbf{Q}^{(n)}=[\mathbf{q}_1^{(n)},\dots,\mathbf{q}_{K}^{(n)}]$ with $\mathbf{q}_k^{(n)}=\sum_j\lambda_j\boldsymbol{\nu},\,\,\forall k$, where $\lambda_j,\,\,\forall j$, are the non-zero eigenvalues of $\mathbf{Q}_k^{(n)}$ and $\boldsymbol{\nu}$ is the eigenvector corresponding to the largest eigenvalue.
				\STATE Select user $k^*$ based on \eqref{Eq:UserSelection}.
				\STATE Select the transmission mode of the $n$-th tile as $m^*$ based on \eqref{Eq:ModeSelection} and set $[\mathbf{S}]_{n,m^*}=1$.
				\ENDFOR		
				\STATE Set $\mathbf{S}^*=\mathbf{S}$ and find $\mathbf{Q}_j^*,\forall j$, as the solution of P2 in \eqref{Eq:ProbRelaxedrank} for $\mathbf{S}^*$.
				\STATE  Set $\mathbf{Q}^*=[\mathbf{q}_1^*,\dots,\mathbf{q}_{K}^*]$ with $\mathbf{q}_k^{*}=\sum_j\lambda_j\boldsymbol{\nu},\,\,\forall k$, where $\lambda_j,\,\,\forall j$, are the non-zero eigenvalues of $\mathbf{Q}_k^*$ and $\boldsymbol{\nu}$ is the eigenvector corresponding to the largest eigenvalue.
				\STATE \textbf{output:} $\mathbf{Q}^*$ and $\mathbf{S}^*$.
			\end{algorithmic}
			\end{spacing}
		\end{algorithm}
	\end{minipage}
	\begin{minipage}{0.01\linewidth}
		\quad
	\end{minipage}%
\end{figure} 

The proposed AO-based online optimization is summarized in Algorithm~\ref{Alg:AOdesign}. Note that the proof of Lemma~\ref{Lem:Rank} shows that $\mathbf{Q}_k$ as a solution of  \eqref{Eq:ProbRelaxedrank} has either a rank equal to one or facilitates the construction of another feasible solution with identical objective function value that has rank one, see line~6 of Algorithm~\ref{Alg:AOdesign}. {\BLUE Moreover, since Lemmas~\ref{Lem:ProbP1} and \ref{Lem:Rank} provide the globally-optimal solution of Subproblems P1 and P2, respectively, Algorithm~\ref{Alg:AOdesign} produces a sequence of non-increasing transmit powers $p$ that converges to a locally-optimal solution of \eqref{Eq:Prob} \cite{tseng2001convergence}. We illustrate the  convergence behavior of Algorithm~\ref{Alg:AOdesign} in Fig.~\ref{Fig:PowerAO} in Section~\ref{Sec:Sim}.} {\RED However, for any AO-based algorithm, different locally-optimal solutions may yield different performances.  Therefore, starting from a ``good'' initial point is crucial. To address this issue, in the following, we propose a greedy \textit{iterative} algorithm for the optimization of $\mathbf{S}$ that is not based on AO and always terminates after $N$ iterations. The solution of the greedy algorithm can provide an initial point for  Algorithm~\ref{Alg:AOdesign}.}


\subsubsection{\textbf{Greedy Iterative Solution}}
The proposed greedy algorithm is  performed in $N$ iterations where \textit{one tile} is configured in each iteration. In particular, this algorithm involves the following stages:

\textbf{Stage~1 (Precoder Design):} In each iteration, we first solve P2 in \eqref{Eq:ProbRelaxedrank} for a given $\mathbf{S}$ obtained from the previous iteration. Note that in the first iteration, we use $[\mathbf{S}^{(0)}]_{0,1}=1$ and $[\mathbf{S}^{(0)}]_{n,m}=0,\,\,\forall\,(n,m)\neq(0,1)$, i.e., only  the direct link exists. Let $\mathbf{Q}^{(n)}=[\mathbf{q}_1^{(n)},\dots,\mathbf{q}_{K}^{(n)}]$ denote the precoder designed in the $n$-th iteration.

\textbf{Stage~2 (User Selection):}  In the $n$-th iteration, we configure the $n$-th tile such that it improves the channel of one user. Since our goal is to minimize the BS's transmit power, we choose the user which contributes most to the power consumption of the BS, i.e.,  
\begin{IEEEeqnarray}{cll}\label{Eq:UserSelection}
	k^*=\underset{k\in\{1,\dots,K\}}{\mathrm{argmax}} \,\,\|\mathbf{q}_k^{(n)}\|_2.
\end{IEEEeqnarray}

\textbf{Stage~3 (IRS Configuration):} Now, from the tile codebook $\mathcal{M}$, we select that element which improves the end-to-end channel gain of the selected user $k^*$ the most, i.e.,
\begin{IEEEeqnarray}{cll}\label{Eq:ModeSelection}
	m^*=\underset{m\in\mathcal{M}}{\mathrm{argmax}} \,\,\Big\|\mathbf{h}_{n,m,k^*}+\sum_{n'=1}^{n-1} \sum_{m'\in\mathcal{M}} s_{n',m'} \mathbf{h}_{n',m',k^*}\Big\|_2.
\end{IEEEeqnarray}
Then, we set $[\mathbf{S}]_{n,m^*}=1$.

After $N$ iterations, we have readily determined $\mathbf{S}$. Thereby, we solve P2 in \eqref{Eq:ProbRelaxedrank} for the obtained $\mathbf{S}$ to optimize $\mathbf{Q}$. Algorithm~\ref{Alg:GIdesign} summarizes the proposed greedy algorithm.

{\BLUE \subsubsection{Complexity Analysis} In the following, we  analyze the complexities of Algorithms~1 and 2 and then compare them with the complexities of  benchmark schemes from the literature.  The complexity of Subproblem P1 in Algorithm~1 is linear in $|\mathcal{M}|$, i.e., $\mathcal{O}(|\mathcal{M}|)$, where $\mathcal{O}(\cdot)$ is the big-O notation.  Subproblem P2 in Algorithm~1 is a semidefinite programming (SDP) problem.  The complexity of an SDP problem is $\mathcal{O}(\sqrt{n}(mn^3+m^2n^2+m^3))$, where $n$ and $m$ denote the number of semidefinite cone constraints and the	dimension of the semidefinite cone, respectively \cite{najafi2019cran}. For Subproblem P2, we have $m=K$ and $n=N_t$, which assuming $N_t\geq K$ (needed to separate the  signals of single-antenna users via linear precoding  at the BS), leads to a complexity order of $\mathcal{O}(K N_t^{3.5})$. Hence, the overall complexity order of Algorithm~1 is $\mathcal{O}(N^{\rm itr}(N|\mathcal{M}|+K N_t^{3.5}))$. Using a similar analysis, the complexity order of Algorithm~2 is obtained as $\mathcal{O}(N(|\mathcal{M}|+K N_t^{3.5}))$.

The complexity of most IRS online  optimization algorithms for IRS-assisted systems in the literature that rely on the phase-shift model scales with the number of IRS unit cells $Q$. Thereby, regardless of the specific algorithm used for IRS optimization, the scalability of the  optimization framework developed in this paper is readily reflected in reducing the size of the search space from $|\mathcal{B}|^Q$ for the phase-shift model to $|\mathcal{M}|^N$ for the proposed transmission-mode model, where $|\mathcal{M}|$ and $N$ are design parameters that can be chosen to trade performance for complexity. Moreover, for the phase-shift model, various suboptimal IRS optimization algorithms have been developed in the literature whose complexity scales with $Q$ but in a polynomial manner.
 For example, the IRS optimization algorithms proposed in \cite{yu2020robust,pan2020multicell,guo2019weighted,bai2020latency,abeywickrama2020intelligent} have a complexity order which is at least cubic in $Q$, i.e., $\mathcal{O}(Q^3)$. In contrast, the complexities of the proposed online optimization algorithms do not directly depend on $Q$ and instead depend on the number of IRS tiles $N$ and the size of the reduced offline codebook  $|\mathcal{M}|$. In Section~\ref{Sec:Sim}, we consider an IRS with $Q=3600$ unit cells, which the algorithms in \cite{yu2020robust,pan2020multicell,guo2019weighted,bai2020latency,abeywickrama2020intelligent} may not be able to handle. In contrast, we choose $N=9$ and $|\mathcal{M}|=32$, which can be comfortably handled by Algorithms 1 and 2.}	

\subsection{Channel Estimation}\label{Sec:ChannelEstimation}

A detailed treatment of the channel estimation problem is beyond the scope of this paper and is left for future work, {\BLUE see e.g. \cite{zheng2019intelligent,alwazani2020intelligent} for channel estimation schemes for IRS-assisted systems.} Nevertheless, we briefly explain how the proposed offline codebook design and the end-to-end channel model in \eqref{Eq:E2Emodelsimple} can be exploited to simplify the channel acquisition problem. In particular, considering the model in \eqref{Eq:E2Emodelsimple}, the end-to-end channel  has to be estimated only for the transmission modes contained in the offline codebook. In other words, for each transmission mode, the IRS sets the unit cells according to the transmission mode, the BS sends a pilot signal, and the user estimates the end-to-end channel (or the related physical parameters). {\BLUE As a consequence, the channel estimation overhead and complexity scale with the number of transmission modes.  Therefore, to reduce the channel estimation overhead and to improve the channel estimation quality, the design of small-size offline codebooks is of high importance. Preliminary work in this direction has been reported in \cite{jamali2020intelligent}.} In practice, for the proposed model, it is not necessary to estimate the channel for all transmission modes and the channel estimation procedure can be terminated as soon as a few satisfactory transmission modes have been identified, according to e.g. \eqref{Eq:SelectedMode}.  {\BLUE Furthermore,  the overhead can be further reduced if instead of a naive brute-force search, the relevant transmission modes are identified using an efficient search over  hierarchical multi-resolution codebooks, see e.g. \cite{alkhateeb2014channel} for the application of such codebooks for channel estimation in millimeter wave communication systems.  
Nevertheless, as in conventional communication systems,  channel estimation errors are unavoidable. Therefore, the study of the  minimum channel estimation quality required to achieve a given QoS requirements and the design of robust transmission schemes in the presence of channel estimation errors are interesting directions for future~research, see e.g. \cite{Roth2018hybrid,jamali2016csi,yu2020robust} for communication system design under imperfect channel state information.}

\section{Simulation Results}\label{Sec:Sim}

\begin{table*}\vspace{-0.5cm}
	\caption{Default Values of System Parameters.\vspace{-0.7cm}} 	\label{Table:Parameter}
	\begin{center}
		\scalebox{0.6}{
			\begin{tabular}{|| c | c | c | c | c | c | c | c | c | c | c | c | c | c | c ||}\hline
				Parameter & $N_t$ & $K$ & $N$ & $L_\x,L_\y$ & $d_\x,d_\y$ & $Q$& $L_{\rm uc}$& $\tau $ & $(|\mathcal{B}_\x|,|\mathcal{B}_\y|,|\mathcal{B}_0|)$ & $|\mathcal{M}|$ &  $\gamma_j^{\mathrm{thr}}$ & $N_0$ & $N_F$ & $W$  \\ \hline
				Value & $16 $ $(4\times 4)$  & $2$ & $9$ $(3\times 3)$ &  $10\lambda$ &  $\lambda/2$& $3600$ & $0.8d_\x$ & $0.8$ & $(10,10,4)$ & $32$ &  $10$ dB & $-174$~dBm/Hz & $6$ dB & $20$ MHz 
				\\ \hline 
			\end{tabular}
		} 
	\end{center}
	\vspace{-0.6cm}
	\begin{center}
		\scalebox{0.6}{
			\begin{tabular}{|| c | c | c | c | c | c ||}\hline
				  &  & & & & \vspace{-0.45cm} \\
				Parameter & $(\rho_d,\rho_t,\rho_r)$  & $(L_d,L_t,L_r)$ & $(\hat{h}_d,\hat{h}_t,\hat{h}_r)$  & Range of $\phi$ and $\varphi$ & Range of $\theta$ 
				\\ \hline
				Value & $(4,3.2,0.8)\times 10^3\lambda$ (e.g. $(240,192,48)$~m at $5$~GHz)  & $(1,2,2)$ & $(-40,0,0)$ dB  & $[0,2\pi]$ & $[0,\pi/2]$ 
				\\ \hline 
			\end{tabular}
		} \vspace{-1cm}
	\end{center}
	
\end{table*}

In this section, we evaluate the performance of the resource allocation schemes derived in Section~\ref{Sec:ExampleOpt}.  We assume that both the direct and IRS-assisted links exist; however, the direct link is severely shadowed, which motivates the deployment of the IRS.  The channel gain for each effective path (i.e., the diagonal elements of $\boldsymbol{\Sigma}_d^{(j)}$, $\boldsymbol{\Sigma}_t$, and $\boldsymbol{\Sigma}_r^{(j)}$) is modeled as $h_i=\sqrt{\bar{h}_i\hat{h}_i}\tilde{h}_i,\,\,i\in\{d,r,t\},$ where $\bar{h}_i = \big(\frac{\lambda}{4\pi \rho_i}\big)^{2}$, $\hat{h}_i$, and $\tilde{h}_i\sim\mathcal{CN}(0,1)$ represent the free-space path-loss, large-scale shadowing/blockage, and small-scale Rayleigh fading, respectively, and the subscripts $d$, $t$, and $r$ refer to the BS-to-user, BS-to-IRS, and IRS-to-user paths, respectively.  The AoAs and AoDs at  BS and  IRS are generated as uniformly distributed random variables. We assume that the noise power at the receiver is given by $\sigma^2=WN_0N_F$ where $W$ is the bandwidth, $N_0$ represents the noise power spectral density, and $N_F$ denotes the noise figure. The simulation results shown in this	section have been averaged over  $10^3$ random channel realizations. We focus on IRSs with discrete tiles and the phase shift design in Proposition~\ref{Prop:Discrete}.  
The offline codebook is designed based on uniform discretization of $\bar{\beta}_\x$, $\bar{\beta}_\y$, and $\bar{\beta}_0$, as discussed in Section~\ref{Sec:Codebook}. For online optimization, we select a fixed number of transmission modes from the reflection codebook, $\mathcal{B}_\x\times\mathcal{B}_\y$, for each user $k$ based on the strength of the corresponding end-to-end channel gain $\|\mathbf{h}_{n,m,k}\|_2$.  Moreover, we include the entire wavefront phase codebook $\mathcal{B}_0$ for online optimization to allow for an efficient superposition of the signals arriving at the users from different tiles and from the direct link. Unless otherwise stated, we adopt the default values of the system parameters  provided in Table~\ref{Table:Parameter}.

\begin{figure}\vspace{-0.5cm}
	\begin{minipage}[c]{0.48\linewidth}
		\centering
		\includegraphics[width=1.05\textwidth]{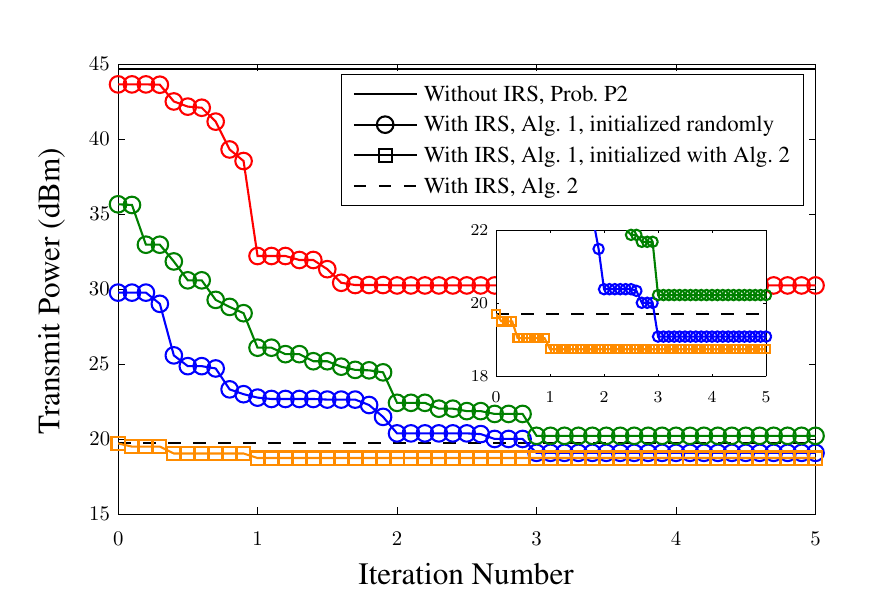}\vspace{-0.3cm}
		\caption{\BLUE Transmit power vs.~iteration number of Algorithm 1. Each iteration is further divided into ten sub-iterations (nine points for configuring all IRS tiles and one point for the BS precoder design) where the solutions obtained in lines 4 and 6  of Algorithm~1~are~shown. \vspace{-0.5cm}} 
		\label{Fig:PowerAO}
	\end{minipage}
	\begin{minipage}[c]{0.02\linewidth}
		\quad
	\end{minipage}
	\begin{minipage}[c]{0.48\linewidth}
		\centering
		\includegraphics[width=1.05\textwidth]{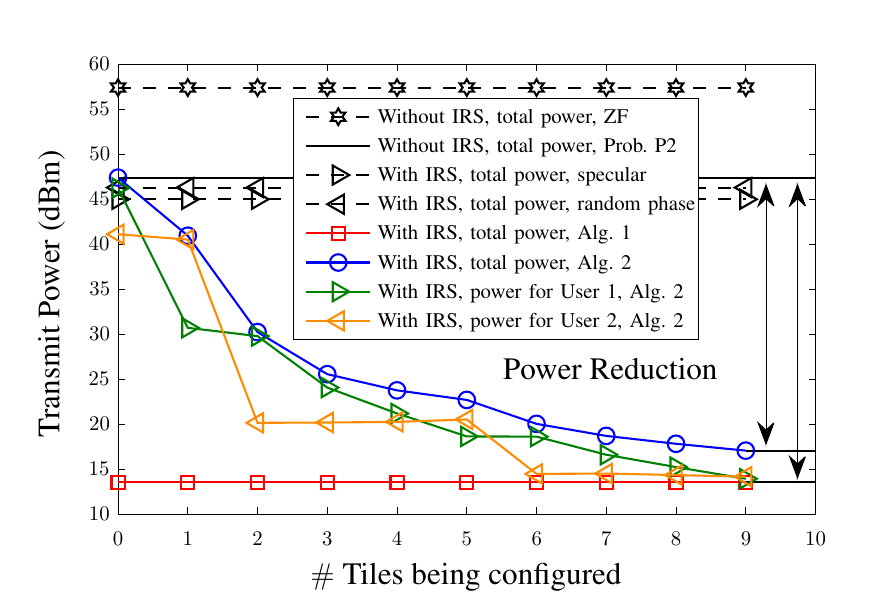}\vspace{-0.3cm}
		\caption{\BLUE Transmit power for Algorithm 2 vs. the number of configured tiles (iterations) for one channel realization. We note that different channel realizations are used for Figs.~\ref{Fig:PowerAO} and \ref{Fig:PowerConcept}. \vspace{-0.5cm}} 
		\label{Fig:PowerConcept}
	\end{minipage}
\end{figure}

{\BLUE First, in Fig.~\ref{Fig:PowerAO}, we study the convergence behavior of  Algorithm~1, where  for one channel realization,  the BS's transmit power is shown vs. the iteration number. For each iteration, the intermediate solutions obtained via Subproblem~P1 (nine points corresponding to alternatingly configuring the nine IRS tiles, cf. line~4 of Algorithm~1) and Subproblem~P2 (one point corresponding to the BS precoder design, cf. line ~6  of Algorithm~1) are also shown.  Moreover, we consider two different initializations for Algorithm~1, namely \textit{i)} random initialization of $\mathbf{S}^{(0)}$ and   initializing $\mathbf{Q}^{(0)}$ as the optimal precoder obtained from problem P2 for the given $\mathbf{S}^{(0)}$, and \textit{ii)} initializing $\mathbf{S}^{(0)}$ and $\mathbf{Q}^{(0)}$ as the output of Algorithm~2.  For comparison, we also show the performance of Algorithm~2 and a benchmark scheme which does not employ an IRS but uses the optimal precoder from Subproblem P2. As can be observed from Fig.~\ref{Fig:PowerAO}, while in general, the convergence behavior of Algorithm~1 depends on the initialization, Algorithm~1 converges within $1$-$3$ iterations for all considered initializations. In fact, extensive computer simulations revealed that Algorithm~1 typically converges within $1$-$5$ iterations. }
	

Next, to illustrate how the proposed greedy Algorithm~\ref{Alg:GIdesign} behaves, we show in Fig.~\ref{Fig:PowerConcept} the corresponding transmit power as a function of the number of iterations for one channel realization (note that different channel realizations are used for Figs.~\ref{Fig:PowerAO} and \ref{Fig:PowerConcept}). Fig.~\ref{Fig:PowerConcept} confirms that a considerable  power reduction is achieved in each iteration by configuring an additional tile. Moreover, we observe from Fig.~\ref{Fig:PowerConcept}  that, in each iteration, the power needed to serve the user that requires the higher BS transmit power to satisfy its QoS is significantly reduced which is in-line with the user selection strategy in line~6 of Algorithm~\ref{Alg:GIdesign}. For the considered channel realization, with the help Algorithm~2, an IRS with nine tiles of size  $10\lambda\times10\lambda$ can reduce the required transmit power by approximately $30$ dBm compared to the case when the IRS is absent.  {\BLUE For comparison, in Fig.~\ref{Fig:PowerConcept}, we also show the transmit power required for AO-based Algorithm~\ref{Alg:AOdesign}, which employs the solution of the greedy scheme for initialization, as well as two benchmark schemes, which do not use an IRS, namely \textit{i)} a zero-forcing precoder at the BS \cite{hu2017generalized} and \textit{ii)} the optimal precoder at the BS (according to problem P2), and two IRS-assisted benchmark schemes employing \textit{iii)} an IRS with random phase shifts \cite{wan2020broadband} and \textit{iv)} an IRS partitioned into tiles (similar to our setup) but with identical phase shifts for unit cells belonging to the same tile \cite{zheng2019intelligent}. For both IRS-assisted benchmark schemes \textit{iii)} and \textit{iv)}, the optimal precoder is used at the BS and for benchmark scheme \textit{iv)}, Algorithm~2 is applied for optimization of the tile phase shifts.  Moreover, for benchmark scheme \textit{iv)}, since each tile is able to only change the phase of the reflected wavefront but not the direction of the reflected wave, we refer to it as ``\textit{specular}".} We observe from Fig.~\ref{Fig:PowerConcept} that  to meet the users' QoS requirements without an IRS, the proposed optimal precoder requires a transmit power which is more than $10$ dBm lower than that needed by the ZF precoder. {\BLUE Moreover,  Fig.~\ref{Fig:PowerConcept} shows that an IRS-assisted system employing random phase shifts or  identical phase shifts for all unit-cells of a  tile can reduce the transmit power by less than $2$ and $1$ dB, respectively,  compared to a system without IRS. The reason for these small gains is that scattering the incident wave in all directions (as in the case of random unit-cell phase shifts) or reflecting it along the specular direction where the receiver may not be located yields negligible received power at the receivers.}  Finally, for this channel realization, less than $3$~dB power reduction is achieved by Algorithm \ref{Alg:AOdesign} with respect to Algorithm~\ref{Alg:GIdesign}. Nevertheless, extensive computer simulations have revealed that, for most channel realizations, the performance gain of the AO-based algorithm over the greedy algorithm is small.
Therefore, in the remainder of this section, we adopt the proposed greedy algorithm.

\begin{figure}\vspace{-0.5cm}
	\begin{minipage}[c]{0.48\linewidth}
		\centering\vspace{-0.2cm}
\includegraphics[width=1.05\textwidth]{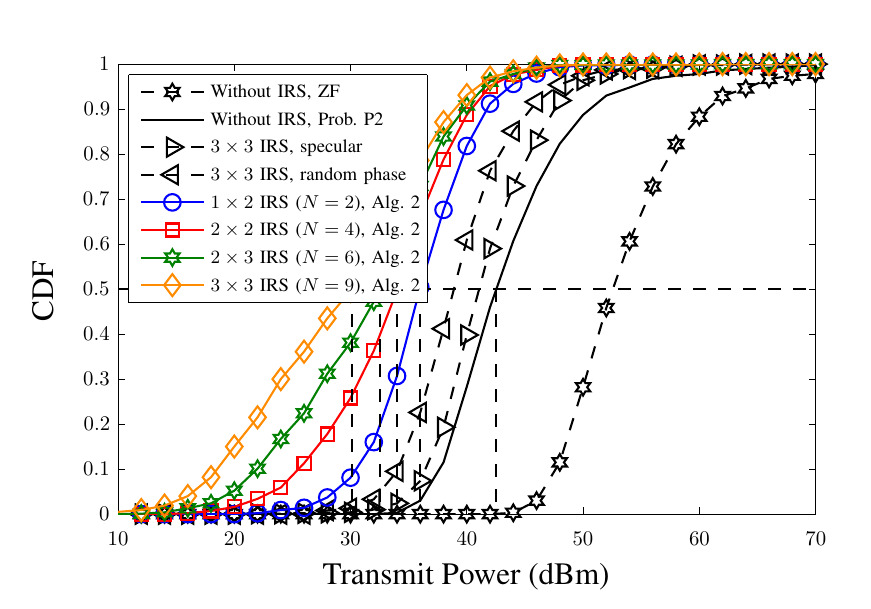}\vspace{-0.3cm}
\caption{\BLUE CDF of the required transmit power for different IRS sizes and benchmark schemes.\vspace{-0.5cm}} 
\label{Fig:PowerPDF}
	\end{minipage}
	\begin{minipage}[c]{0.02\linewidth}
		\quad
	\end{minipage}
	\begin{minipage}[c]{0.48\linewidth}
		\centering\vspace{-0.2cm}
		\includegraphics[width=1.05\textwidth]{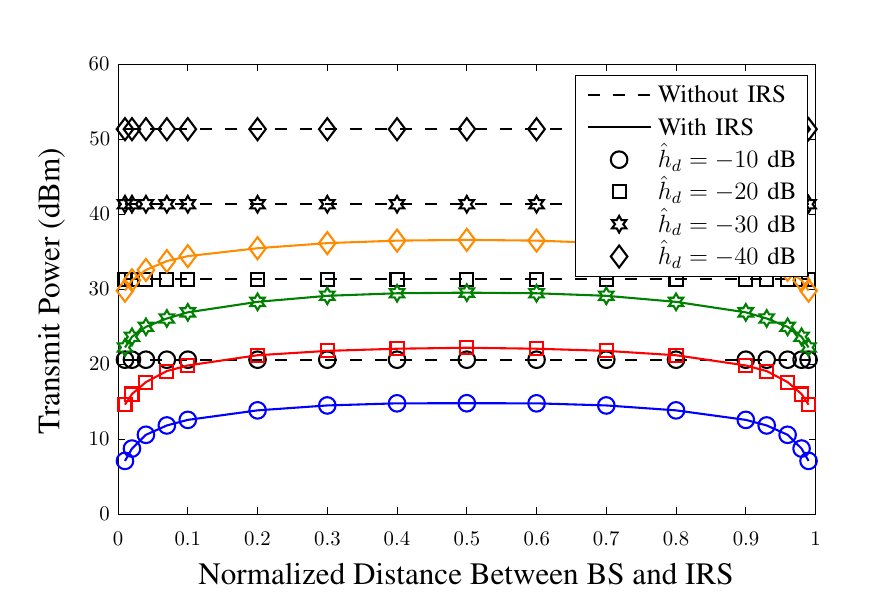}\vspace{-0.3cm}
		\caption{Required transmit power vs. the distance between the BS and the IRS normalized to the distance between the BS and the users.\vspace{-0.5cm}} 
		\label{Fig:PowerDistance}
	\end{minipage}
\end{figure}

In Fig.~\ref{Fig:PowerPDF}, we show the cumulative density function (CDF) of the required BS transmit power. This figure shows that the probability that the IRS can satisfy the users' QoS requirements with a small transmit power increases as the size of the IRS increases. For example, if the IRS is equipped with $N=0,2,4,6,9$ tiles (i.e., $Q=0,800,1600,2400,3600$ elements), for $50\%$  of the channel realizations, the BS needs a transmit power of less than $42$, $36$, $34$, $32$, and $30$ dBm, respectively, to satisfy the users' QoS requirements.  Nevertheless, Fig.~\ref{Fig:PowerPDF} also suggests that  there still exist channel realizations which require a high transmit power to meet the users' QoS requirements due to faded IRS-assisted links. This issue could be remedied by deploying more than one IRS and associating the users with the IRS that yields the strongest end-to-end channel. {\BLUE Similar to Fig.~\ref{Fig:PowerConcept}, we observe from Fig.~\ref{Fig:PowerPDF}  that an IRS employing random phase shifts and identical  phase shifts for all unit-cells of a tile can only marginally reduce the required transmit power compared to a system without an IRS.}


Next, we investigate the impact of the shadowing loss of the direct link, $\hat{h}_d$, and the position of the IRS on performance, where we assume $\rho_t+\rho_r=\rho_d=4\times 10^3\lambda$ (e.g., $400,240,120,42$~m at carrier frequencies of $3,5,10,28$~GHz, respectively). In Fig.~\ref{Fig:PowerDistance}, we show the required transmit power vs. the distance between the BS and the IRS normalized to the distance between the BS and the users, i.e., $\rho_t/\rho_d$, for several values of $\hat{h}_d$. We observe from this figure that the required transmit power is minimized when the IRS is either close to the BS or close to the users, see \cite{bjornson2019intelligent} for a  similar observation.  Moreover, Fig.~\ref{Fig:PowerDistance} shows that the relative amount of power saved by the IRS-assisted system compared to the system without IRS decreases as the direct link becomes stronger. 

\begin{figure}[t]\vspace{-0.5cm}
	\begin{minipage}[t]{0.48\linewidth}
	\centering
	\includegraphics[width=1.05\textwidth]{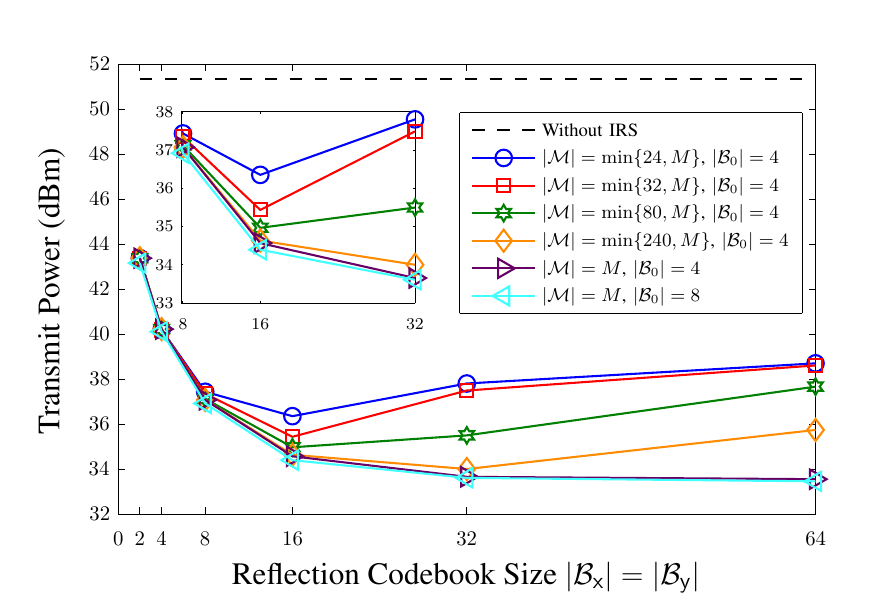}\vspace{-0.3cm}
	\caption{Required transmit power vs. the size of offline reflection codebooks $|\mathcal{B}_\x|=|\mathcal{B}_\y|$ for  different sizes of the selected online codebook  $|\mathcal{M}|$, different sizes of wavefront phase codebook~$|\mathcal{B}_0|$, and an IRS with $N=9$ tiles each of size $10\lambda\times10\lambda$. \vspace{-0.5cm}} 
	\label{Fig:PowerCodebookM}
\end{minipage}
	\begin{minipage}[c]{0.02\linewidth}
		\quad
	\end{minipage}
	\begin{minipage}[t]{0.48\linewidth}
	\centering
\includegraphics[width=1.05\textwidth]{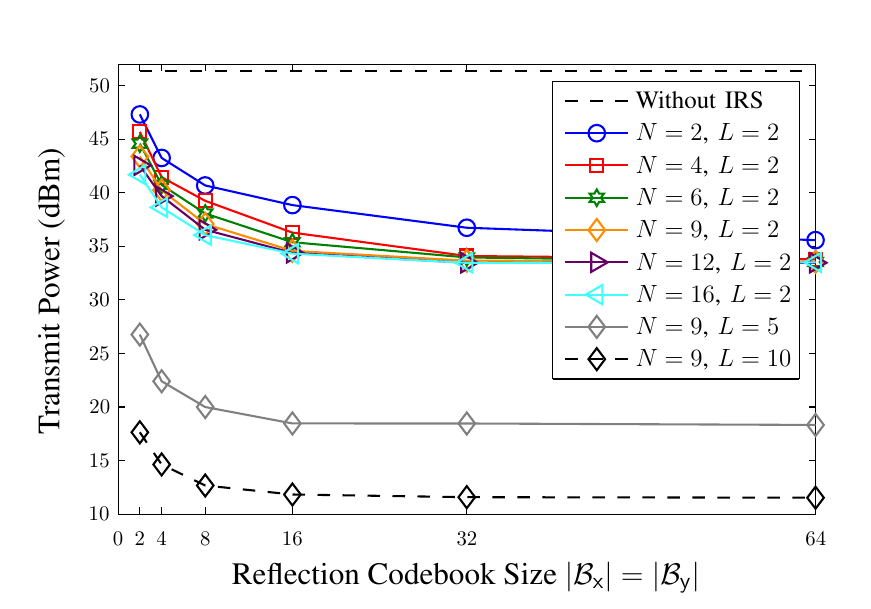}\vspace{-0.3cm}
\caption{Required transmit power vs. the size of offline reflection codebooks $|\mathcal{B}_\x|=|\mathcal{B}_\y|$ for $|\mathcal{B}_0|=4$, $|\mathcal{M}|=M=|\mathcal{B}_\x|\times|\mathcal{B}_\y|\times|\mathcal{B}_0|$, different numbers of IRS tiles $N$ while the total IRS area  is $30\lambda\times30\lambda$, and different numbers of scatterers $L_t=L_r=L$. \vspace{-0.5cm}} 
\label{Fig:PowerCodebookN}
	\end{minipage}
\end{figure}

In Fig.~\ref{Fig:PowerCodebookM},  we show the required transmit power vs. the size of the offline reflection codebooks, $|\mathcal{B}_\x|=|\mathcal{B}_\y|$, for different values of $|\mathcal{M}|$ (number of transmission modes used for online optimization) and different sizes of wavefront phase codebook $|\mathcal{B}_0|$. We observe from Fig.~\ref{Fig:PowerCodebookM}  that for fixed $|\mathcal{M}|$, as the size of the reflection codebook increases, the required transmit power first decreases and then increases. The reason for this behavior is that when $|\mathcal{M}|$ is fixed but $|\mathcal{B}_\x|$ and $|\mathcal{B}_\y|$ are large, only the transmission modes that reflect the incident beam along angles around the AoD with the strongest path to the users are selected and the transmission modes reflecting along other AoDs are removed by the pre-selection of $\mathcal{M}$ and do not participate in the online optimization. This has a negative impact on the system performance for large $|\mathcal{B}_\x|$ and $|\mathcal{B}_\y|$. However, if the pre-selection is removed, i.e., $|\mathcal{M}|=M$, then the required transmit power monotonically decreases as $|\mathcal{B}_\x|,|\mathcal{B}_\y|$ increase. 
On the other hand, recall that the pre-selection of $\mathcal{M}$ reduces the complexity of online optimization, cf. \eqref{Eq:ProbRelaxedrank} and \eqref{Eq:ModeSelection}. In fact, for $|\mathcal{B}_\x|=|\mathcal{B}_\y|=8$ and $|\mathcal{B}_0|=4$,  using $|\mathcal{M}|=24$ transmission modes (out of the $M=8\times 8\times 4=256$ entries of the offline codebook) causes no performance degradation and for $|\mathcal{B}_\x|=|\mathcal{B}_\y|=16$ and $|\mathcal{B}_0|=4$,  using $|\mathcal{M}|=24$ transmission modes (out of the $M=16\times 16\times 4=1024$ entries of the offline codebook) requires   an approximately only two dB higher transmit power compared to the case when all $M=1024$ elements of the offline codebook are used  for online optimization. 
Moreover, Fig.~\ref{Fig:PowerCodebookM} suggests that an additional power reduction can be achieved by increasing the size of the wavefront phase codebook from 4 to 8; nevertheless, this gain~is~small. 


Finally, we study the impact of the tile size and the number of channel scatterers on the system performance.  Fig.~\ref{Fig:PowerCodebookN} shows the required transmit power vs. the size of offline reflection codebooks $|\mathcal{B}_\x|=|\mathcal{B}_\y|$ for different numbers of tiles $N$ while the total  IRS  size is kept fixed as $30\lambda\times30\lambda$ and different numbers of channel scatterers for the BS-to-IRS link, $L_t$, and IRS-to-user links, $L_r$, where we assumed $L_t=L_r=L$. As can be seen from Fig.~\ref{Fig:PowerCodebookN}, the  required transmit power decreases as the number of tiles increases; however, increasing the number of tiles beyond $N=9$ yields only a small power reduction particularly for large reflection codebook sizes. In other words, this figure confirms that online optimization of the large number of sub-wavelength unit cells on the IRS (e.g., $3600$ unit cells for the example in Fig.~\ref{Fig:PowerCodebookN}) is not necessary and a much smaller number of  tiles (e.g., $N=9$ in Fig.~\ref{Fig:PowerCodebookN}) is sufficient for efficient online optimization. Nevertheless, we emphasize that, in general, the  number of tiles required to achieve a certain performance depends on various system parameters including the number of transmitters/receivers and the number of scatterers in the channel. Furthermore, we observe from Fig.~\ref{Fig:PowerCodebookN} that as the number of channel scatterers in the environment increases, the required transmit power significantly decreases. This is due to the fact that as the number of scatterers increases, there are more  AoAs and AoDs for the IRS to exploit for selecting strong BS-to-IRS and IRS-to-user paths.

\section{Conclusions}
\label{Sec:Conclusions}

In this paper, we have presented a scalable optimization framework for large IRSs based on a physics-based channel model for IRS-assisted wireless systems. In particular, we partitioned the IRS unit cells into several subsets, i.e., tiles, and derived the tile response function for both continuous and discrete tiles. Furthermore, we have developed a physics-based end-to-end channel model for IRS-assisted communications and proposed to optimize the IRS  in two stages: an offline design stage and an online optimization stage.  For offline design, we jointly designed the unit cells of a tile  for the support of different transmission modes.
For online optimization, we formulated a mixed-integer programming problem, where for each fading realization, the objective is to select the best transmission mode for each tile such that a desired QoS is maximized. Moreover, for an exemplary downlink system, we have studied the minimization of the BS transmit power subject to QoS constraints for the users and derived two corresponding algorithms employing  AO and a greedy approach, respectively. 
Computer simulation results have shown that the proposed modeling and optimization framework can be used to efficiently configure large IRSs containing thousands of elements. Furthermore, we have shown that increasing the number of tiles for a given IRS size beyond a certain number  yields a negligible performance improvement. 


 \appendices
 
 {\BLUE 
 \section{Proof of Lemma~\ref{Lem:Pathloss}}
 \label{App:LemPathloss}
 
 The radiation power intensity (Watt/m$^2$) at the IRS is obtained as $S_{t}=\frac{U_{\rm tx}}{\rho_t^2}=\frac{P_{\rm tx}D_{\rm tx}}{4\pi\rho_t^2}$, where $U_{\rm tx}$ is the radiation power intensity in Watt/solid angle \cite{balanis2005antenna}. Moreover, for plane waves, $S_{t}$ is related to the electric and magnetic fields through the time-averaged Poynting vector $\mathbf{S}_t=\frac{1}{2}\Re\{\mathbf{E}_t\times\mathbf{H}_t^*\}$ by $S_{t}=\|\mathbf{S}_t\|=\frac{|E_t|^2}{2\eta}$, where $\mathbf{E}_t$ and $\mathbf{H}_t$ denote the incident electric and magnetic vector fields, respectively, $\eta$ is the characteristic impedance, $\Re\{\cdot\}$ is the real part of a complex number, $\times$ denotes the cross product between two vectors, and $(\cdot)^*$ is the complex conjugate \cite{balanis2012advanced}. Furthermore,  \eqref{Eq:Gdef} relates the strength of the incident electric field at the receiver to that at the IRS via tile response function $g$ according to $|E_r|^2=\frac{g^2}{4\pi\rho_r^2}|E_t|^2$. Finally, the  power  collected by the receiver can be obtained as $P_{\rm rx}=S_rA_{\rm rx}$, where $S_r=\frac{|E_r|^2}{2\eta}$ is the radiation power intensity (Watt/m$^2$) at the receiver and  $A_{\rm rx}=\frac{D_{\rm rx}\lambda^2}{4\pi}$ is the effective area of the receive antenna  \cite{balanis2005antenna}. Combining these results yields $\frac{P_{\rm rx}}{P_{\rm tx}}=D_{\rm tx}D_{\rm rx}{\rm PL}_{\rm IRS}$ with ${\rm PL}_{\rm IRS}$ given in \eqref{Eq:Pathloss}. This completes the proof.
}

 \section{Proof of Proposition~\ref{Prop:Continuous}}
 \label{App:PropContinuous}

 The electric field $\mathbf{E}_r$ is a vector that is defined for each point in space by \eqref{Eq:HEV} and \eqref{Eq:EintegralFF}. Using a spherical coordinate system, at any point $(\rho,\theta,\phi)$, the electric field $\mathbf{E}_r$ (and the magnetic field $\mathbf{H}_r$) is a vector that in general has components in the radial $\mathbf{e}_{\rho}$, elevation $\mathbf{e}_{\theta}$, and azimuth $\mathbf{e}_{\phi}$ directions. However, the elevation and azimuth components decay with $1/\rho$ whereas the radial component decays faster. Therefore, in the far field, the impact of the radial component becomes negligible and the electric and magnetic fields become perpendicular to the propagation direction  \cite[Ch. 6.7]{balanis2012advanced}. In addition, for the elevation and azimuth components, $\mathbf{E}_r$ and $\mathbf{H}_r$ in \eqref{Eq:HEV} simplify~to
 \begin{IEEEeqnarray}{lll}\label{Eq:EHV_farfield}
 	\mathbf{E}_r = -\jj\omega \mathbf{V} \quad\text{and}\quad \mathbf{H}_r  = -\jj\frac{\omega}{\eta} \mathbf{e}_{\rho} \times \mathbf{V}.
 \end{IEEEeqnarray}
 To find the elevation and azimuth components of the electric and magnetic fields at the observation point specified by $(\rho_r,\theta_r,\phi_r)$, we first find the corresponding generating components of $\mathbf{J}_r$ using  \eqref{Eq:Current_r} as
 \begin{IEEEeqnarray}{lll}
 	J_\theta &= \tilde{J}_\theta \e^{\jj \kappa [A_\x(\boldsymbol{\Psi}_t)x+A_\y(\boldsymbol{\Psi}_t)y]}\e^{\jj\beta(x,y)}  
 	\quad\text{and}\quad
 	J_\phi &=\tilde{J}_\phi \e^{\jj \kappa[A_\x(\boldsymbol{\Psi}_t)x+A_\y(\boldsymbol{\Psi}_t)y]}\e^{\jj\beta(x,y)}, 
 \end{IEEEeqnarray}
 where $\tilde{J}_\theta=J_\x\cos(\theta_r)\cos(\phi_r)+J_\y\cos(\theta_r)\sin(\phi_r)$ and $\tilde{J}_\phi=-J_\x\sin(\phi_r)+J_\y\cos(\phi_r)$.
 Therefore, the elevation and azimuth components of $\mathbf{V}(\boldsymbol{\Psi}_r)$ in \eqref{Eq:EintegralFF}, denoted by $V_{\theta}(\theta_r)$ and $V_{\phi}(\phi_r)$, respectively, are found as 
 \begin{IEEEeqnarray}{lll}
 	V_{\theta}(\theta_r) =   \frac{\mu \e^{-\jj \kappa\rho_r}}{4\pi \rho_r}  \int_{x=-L_\x/2}^{L_\x/2}\int_{y=-L_\y/2}^{L_\y/2} \tilde{J}_\theta  \e^{\jj \kappa[A_\x(\boldsymbol{\Psi}_t)x+A_\y(\boldsymbol{\Psi}_t)y]} \e^{\jj\beta(x,y)} \e^{\jj \kappa \sqrt{x^2+y^2} \cos(\alpha)} \mathrm{d}x\mathrm{d}y \qquad \IEEEyesnumber\IEEEyessubnumber \label{Eq:Vtheta} \\
 	V_{\phi}(\phi_r)  =   \frac{\mu \e^{-\jj \kappa\rho_r}}{4\pi \rho_r}  \int_{x=-L_\x/2}^{L_\x/2}\int_{y=-L_\y/2}^{L_\y/2} \tilde{J}_\phi \e^{\jj \kappa[A_\x(\boldsymbol{\Psi}_t)x+A_\y(\boldsymbol{\Psi}_t)y]} \e^{\jj\beta(x,y)} \e^{\jj \kappa \sqrt{x^2+y^2} \cos(\alpha)} \mathrm{d}x\mathrm{d}y.  \IEEEyessubnumber
 \end{IEEEeqnarray}
 In the far field, we obtain $\cos(\alpha)= \frac{(A_\x(\boldsymbol{\Psi}_r),A_\y(\boldsymbol{\Psi}_r),A_\z(\boldsymbol{\Psi}_r)\cdot(x,y,0)}{\|(x,y,0)\|}
 = \frac{A_\x(\boldsymbol{\Psi}_r)x+A_\y(\boldsymbol{\Psi}_r)y}{\sqrt{x^2+y^2}}$.
 Using \eqref{Eq:EHV_farfield}, \eqref{Eq:Vtheta}, and the phase shift $\beta(x,y)$ in \eqref{Eq:BetaContinuous}, the elevation component of the electric vector field, denoted by $E_r^{\theta}(\theta_r)$, is obtained as
 \begin{IEEEeqnarray}{lll}\label{Eq:Etheta_int} 
 	E_r^{\theta}(\theta_r) \!=\!  \widetilde{E}_{\theta} \! \int_{x=-L_\x/2}^{L_\x/2} \e^{\jj \kappa [A_\x(\boldsymbol{\Psi}_t,\boldsymbol{\Psi}_r) - A_\x(\boldsymbol{\Psi}_t^*,\boldsymbol{\Psi}_r^*)] x} \mathrm{d}x \int_{y=-L_\y/2}^{L_\y/2} \!  \e^{\jj \kappa [A_\y(\boldsymbol{\Psi}_t,\boldsymbol{\Psi}_r) - A_\y(\boldsymbol{\Psi}_t^*,\boldsymbol{\Psi}_r^*)] y} \mathrm{d}y,  
 \end{IEEEeqnarray}
 where $\widetilde{E}_{\theta}=\frac{-\jj \kappa\eta \e^{-\jj k\rho_r+\jj\beta_0}}{4\pi \rho_r} \tilde{J}_\theta$. To solve the integrals in \eqref{Eq:Etheta_int}, we use the integral identity $\int_{z=-c/2}^{c/2} \e^{\jj Az} \mathrm{d}z= c\,\mathrm{sinc}\left(\frac{Ac}{2}\right)$ \cite[p. 593]{balanis2012advanced}.
 This leads to
 \begin{IEEEeqnarray}{lll} 
 	\!\!E_r^{\theta}(\theta_r) \!\!=\!\!  \widetilde{E}_{\theta}L_\x L_\y\, \mathrm{sinc}\!\left(\!\frac{\kappa L_\x [A_\x(\boldsymbol{\Psi}_t,\boldsymbol{\Psi}_r) \!-\! A_\x(\boldsymbol{\Psi}_t^*,\boldsymbol{\Psi}_r^*)]}{2}\!\right) \!\mathrm{sinc}\!\left(\!\frac{\kappa L_\y [A_\y(\boldsymbol{\Psi}_t,\boldsymbol{\Psi}_r) \!-\! A_\y(\boldsymbol{\Psi}_t^*,\boldsymbol{\Psi}_r^*)]}{2}\!\right)\!\!. \quad\,\,\,\,
 \end{IEEEeqnarray}
 The azimuth component of the electric vector field, denoted by $E_r^\phi(\phi_r)$, is obtained in a similar manner as follows
 \begin{IEEEeqnarray}{lll} 
 	E_r^\phi(\phi_r) \!=\!  \widetilde{E}_{\phi}L_\x L_\y\, \mathrm{sinc}\!\left(\!\frac{\kappa L_\x [A_\x(\boldsymbol{\Psi}_t,\boldsymbol{\Psi}_r) \!-\! A_\x(\boldsymbol{\Psi}_t^*,\boldsymbol{\Psi}_r^*)]}{2}\!\right) \!\mathrm{sinc}\!\left(\!\frac{\kappa L_\y [A_\y(\boldsymbol{\Psi}_t,\boldsymbol{\Psi}_r) \!-\! A_\y(\boldsymbol{\Psi}_t^*,\boldsymbol{\Psi}_r^*)]}{2}\!\right)\!\!,
 \end{IEEEeqnarray}
 where $\widetilde{E}_{\phi}=\frac{-\jj \kappa\eta \e^{-\jj k\rho_r+\jj\beta_0}}{4\pi \rho_r}\tilde{J}_\phi$. Recalling that the radial component is negligible,  the magnitude of  the electric field (up to a sign since  the sinc functions can be negative) is found~as 
 \begin{IEEEeqnarray}{lll} 
 	\|\mathbf{E}_r(\boldsymbol{\Psi}_r)\|=\sqrt{|E_r^\theta(\theta_r)|^2+|E_r^\phi(\phi_r)|^2} \nonumber \\
 	= C L_\x L_\y \, \Big|\mathrm{sinc}\left(\frac{\kappa L_\x [A_\x(\boldsymbol{\Psi}_t,\boldsymbol{\Psi}_r) - A_\x(\boldsymbol{\Psi}_t^*,\boldsymbol{\Psi}_r^*)]}{2}\right) \mathrm{sinc}\left(\frac{\kappa L_\y [A_\y(\boldsymbol{\Psi}_t,\boldsymbol{\Psi}_r) - A_\y(\boldsymbol{\Psi}_t^*,\boldsymbol{\Psi}_r^*)]}{2}\right)\!\Big|, \quad\,\,\,
 \end{IEEEeqnarray}
 where $C = \sqrt{|\widetilde{E}_{\theta}|^2+|\widetilde{E}_{\phi}|^2} 
 = \frac{\kappa \eta}{4\pi \rho_r} \sqrt{\tilde{J}_\theta^2+\tilde{J}_\phi^2} 
 =\frac{E_0}{\sqrt{4\pi \rho_r^2}} \frac{\sqrt{4\pi}\rho_\mathrm{eff}}{\lambda} \widetilde{g}(\boldsymbol{\Psi}_t,\boldsymbol{\Psi}_r)$
 and $\widetilde{g}(\boldsymbol{\Psi}_t,\boldsymbol{\Psi}_r)$ is given in \eqref{Eq:Gtilde}. Substituting $\|\mathbf{E}_r(\boldsymbol{\Psi}_r)\|$ into \eqref{Eq:Gdef} leads to the magnitude of $g(\boldsymbol{\Psi}_t,\boldsymbol{\Psi}_r)$ given in \eqref{Eq:AmplContinuous}. Moreover, the phase of $g(\boldsymbol{\Psi}_t,\boldsymbol{\Psi}_r)$ (up to a sign) is given by $\pi/2+\beta_0$. This completes the proof.

 \section{Proof of Proposition~\ref{Prop:Discrete}} 
 \label{App:PropDiscrete}

 The proof is similar to that provided in \cite[p. 963]{balanis2005antenna} for the analysis of uniform planar arrays. In particular,
 employing identities
 \begin{IEEEeqnarray}{rll}
 	\sum_{n=0}^{N-1} a^n= \frac{1-a^N}{1-a} \quad\text{and} \quad \sin(b)=\frac{\e^{\jj b}-\e^{-\jj b}}{2\jj},
 \end{IEEEeqnarray}
 and setting $a=\e^{\jj A}$, we have
 \begin{IEEEeqnarray}{rll}
 	\sum_{n=0}^{N-1} \e^{\jj An} = \frac{1-\e^{\jj NA}}{1-\e^{\jj A}} 
 	= \frac{\e^{\jj NA/2}}{\e^{\jj A/2}}\times\frac{\e^{-\jj NA/2}-\e^{\jj NA/2}}{\e^{-\jj A/2}-\e^{\jj A/2}}
 	= \e^{\jj(N-1)A/2} \times \frac{\sin(NA/2)}{\sin(A/2)}.
 \end{IEEEeqnarray}
 Now, using the above identity and the $\beta_{n_\x,n_\y}$ given in  \eqref{Eq:Beta_discrete}, we have 
 \begin{IEEEeqnarray}{rll}
 	&\sum_{n_\x=-\frac{Q_\x}{2}+1}^{\frac{Q_\x}{2}} \e^{\jj \kappa d_\x [ A_\x(\boldsymbol{\Psi}_t)+ A_\x(\boldsymbol{\Psi}_r)]n_\x+\jj\beta_{n_\x}} 
 	\overset{(a)}{=}\sum_{\tilde{n}_\x=0}^{Q_\x-1} \e^{\jj \kappa d_\x [ A_\x(\boldsymbol{\Psi}_t,\boldsymbol{\Psi}_r)-A_\x(\boldsymbol{\Psi}_t^*,\boldsymbol{\Psi}_r^*)](\tilde{n}_\x-Q_\x/2+1)+\jj\beta_{0}/2} \nonumber\\
 	&= \e^{\jj\beta_0/2} \e^{-\jj W_\x \big(\frac{Q_\x}{2}-1\big)}  \sum_{n_\x=0}^{Q_\x-1} \e^{\jj W_\x n_\x} = \e^{\jj\beta_0/2} \e^{\jj W_\x/2} \times \frac{\sin(Q_\x W_\x/2)}{\sin(W_\x/2)},
 \end{IEEEeqnarray}
 where $W_\x = \kappa d_\x[A_\x(\boldsymbol{\Psi}_t,\boldsymbol{\Psi}_r)-A_\x(\boldsymbol{\Psi}_t^*,\boldsymbol{\Psi}_r^*)]$ and for equality $(a)$, we used $\tilde{n}_\x=n_\x+\frac{Q_\x}{2}-1$. Applying the same approach to simplify to the second sum in \eqref{Eq:Discrete}, we obtain the amplitude of $g^{\mathsf{d}}(\boldsymbol{\Psi}_t,\boldsymbol{\Psi}_r)$ (up to a sign)  as 
 \begin{IEEEeqnarray}{rll} 
 	g^{\mathsf{d}}_{||}(\boldsymbol{\Psi}_t,\boldsymbol{\Psi}_r) = 
 	|g_{\rm uc}(\boldsymbol{\Psi}_t,\boldsymbol{\Psi}_r)|  \times \frac{\sin(Q_\x W_\x/2)}{\sin(W_\x/2)}\times \frac{\sin(Q_\y W_\y/2)}{\sin(W_\y/2)},
 \end{IEEEeqnarray}
 where $W_\y = \kappa d_\y[A_\y(\boldsymbol{\Psi}_t,\boldsymbol{\Psi}_r)-A_\y(\boldsymbol{\Psi}_t^*,\boldsymbol{\Psi}_r^*)]$. The phase of $g^{\mathsf{d}}(\boldsymbol{\Psi}_t,\boldsymbol{\Psi}_r)$ (up to a sign)  can be found~as
 \begin{IEEEeqnarray}{rll}
 	 g^{\mathsf{d}}_{\angle}(\boldsymbol{\Psi}_t,\boldsymbol{\Psi}_r)  = \angle g_{\rm uc}(\boldsymbol{\Psi}_t,\boldsymbol{\Psi}_r)+\frac{W_\x}{2}+\frac{W_\y}{2} + \beta_0 = \frac{\pi}{2}+\frac{W_\x}{2}+\frac{W_\y}{2} + \beta_0.
 \end{IEEEeqnarray}
 This leads to the amplitude and phase functions given in Proposition~\ref{Prop:Discrete} and concludes the proof.

 \section{Proof Of Lemma~\ref{Lem:Rank}}
 \label{App:Lem_Rank}
 
 The  problem in \eqref{Eq:ProbRelaxedrank} is jointly convex with
 respect to the optimization variables $\mathbf{Q}_k$ and satisfies Slater’s
 constraint qualification \cite{boyd2004convex}. Hence, strong duality holds
 and solving the dual problem of \eqref{Eq:ProbRelaxedrank} yields the optimal primal solution. The Lagrangian dual function of the problem in \eqref{Eq:ProbRelaxedrank}
 can be formulated as
 \begin{IEEEeqnarray}{lll}\label{Eq:Lagrange}
 	\mathcal{L}(\mathbf{Q}_k,\lambda_k,\mathbf{X}_k,\forall k) = \sum_{k} \mathrm{tr}\big(\mathbf{Q}_k\big)
 	- \sum_{k} \mathrm{tr}\big(\mathbf{X}_k\mathbf{Q}_k\big) 
 	- \sum_{k}\lambda_k \boldsymbol{\Upsilon}_k(\mathbf{Q}_k,\forall k),
 \end{IEEEeqnarray}
 where $\boldsymbol{\Upsilon}_k(\mathbf{Q}_k,\forall k)=\!\!\!\!\!\!\!\!\displaystyle\sum_{n,m,n',m'}\!\!\!\! \tilde{s}_{n,m}^{n',m'} 
 \big[\mathrm{tr}\big({\mathbf{h}_{n,m,k}}{\mathbf{h}_{n',m',k}^{\mathsf{H}}}\mathbf{Q}_k\big)- \gamma_k^{\mathrm{thr}}\sum_{k'\neq k} \mathrm{tr}\big({\mathbf{h}_{n,m,k}} {\mathbf{h}_{n',m',k'}^{\mathsf{H}}}\mathbf{Q}_{k'}\big)\big]	- \gamma_k^{\mathrm{thr}}\sigma^2$, and $\lambda_k$ and $\mathbf{X}_k$ denote the associated Lagrange multipliers for  constraints $\widetilde{\widetilde{\mathrm{C1}}}$ and $\mathrm{C3}$, respectively. Next, we highlight the following Karush
 Kuhn Tucker (KKT) necessary optimality  conditions that the optimal solution $\mathbf{Q}_k^*$, $\lambda_k^*$, and $\mathbf{X}_k^*$
 has to meet:
 \begin{IEEEeqnarray}{lll}\label{Eq:KKT}
 	\text{K1:} \,\,\, \mathbf{Q}_k^*\succeq\boldsymbol{0}, 
 	\mathbf{X}_k^*\succeq\boldsymbol{0},
 	\lambda_k^*\geq 0, \quad
 	\text{K2:} \,\,\, \mathbf{X}_k^*\mathbf{Q}_k^*=\boldsymbol{0},
 	\lambda_k^* \boldsymbol{\Upsilon}_k(\mathbf{Q}_k^*,\forall k)=0 \nonumber \\
 	\text{K3:} \,\,\, \nabla_{\mathbf{Q}_k} \mathcal{L}(\mathbf{Q}_k,\lambda_k,\mathbf{X}_k,\forall k) = \boldsymbol{0}
 	\quad\Rightarrow\quad
 	\mathbf{X}_k^* = \mathbf{I} - \boldsymbol{\Delta}_k, \quad\,\,
 \end{IEEEeqnarray}
 where $\boldsymbol{\Delta}_k=\displaystyle\sum_{n,m,n',m'} \tilde{s}_{n,m}^{n',m'} \big(\lambda_k^*-\sum_{k'\neq k}\lambda_{k'}^*\gamma_{k'}^{\mathrm{thr}}\big){\mathbf{h}_{n,m,k}}{\mathbf{h}_{n',m',k}^{\mathsf{H}}}$. First, note that for K2 to hold, the columns of $\mathbf{Q}_k^*$ should belong to the null space of $\mathbf{X}_k^*$, which implies that $\mathrm{rank}(\mathbf{Q}_k^*)\leq N_t- \mathrm{rank}(\mathbf{X}_k^*)$. Let us rewrite $\boldsymbol{\Delta}_k=\sum_{i}\delta_i\mathbf{u}_i\mathbf{u}_i^{\mathsf{H}}$, where $\delta_i$ and $\mathbf{u}_i$ are the eigenvalues and  the corresponding eigenvectors of $\boldsymbol{\Delta}_k$, respectively. Moreover, we assume that the $\delta_i$ are arranged in a descending order and define $\delta^{\max}= \max_i \delta_i = \delta_1$. Notice that $\delta^{\max}>1$ cannot hold since  we can conclude from K3 that at least one eigenvalue of $\mathbf{X}_k^*$ is negative which contradicts condition $\mathbf{X}_k^*\succeq\boldsymbol{0}$ in K1. On the other hand, if $\delta^{\max}<1$ holds, we can conclude from K3 that $\mathbf{X}_k^*$ has full rank, i.e., $\mathrm{rank}(\mathbf{X}_k^*)=N_t$, which implies that for K2 to hold,  we must have $\mathrm{rank}(\mathbf{Q}_k^*)\leq N_t-N_t =0$, i.e., $\mathbf{Q}_j^*=\boldsymbol{0}$, which violates $\widetilde{\widetilde{\mathrm{C1}}}$ in  \eqref{Eq:ProbRelaxedrank} for $\gamma_k^{\mathrm{thr}}>0$. Finally, for $\delta^{\max}=1$, we study the general case where there are $r$ eigenvalues equal to one, i.e., $\delta_1=\dots=\delta_r=1$. For K2 to hold, we can write $\mathbf{Q}_k^* = \sum_{i=1}^r c_i\mathbf{u}_i\mathbf{u}_i^{\mathsf{H}}$, where $c_i$ are some coefficients. In general, $\mathrm{rank}(\mathbf{Q}_k^*)\leq r$; however, we show in the following that for any $\mathbf{Q}_k^*$ with rank $r>1$, we can construct another solution $\widetilde{\mathbf{Q}}_k^*=\big(\sum_{i=1}^rc_i\big) \mathbf{u}_1\mathbf{u}_1^{\mathsf{H}}$ which has rank one and does not affect the Lagrangian function, i.e., 
 \begin{IEEEeqnarray}{lll}\label{Eq:LagrangeQtilde}
 	\mathcal{L}(\mathbf{Q}_k,\lambda_k,\mathbf{X}_k,\forall k) =  
 	\sum_{k} \mathrm{tr}\big(\mathbf{Q}_k\big)
 	- \sum_{k} \mathrm{tr}\big(\mathbf{X}_k\mathbf{Q}_k\big) 
 	- \sum_{k}\lambda_k \boldsymbol{\Delta}_k(\mathbf{Q}_k,\forall k)
 	+\sum_{k}\lambda_k\gamma_k^{\mathrm{thr}}\sigma^2.
 \end{IEEEeqnarray}
We can readily confirm that $\mathrm{tr}\big(\mathbf{Q}_k^*\big)=\mathrm{tr}\big(\widetilde{\mathbf{Q}}_k^*\big)=\sum_{i=1}^rc_i$, $\mathbf{X}_k\mathbf{Q}_k^*=\mathbf{X}_k\widetilde{\mathbf{Q}}_k^*=\boldsymbol{0}$, and $\mathrm{tr}\big(\boldsymbol{\Delta}_k\mathbf{Q}_k^*\big)=\mathrm{tr}\big(\boldsymbol{\Delta}_k\widetilde{\mathbf{Q}}_k^*\big)=\sum_{i=1}^rc_i$. In summary, there always exists an optimal solution for \eqref{Eq:ProbRelaxedrank} for which  $\mathrm{rank}(\mathbf{Q}_k^*)= 1$ holds. This completes the proof.

\bibliographystyle{IEEEtran}
\bibliography{References}

\begin{thebibliography}{10}
\providecommand{\url}[1]{#1}
\csname url@samestyle\endcsname
\providecommand{\newblock}{\relax}
\providecommand{\bibinfo}[2]{#2}
\providecommand{\BIBentrySTDinterwordspacing}{\spaceskip=0pt\relax}
\providecommand{\BIBentryALTinterwordstretchfactor}{4}
\providecommand{\BIBentryALTinterwordspacing}{\spaceskip=\fontdimen2\font plus
\BIBentryALTinterwordstretchfactor\fontdimen3\font minus
  \fontdimen4\font\relax}
\providecommand{\BIBforeignlanguage}[2]{{%
\expandafter\ifx\csname l@#1\endcsname\relax
\typeout{** WARNING: IEEEtran.bst: No hyphenation pattern has been}%
\typeout{** loaded for the language `#1'. Using the pattern for}%
\typeout{** the default language instead.}%
\else
\language=\csname l@#1\endcsname
\fi
#2}}
\providecommand{\BIBdecl}{\relax}
\BIBdecl

\bibitem{najafi2020asilomar}
M.~Najafi, V.~Jamali, R.~Schober, and H.~V. Poor, ``Physics-based modeling of
  large intelligent reflecting surfaces for scalable optimization,''
  \emph{Accepted for Presentation at Asilomar Conf. Sig., Sys., and Computers},
  pp. 1--5, 2020.

\bibitem{di2019smart}
{M. D. Renzo, et. al}, ``Smart radio environments empowered by {AI}
  reconfigurable meta-surfaces: {An} idea whose time has come,'' \emph{EURASIP
  J. Wireless Commun. Netw.}, vol. 129, May 2019.

\bibitem{liaskos2019novel}
C.~Liaskos, S.~Nie, A.~Tsioliaridou, A.~Pitsillides, S.~Ioannidis, and
  I.~Akyildiz, ``A novel communication paradigm for high capacity and security
  via programmable indoor wireless environments in next generation wireless
  systems,'' \emph{Ad Hoc Netw.}, vol.~87, pp. 1--16, 2019.

\bibitem{di2020smart}
M.~Di~Renzo, A.~Zappone, M.~Debbah, M.-S. Alouini, C.~Yuen, J.~de~Rosny, and
  S.~Tretyakov, ``Smart radio environments empowered by reconfigurable
  intelligent surfaces: {How} it works, state of research, and road ahead,''
  \emph{IEEE J. Select. Areas in Commun.}, 2020.

\bibitem{huang2020holographic}
C.~{Huang}, S.~{Hu}, G.~C. {Alexandropoulos}, A.~{Zappone}, C.~{Yuen},
  R.~{Zhang}, M.~{Di Renzo}, and M.~{Debbah}, ``Holographic {MIMO} surfaces for
  {6G} wireless networks: {Opportunities}, challenges, and trends,'' \emph{IEEE
  Wireless Communications}, pp. 1--8, 2020.

\bibitem{kaina2014shaping}
N.~Kaina, M.~Dupr{\'e}, G.~Lerosey, and M.~Fink, ``Shaping complex microwave
  fields in reverberating media with binary tunable metasurfaces,''
  \emph{Scientific Reports}, vol.~4, no.~1, pp. 1--8, 2014.

\bibitem{zhu2013active}
B.~O. Zhu, J.~Zhao, and Y.~Feng, ``Active impedance metasurface with full 360
  reflection phase tuning,'' \emph{Scientific Reports}, vol.~3, p. 3059, 2013.

\bibitem{wu2019intelligent}
Q.~Wu and R.~Zhang, ``Intelligent reflecting surface enhanced wireless network
  via joint active and passive beamforming,'' \emph{IEEE Trans. Wireless
  Commun.}, 2019.

\bibitem{jamali2018scalable}
V.~Jamali, A.~M. Tulino, G.~Fischer, R.~M{\"u}ller, and R.~Schober, ``Scalable
  and energy-efficient millimeter massive {MIMO} architectures: {R}eflect-array
  and transmit-array antennas,'' in \emph{Proc. IEEE Int. Conf. Commun. (ICC)},
  May 2019, pp. 1--7.

\bibitem{tang2019wireless}
W.~Tang, M.~Z. Chen, X.~Chen, J.~Y. Dai, Y.~Han, M.~Di~Renzo, Y.~Zeng, S.~Jin,
  Q.~Cheng, and T.~J. Cui, ``Wireless communications with reconfigurable
  intelligent surface: {Path loss} modeling and experimental measurement,''
  \emph{arXiv preprint arXiv:1911.05326}, 2019.

\bibitem{yang2016programmable}
H.~Yang, X.~Cao, F.~Yang, J.~Gao, S.~Xu, M.~Li, X.~Chen, Y.~Zhao, Y.~Zheng, and
  S.~Li, ``A programmable metasurface with dynamic polarization, scattering and
  focusing control,'' \emph{Scientific Reports}, vol.~6, p. 35692, 2016.

\bibitem{dai2020reconfigurable}
L.~{Dai}, B.~{Wang}, M.~{Wang}, X.~{Yang}, J.~{Tan}, S.~{Bi}, S.~{Xu},
  F.~{Yang}, Z.~{Chen}, M.~D. {Renzo}, C.~{Chae}, and L.~{Hanzo},
  ``Reconfigurable intelligent surface-based wireless communications: {Antenna}
  design, prototyping, and experimental results,'' \emph{IEEE Access}, vol.~8,
  pp. 45\,913--45\,923, Mar. 2020.

\bibitem{yu2020robust}
X.~Yu, D.~Xu, Y.~Sun, D.~W.~K. Ng, and R.~Schober, ``Robust and secure wireless
  communications via intelligent reflecting surfaces,'' \emph{IEEE J. Select.
  Areas in Commun.}, 2020.

\bibitem{jamali2019intelligent}
V.~Jamali, A.~Tulino, G.~Fischer, R.~M{\"u}ller, and R.~Schober, ``Intelligent
  reflecting and transmitting surface aided millimeter wave massive {MIMO},''
  \emph{arXiv preprint arXiv:1902.07670}, 2019.

\bibitem{huang2019reconfigurable}
C.~Huang, A.~Zappone, G.~C. Alexandropoulos, M.~Debbah, and C.~Yuen,
  ``Reconfigurable intelligent surfaces for energy efficiency in wireless
  communication,'' \emph{IEEE Trans. Wireless Commun.}, vol.~18, no.~8, pp.
  4157--4170, Aug. 2019.

\bibitem{pan2020multicell}
C.~{Pan}, H.~{Ren}, K.~{Wang}, W.~{Xu}, M.~{Elkashlan}, A.~{Nallanathan}, and
  L.~{Hanzo}, ``Multicell {MIMO} communications relying on intelligent
  reflecting surfaces,'' \emph{IEEE Trans. Wireless Commun.}, vol.~19, no.~8,
  pp. 5218--5233, Aug. 2020.

\bibitem{guo2019weighted}
H.~Guo, Y.-C. Liang, J.~Chen, and E.~G. Larsson, ``Weighted sum-rate
  optimization for intelligent reflecting surface enhanced wireless networks,''
  in \emph{Proc. IEEE Global Commun. Conf. (Globecom)}, Dec. 2019, pp. 1--6.

\bibitem{wang2019intelligent}
P.~Wang, J.~Fang, X.~Yuan, Z.~Chen, H.~Duan, and H.~Li, ``Intelligent
  reflecting surface-assisted millimeter wave communications: {Joint} active
  and passive precoding design,'' \emph{arXiv preprint arXiv:1908.10734}, 2019.

\bibitem{ozdogan2019intelligent}
{\"O}.~{\"O}zdogan, E.~Bj{\"o}rnson, and E.~G. Larsson, ``Intelligent
  reflecting surfaces: {Physics}, propagation, and pathloss modeling,''
  \emph{{IEEE} Commun. Lett.}, 2019.

\bibitem{bjornson2020power}
E.~Bj{\"o}rnson and L.~Sanguinetti, ``Power scaling laws and near-field
  behaviors of massive {MIMO} and intelligent reflecting surfaces,'' \emph{IEEE
  Open Access J. Commun. Society}, 2020.

\bibitem{di2020analytical}
M.~{Di Renzo}, F.~{Habibi Danufane}, X.~{Xi}, J.~{de Rosny}, and
  S.~{Tretyakov}, ``Analytical modeling of the path-loss for reconfigurable
  intelligent surfaces – {Anomalous} mirror or scatterer ?'' in \emph{Proc.
  IEEE Int. Workshop Sig. Process. Advances in Wireless Commun. (SPAWC)}, 2020,
  pp. 1--5.

\bibitem{bai2020latency}
T.~Bai, C.~Pan, Y.~Deng, M.~Elkashlan, and A.~Nallanathan, ``Latency
  minimization for intelligent reflecting surface aided mobile edge
  computing,'' \emph{IEEE J. Select. Areas in Commun.}, 2019.

\bibitem{cao2019delay}
Y.~Cao, T.~Lv, Z.~Lin, W.~Ni, and N.~C.~Beaulieu, ``Delay-constrained joint
  power control, user detection and passive beamforming in intelligent
  reflecting surface assisted uplink {mmWave} system,'' \emph{arXiv preprint
  arXiv:1912.10030}, 2019.

\bibitem{balanis2012advanced}
C.~A. Balanis, \emph{Advanced Engineering Electromagnetics}.\hskip 1em plus
  0.5em minus 0.4em\relax John Wiley \& Sons, 2012.

\bibitem{salem2014manipulating}
M.~A. Salem and C.~Caloz, ``Manipulating light at distance by a metasurface
  using momentum transformation,'' \emph{Optics express}, vol.~22, no.~12, pp.
  14\,530--14\,543, 2014.

\bibitem{najafi2019intelligent}
M.~Najafi and R.~Schober, ``Intelligent reflecting surfaces for free space
  optical communications,'' in \emph{Proc. IEEE Global Commun. Conf.
  (Globecom)}, Dec. 2019, pp. 1--7.

\bibitem{lee2011mixed}
J.~Lee and S.~Leyffer, \emph{Mixed Integer Nonlinear Programming}.\hskip 1em
  plus 0.5em minus 0.4em\relax Springer Science \& Business Media, 2011, vol.
  154.

\bibitem{ng2014robust}
D.~W.~K. Ng, E.~S. Lo, and R.~Schober, ``Robust beamforming for secure
  communication in systems with wireless information and power transfer,''
  \emph{IEEE Trans. Wireless Commun.}, vol.~13, no.~8, pp. 4599--4615, Aug.
  2014.

\bibitem{ghanem2019resource}
W.~R. Ghanem, V.~Jamali, Y.~Sun, and R.~Schober, ``Resource allocation for
  multi-user downlink {URLLC-OFDMA} systems,'' in \emph{Proc. IEEE Int. Conf.
  Commun. (ICC)}, May 2019, pp. 1--6.

\bibitem{liu2019intelligent}
F.~Liu, O.~Tsilipakos, A.~Pitilakis, A.~C. Tasolamprou, M.~S. Mirmoosa, N.~V.
  Kantartzis, D.-H. Kwon, M.~Kafesaki, C.~M. Soukoulis, and S.~A. Tretyakov,
  ``Intelligent metasurfaces with continuously tunable local surface impedance
  for multiple reconfigurable functions,'' \emph{Physical Review Applied},
  vol.~11, no.~4, p. 044024, 2019.

\bibitem{abeywickrama2020intelligent}
S.~{Abeywickrama}, R.~{Zhang}, Q.~{Wu}, and C.~{Yuen}, ``Intelligent reflecting
  surface: {Practical} phase shift model and beamforming optimization,''
  \emph{IEEE Trans. Wireless Commun.}, vol.~68, no.~9, pp. 5849--5863, Sep.
  2020.

\bibitem{estakhri2016wave}
N.~M. Estakhri and A.~Al{\`u}, ``Wave-front transformation with gradient
  metasurfaces,'' \emph{Physical Review X}, vol.~6, no.~4, p. 041008, 2016.

\bibitem{asadchy2016perfect}
V.~S. Asadchy, M.~Albooyeh, S.~N. Tcvetkova, A.~D{\'\i}az-Rubio, Y.~Ra'di, and
  S.~Tretyakov, ``{Perfect control of reflection and refraction using spatially
  dispersive metasurfaces},'' \emph{Physical Review B}, vol.~94, no.~7, p.
  075142, 2016.

\bibitem{balanis2005antenna}
C.~A. Balanis, \emph{Antenna Theory: Analysis and Design}.\hskip 1em plus 0.5em
  minus 0.4em\relax John Wiley \& Sons, 2005.

\bibitem{lau2012reconfigurable}
J.~Y. Lau, ``Reconfigurable transmitarray antennas,'' Ph.D. dissertation,
  University of Toronto, 2012.

\bibitem{zhang2020joint}
Z.~Zhang and L.~Dai, ``A joint precoding framework for wideband reconfigurable
  intelligent surface-aided cell-free network,'' \emph{arXiv preprint
  arXiv:2002.03744}, 2020.

\bibitem{alkhateeb2014channel}
A.~Alkhateeb, O.~El~Ayach, G.~Leus, and R.~W. Heath, ``Channel estimation and
  hybrid precoding for millimeter wave cellular systems,'' \emph{IEEE J.
  Select. Topics Sig. Process.}, vol.~8, no.~5, pp. 831--846, Oct. 2014.

\bibitem{ghanaatian2019feedback}
R.~Ghanaatian, V.~Jamali, A.~Burg, and R.~Schober, ``Feedback-aware precoding
  for millimeter wave massive {MIMO} systems,'' in \emph{Proc. IEEE Int. Symp.
  Personal, Indoor and Mobile Radio Commun. (PIMRC)}, Sep. 2019, pp. 1--7.

\bibitem{tseng2001convergence}
P.~Tseng, ``Convergence of a block coordinate descent method for
  nondifferentiable minimization,'' \emph{J. Optimization theory and
  Applications}, vol. 109, no.~3, pp. 475--494, 2001.

\bibitem{najafi2019cran}
M.~Najafi, V.~Jamali, D.~W.~K. Ng, and R.~Schober, ``{C-RAN} with hybrid
  {RF/FSO} fronthaul links: {Joint} optimization of fronthaul compression and
  {RF} time allocation,'' \emph{IEEE Trans. Commun.}, vol.~67, no.~12, pp.
  8678--8695, Dec. 2019.

\bibitem{zheng2019intelligent}
B.~Zheng and R.~Zhang, ``Intelligent reflecting surface-enhanced {OFDM}:
  {Channel} estimation and reflection optimization,'' \emph{{IEEE} Wireless
  Commun. Lett.}, 2019.

\bibitem{alwazani2020intelligent}
H.~Alwazani \emph{et~al.}, ``Intelligent reflecting surface-assisted multi-user
  {MISO} communication: {Channel} estimation and beamforming design,''
  \emph{IEEE Open Access J. Commun. Society}, vol.~1, pp. 661--680, Jun. 2020.

\bibitem{jamali2020intelligent}
\BIBentryALTinterwordspacing
V.~Jamali, M.~Najafi, R.~Schober, and H.~V. Poor, ``Power efficiency, overhead,
  and complexity tradeoff in {IRS}-assisted communications -- {Quadratic}
  phase-shift design,'' \emph{arXiv preprint}, 2020. [Online]. Available:
  \url{https://arxiv.org/abs/2009.05956}
\BIBentrySTDinterwordspacing

\bibitem{Roth2018hybrid}
K.~{Roth}, H.~{Pirzadeh}, A.~L. {Swindlehurst}, and J.~A. {Nossek}, ``A
  comparison of hybrid beamforming and digital beamforming with low-resolution
  {ADCs} for multiple users and imperfect {CSI},'' \emph{IEEE J. Select. Areas
  in Sig. Process.}, vol.~12, no.~3, pp. 484--498, Jun. 2018.

\bibitem{jamali2016csi}
V.~{Jamali}, N.~{Waly}, N.~{Zlatanov}, and R.~{Schober}, ``Optimal buffer-aided
  relaying with imperfect {CSI},'' \emph{{IEEE} Commun. Lett.}, vol.~20, no.~7,
  pp. 1309--1312, Jul. 2016.

\bibitem{hu2017generalized}
S.~Hu and F.~Rusek, ``A generalized zero-forcing precoder with successive
  dirty-paper coding in miso broadcast channels,'' \emph{IEEE Trans. Wireless
  Commun.}, vol.~16, no.~6, pp. 3632--3645, Jun. 2017.

\bibitem{wan2020broadband}
Z.~{Wan}, Z.~{Gao}, and M.~{Alouini}, ``Broadband channel estimation for
  intelligent reflecting surface aided {mmWave} massive {MIMO} systems,'' in
  \emph{Proc. IEEE Int. Conf. Commun. (ICC)}, 2020, pp. 1--6.

\bibitem{bjornson2019intelligent}
E.~Bj{\"o}rnson, {\"O}.~{\"O}zdogan, and E.~G. Larsson, ``Intelligent
  reflecting surface vs. decode-and-forward: {How} large surfaces are needed to
  beat relaying?'' \emph{{IEEE} Wireless Commun. Lett.}, vol.~2, no.~9, pp.
  244--248, Feb. 2020.

\bibitem{boyd2004convex}
S.~Boyd and L.~Vandenberghe, \emph{Convex Optimization}.\hskip 1em plus 0.5em
  minus 0.4em\relax Cambridge University Press, 2004.

\end{thebibliography}

\end{document}